 \documentclass[journal]{IEEEtran}

\usepackage{amsmath,graphicx}
\usepackage[english]{babel}
\usepackage{caption}
\usepackage[skip=4pt]{subcaption}
\usepackage[english]{babel} 
\usepackage{amsfonts}

\usepackage{soul, color}
\usepackage{amssymb}
\usepackage{psfrag}
\usepackage{amsthm}
\usepackage{fixmath}

\usepackage{mathrsfs}
\usepackage{bm}
\usepackage{bbm} 
\usepackage{enumerate}
\usepackage{dsfont}
\usepackage{hyperref}
\hypersetup{bookmarks={false}}
\usepackage{cite}
\usepackage{mathtools}
\DeclarePairedDelimiter{\ceil}{\lceil}{\rceil}

\usepackage[thinlines]{easytable}
\newcommand{\mycolor}{\textcolor{black}} 
\newcommand{\anothercolor}{\textcolor{black}} 

\usepackage{titling}
\usepackage{authblk}
\usepackage[T1]{fontenc}

\makeatletter
\renewcommand{\emptythanks}{%
\renewcommand{\thanks}[1]{}
\renewcommand{\@thanks}{}}
\makeatother

\begin{document}

\title{Weakly-supervised Dictionary Learning}
\author{Zeyu You,~\IEEEmembership{Student member,~IEEE,}
        Raviv Raich,~\IEEEmembership{Senior Member,~IEEE,}\\
        Xiaoli Z. Fern,~\IEEEmembership{Member,~IEEE,}
        and Jinsub Kim,~\IEEEmembership{Member,~IEEE,}
\thanks{All authors are with the School
of Electrical Engineering and Computer Science, Oregon State University, Corvallis,
OR, 97331-5501, USA e-mail:\{youz,raich,xfern,kimjinsu\}@oregonstate.edu. This work is partially supported by the National Science Foundation grants CCF-1254218, DBI-1356792 and IIS-1055113.}
}
\date{\vspace{-5ex}}


\maketitle
\thispagestyle{empty}

\begin{abstract}
We present a probabilistic modeling and inference framework for discriminative analysis dictionary learning under a weak supervision setting. Dictionary learning approaches have been widely used for tasks such as low-level signal denoising and restoration as well as high-level classification tasks, which can be applied to audio and image analysis. Synthesis dictionary learning aims at jointly learning a dictionary and corresponding sparse coefficients to provide accurate data representation. This approach is useful for denoising and signal restoration, but may lead to sub-optimal classification performance. By contrast, analysis dictionary learning provides a transform that maps data to a sparse discriminative representation suitable for classification. We consider the problem of analysis dictionary learning for time-series data under a weak supervision setting in which signals are assigned with a global label instead of an instantaneous label signal. We propose a discriminative 
probabilistic model that incorporates both label information and sparsity constraints on the underlying latent instantaneous label signal using cardinality control. We present the expectation maximization (EM) procedure for maximum likelihood estimation (MLE) of the proposed model. To facilitate a computationally efficient E-step, we propose both a chain and a novel tree graph reformulation of the graphical model. The performance of the proposed model is demonstrated on both synthetic and real-world data. 
\end{abstract}


%
\IEEEpeerreviewmaketitle

\section{Introduction}
\label{sec:intro}
Synthesis dictionary learning, in which the dictionary and a corresponding sparse representation are jointly learned from data, aims at minimizing a reconstruction error. 
An over-complete synthesis dictionary learning with sparse coding is introduced in \cite{kreutz2003dictionary}. Various state-of-the-art approaches have been introduced to solve the dictionary learning problem including K-SVD~\cite{aharon2006rm}, matrix factorization~\cite{mairal2010online}, Lagrangian dual gradient descent and feature-sign search~\cite{lee2006efficient}. For audio or spectral image analysis, convolutive dictionary learning has been proposed~\cite{vipperla2011robust, wang2011online, barchiesi2011dictionary}. 
Analysis dictionary learning and transform learning offer an alternative to dictionary learning~\cite{rubinstein2013analysis, ravishankar2013learning}, which produces a sparsified outcome after applying the analysis dictionary to the original data. \mycolor{As stated in \cite{rubinstein2013analysis}, the analysis approach shows a significant advantage over synthesis and other denoising approach in terms of signal recovery for random, piecewise-constant and natural signal data. Having linear dependencies between sets of rows in the dictionary improves the recovery quality of the pursuit algorithm. The linear dependencies can be incorporated by forcing sparse and zero-mean rows during the training. Both synthesis and analysis dictionary learning are used for solving problems such as reconstruction, denoising, and sparse coding. However, without an additional supervision component, such methods have been reported to perform sub-optimally in classification tasks~\cite{gangeh2015supervised}. This is due to the fact that both approaches aim at reconstruction rather than classification. Nevertheless, both approaches can be applied to classification by modifying the objective to include a label fit term that renders the learned dictionary as discriminative as possible. A more detailed discussion on supervised learning approaches is provided in subsection~\ref{subsec:DDL}. 
} 

In this paper, we consider a weak-supervision setting for analysis dictionary learning that is suitable for  classification. In the weak-supervision setting, for each data portion containing multiple data points, a label set describing the classes present or absent is provided while the individual label of each instance remains unavailable. 
A motivating example for this setting is the problem of in-situ bio-acoustic monitoring, where audio recordings of varying signal-to-noise ratio (SNR) are collected by unattended microphones. 
Even in fairly short intervals, it is common to have multiple simultaneous vocalizations from multiple species in addition to other noise sources such as wind, rain, stream, and nearby vehicles. Consequently, intervals of audio data are associated with multiple class labels. \mycolor{Since the audio signals are multi-labeled at the interval level, the precise location and class of each pattern contained in the signal are unknown. Hence, isolating an individual pattern and assigning the appropriate label as a training example in the traditional supervised learning setting is not trivial.}

Discriminative dictionary learning approaches have been considered when the full label information is available, i.e. each data example is associated with a label (binary or multiclass) ~\cite{mairal2009supervised,mairal2012task}.  For weakly-supervised setting, learning a dictionary given a set of observed signals/images and their binary labels has been considered where non-convolutive, synthesis dictionary learning approaches are introduced~\cite{nguyen2009weakly,wang2013max,wang2015relaxed}. To the best of our knowledge, such approaches cannot be easily extended to the convolutive analysis dictionary learning setting. 

This work focuses on dictionary learning under weak supervision, where we learn a dictionary given a set of signals and their label sets. Focusing on time-series, we propose an algorithm for weakly-supervised convolutive analysis dictionary learning. 
Our contributions are as follows. (i) We develop a novel discriminative probabilistic model for analysis dictionary learning under the {\em weak-supervision setting}. (ii) We present an alternative approach for cardinality (or sparsity) constraints as implicit observations in a graphical model as opposed to commonly used norm regularization. This approach allows for localization of the patterns-of-interest. (iii) We introduce a novel framework for efficient message passing using a reformulation of the proposed graphical model both as a chain and as a tree. This reformulation yields a near-linear exact probability calculation that alleviates the need for approximate inference. \mycolor{Our preliminary work on this topic was introduced in \cite{you2016weakly}.
\footnote{\mycolor{The model in this paper provides a simplification of the model in \cite{you2016weakly} to make the new model elegant and clear. We present the derivations in the new model in detail in this paper.
Additionally, a novel tree reformulation of the graphical model (unavailable in \cite{you2016weakly}) is provided in this paper. Thirdly, due to the modeling variation, all derivations have been redone. Results on both synthetic dataset and real-world datasets are new as well.}}}




\section{Background and related work}
\label{sec:relwork}

\subsection{Generative synthesis dictionary learning}
In {\em synthesis dictionary learning}
~\cite{olshausen1997sparse,kreutz2003dictionary,aharon2006rm}, the goal is to simultaneously find a dictionary and corresponding coefficients to represent a set of $n$ signals ${\bf x}_1,{\bf x}_2,\ldots,{\bf x}_n\in \mathds{R}^m$. A dictionary is a collection of atoms ${\bf D}=[{\bf d}_1^T, \ldots, {\bf d}_K^T]^T$,  where ${\bf d}_k \in \mathds{R}^m$ is the $k$th dictionary atom (or word). The $i$th signal can be approximated by a linear combination over the dictionary ${\bf D}$ by
\[ 
{\bf x}_i \approx {\bf D} {\bm \alpha}^i = \sum_{k=1}^K {\bf d}_k {\alpha}_{k}^i,\qquad \textrm{for}~i=1,2,\ldots,n,
\] 
where  ${\bm \alpha}^i=[\alpha^i_1,\ldots,\alpha^i_K]^T$ is the coefficient vector associated with the $i$th signal.
Synthesis dictionary learning is typically formulated as an optimization problem, where the goal is to find ${\bf D}$ and sparse coefficients $\{{\bm \alpha}^i\}_{i=1}^n$ that minimize the reconstruction error. 


{\em Convolutive dictionary learning} is often considered~\cite{zibulevsky2001blind,jafari2011fast} for time invariant signals such as speech and audio. In convolutive dictionary learning, the $i$th signal ${\bf x}_i$ is assumed to be formed by combining the convolution of dictionary words ${\bf d}_1,\ldots,{\bf d}_K$, where ${\bf d}_k \in \mathds{R}^m$, with their corresponding sparse activation signals ${\bm \alpha}_1^i,\ldots,{\bm \alpha}_K^i$: 
\[ {\bf x}_i \approx \sum_{k=1}^K {\bf d}_k \ast {\bm \alpha}_{k}^i, \quad \textrm{for}~ i=1,2,\ldots n. \] 
In this approach, a signal may contain multiple time-shifted copies of the same dictionary signal, 
convolutive dictionary learning  eliminates the need to use additional dictionary words to model multiple shifts of the same dictionary word. The joint recovery of the dictionary and the sparse activation signals can be achieved by minimizing the reconstruction error subject to sparsity constraints on the activation signals.

\subsection{Analysis dictionary learning}
In {\em analysis dictionary learning}~\cite{rubinstein2013analysis, ravishankar2013learning}, given a signal ${\bf x}_i$, we are looking for an analysis dictionary ${\bf W}=[{\bf w}_1, {\bf w}_2,\ldots, {\bf w}_K] \in \mathds{R}^{m\times K}$ and an estimated noiseless signal ${\bf x}'_i$ (${\bf x}_i \approx {\bf x}'_i$) such that the resulting analyzed signal
\[{\bf W}^T{\bf x}'_i = [{\bf w}^T_1{\bf x}'_i,{\bf w}^T_2{\bf x}'_i,\ldots,{\bf w}^T_K{\bf x}'_i]^T\] is sparse.
For a noisy signal model, the analysis dictionary learning can be formulated as minimizing a quadratic error between all ${\bf x}_i$'s and ${\bf x}_i^{'}$'s subject to the resulting analyzed signals are sparse.

{\em Convolutive analysis dictionary learning} convolves the analysis dictionary ${\bf W}$ with estimated noiseless signals ${\bf x}'_1,{\bf x}'_2,\ldots, {\bf x}'_n$ so that the resulting signals ${\bf w}_1*{\bf x}'_1,\ldots,{\bf w}_K*{\bf x}'_1,\ldots,{\bf w}_1*{\bf x}'_n,\ldots, {\bf w}_K*{\bf x}'_n$ are sparse~\cite{peyre2011learning,pfister2015learning}. 

\subsection{Discriminative dictionary learning}
\label{subsec:DDL}
Several approaches have been proposed for dictionary learning in the presences of labels. For supervised dictionary learning, the following approaches have been proposed: (i) Learn one dictionary per class~\cite{varma2005statistical,varma2009statistical,wright2009robust,yang2010metaface,ramirez2010classification}; (ii) Prune large dictionaries~\cite{fulkerson2008localizing,winn2005object}; (iii) Jointly learn dictionary and classifier~\cite{mairal2009supervised,zhang2010discriminative,jiang2011learning,babagholami2013bayesian}; (iv) Embed class labels into the learning of sparse coefficients~\cite{moosmann2006fast,gangeh2013kernelized,zhang2013simultaneous,lazebnik2009supervised,yang2011fisher}, and (v) Learn a histogram of dictionary elements over signal constituents~\cite{varma2009statistical,julesz1981textons,leung2001representing,cula20043d,gangeh2011dictionary,xie2010texture,lian2010probabilistic,perronnin2008universal}.

For weakly supervised dictionary learning, several max margin based, non-convolutive, synthesis dictionary learning approaches are proposed~\cite{nguyen2009weakly,wang2013max,wang2015relaxed}.
Other approaches propose to learn a discriminative synthesis dictionary by fully exploiting visual attribute correlations rather than label priors~\cite{gao2014weakly,wu2016exploiting}. 
 In our preliminary work \cite{you2016weakly}, a convolutive analysis dictionary learning approach under weak-supervision has been proposed.  Here, we expand on the approach of \cite{you2016weakly}. We simplify the previous graphical model and re-derive all formulations. In addition, we propose a novel inference approach with a tree reformulation of the graphical model allowing for a near linear processing time of the probabilistic calculations.


\section{Problem Formulation and Solution Approach}
\label{sec:prob}
 Throughout the paper, we denote signals in lower case, e.g., $x(t)$ or $y(t)$. Similarly, we use lower case letters to represent indexes such as $i$ or $j$. For simplicity, we omit the dependence on time for signals, e.g., we also use $x$ to denote signal $x(t)$. In some cases, we use the time-evaluation operator $|_t$ to denote evaluation of a signal at time $t$, e.g., $x|_t=x(t)$. We denote vectors by boldfaced lower case, e.g., ${\bf x}$ or ${\bf y}$.  We use upper case letters to denote sets, e.g,. $Y$, or constants such as $C$. All signals in this paper are assumed to be in discrete time. Consequently, we use the convolution operator $*$ to denote discrete time convolution  $w*x|_t=\sum_{u=-\infty}^{\infty} x(t-u)w(u)$.

\begin{figure}[!t]
  \centering
  \resizebox{0.5\textwidth}{!}{
  \includegraphics{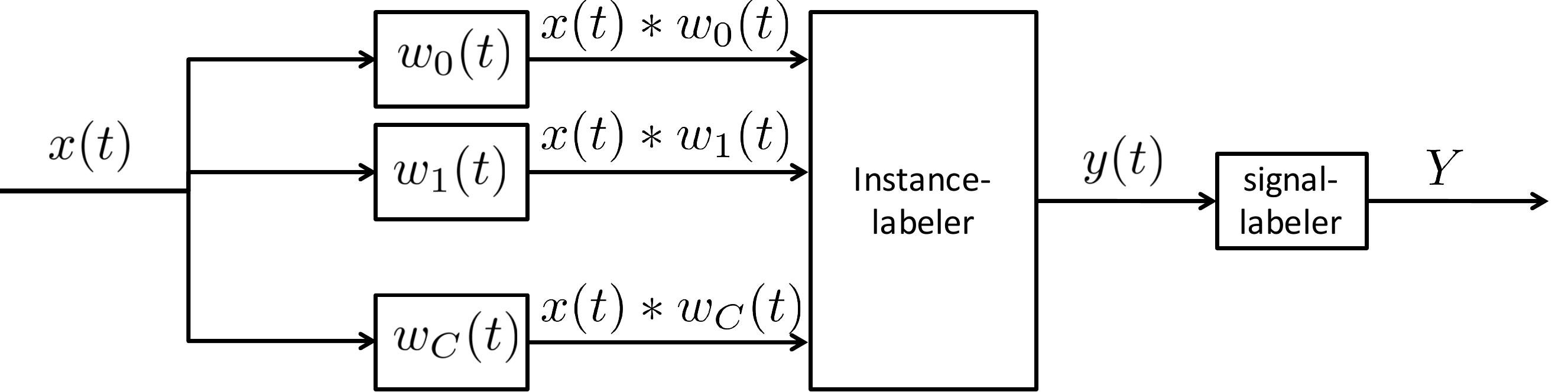}}
  \caption{An illustration of the setting of weakly supervised analysis dictionary learning}
  \label{fig:prob}
 \end{figure}
 

\subsection{Problem statement}
\label{sec:probstate} 
In the sparsity-driven setting of {\em convolutive analysis dictionary learning}, we are given a set of discrete-time signals $\{ x_1,x_2,\ldots, x_N \}$ and look for an analysis dictionary  $\{w_1,w_2,\ldots,w_C\}$ such that for each $n\in \{1,2,\ldots,N\}$, each analysis signal of $x_n$ in $\{w_1*x_n,w_2*x_n,\ldots, w_C*x_n\}$ is sparse~\cite{peyre2011learning,pfister2015learning}. Without loss of generality, we assume that the support of $x_n$ is included in $\{0,\ldots, T_n-1\}$, i.e., $x_n(t)$ $=$ $0$ if $t\notin \{0,\ldots, T_n-1\}$.  

In the {\em fully-supervised setting}, for each $x_n$, a potentially sparse time-instance label signal $y_n$ is provided and the observed data is of the form $\{ (x_1,y_1), (x_2,y_1),\ldots, (x_N,y_N) \}$. The signal $y_n$ can be viewed as a fine-grain label indicating the presence of particular  class patterns at each point in time. In this setting, location of a pattern from a given class can be used to extract examples for that class to train a classifier.

In the {\em weak supervision setting}, $y_n$ is unknown and we are interested in learning a discriminative version of the convolutive analysis dictionary given the observed data. Under this setting, every signal $x_n$ is accompanied with a single label set $Y_n$ containing the classes that are present in signal $x_n$, e.g., $Y_n=\{2,6\}$ if $x_n$ contains patterns from only classes 2 and 6. Hence the data provided in our setting is of the form $\{ (x_1,Y_1), (x_2,Y_2),\ldots, (x_N,Y_N) \}$. Our goal in this setting, is two-fold: (i) to develop a time-instance-level classifier that predicts the latent label signal $y(t)$ value for a previously unseen signal $x(t)$ based on training data ${\cal D}$; and (ii) to develop a signal-level classifier that predicts the label set $Y$ for a previously unseen signal $x(t)$ based on training data ${\cal D}$. We proceed with the proposed formulation of this problem using a probabilistic model approach.

\subsection{Probabilistic graphical model}
\label{sec:model} 
To solve the weakly-supervised dictionary learning problem, we present a probabilistic model 
for learning a discriminative convolutional dictionary. We begin by introducing the convolution used in our model and proceed with a graphical representation of the proposed discriminative convolutional dictionary learning model.

\vspace{0.2cm}
\noindent{\bf Convolutional model:}
To simplify the exposition of the model using vectors instead of signals, we convert each signal $x_n$ to a set of $T_n+T_w-1$ vectors such that each vector is a $T_w$ width windowed portion of the signal. 
For simplicity, we assume $T_w$ to be odd and denote $\Delta = (T_w-1)/2$.
This notation allows us to replace the convolution $w_c*x_n~|_{t}$  with ${\bf w}_c^T{\bf x}_{nt}$ for $t=-\Delta, -\Delta+1,\ldots,T_n-1+\Delta$ such that
\[
x_n * w_c~|_t=\sum_{\tau=-\Delta}^{\Delta} x_n(t-\tau) w_c(\tau)={\bf x}_{nt}^T{\bf w}_c,
\] 
where ${\bf x}_{nt} \in \mathds{R}^{T_w}$ is defined as
\[ {\bf x}_{nt}=[x_n(t+\Delta),x_n(t+\Delta-1), \ldots, x_n(t-\Delta)]^T,\]
and ${\bf w}_c \in \mathds{R}^{T_w}$ is given by 
\[
{\bf w}_c=[w_c(-\Delta),w_c(-\Delta+1),\ldots,w_c(\Delta)]^T.
\] 
The aforementioned one-dimensional signal model can be extended to a \textit{two-dimensional} signal model in which convolution analysis dictionary is applied only on the time dimension. In the two-dimensional setting, $x_n$ denotes $x_n(f,t)$ and the $c$-th analysis dictionary word signal $w_c$ denotes $w_c(f,t)$. Using a 2-D window with size $F\times T_w$, the analysis signal $x_n * w_c$ with the time-convolution of the two signal is given by 
\[
x_n * w_c~|_t=\sum_{f=1}^F\sum_{\tau=-\Delta}^{\Delta} x_n(f,t-\tau) w_c(f,\tau)={\bf x}_{nt}^T{\bf w}_c,
\] 
where each windowed portion of the signal is 
\[{\bf x}_{nt}=[x_n(1,t+\Delta),x(1,t+\Delta-1),\ldots,x(F,t-\Delta)]^T,\]
and 
\[{\bf w}_c=[w_c(1,-\Delta),w_c(1,-\Delta+1), \ldots,w_c(F,\Delta)]^T.\] 
While it is possible to develop a model that can handle boundary effects, such models are not time-invariant and hence may not benefit from the simplicity of the convolutional model.

\begin{figure}[!ht]
  \centering
  \includegraphics[scale=0.32]{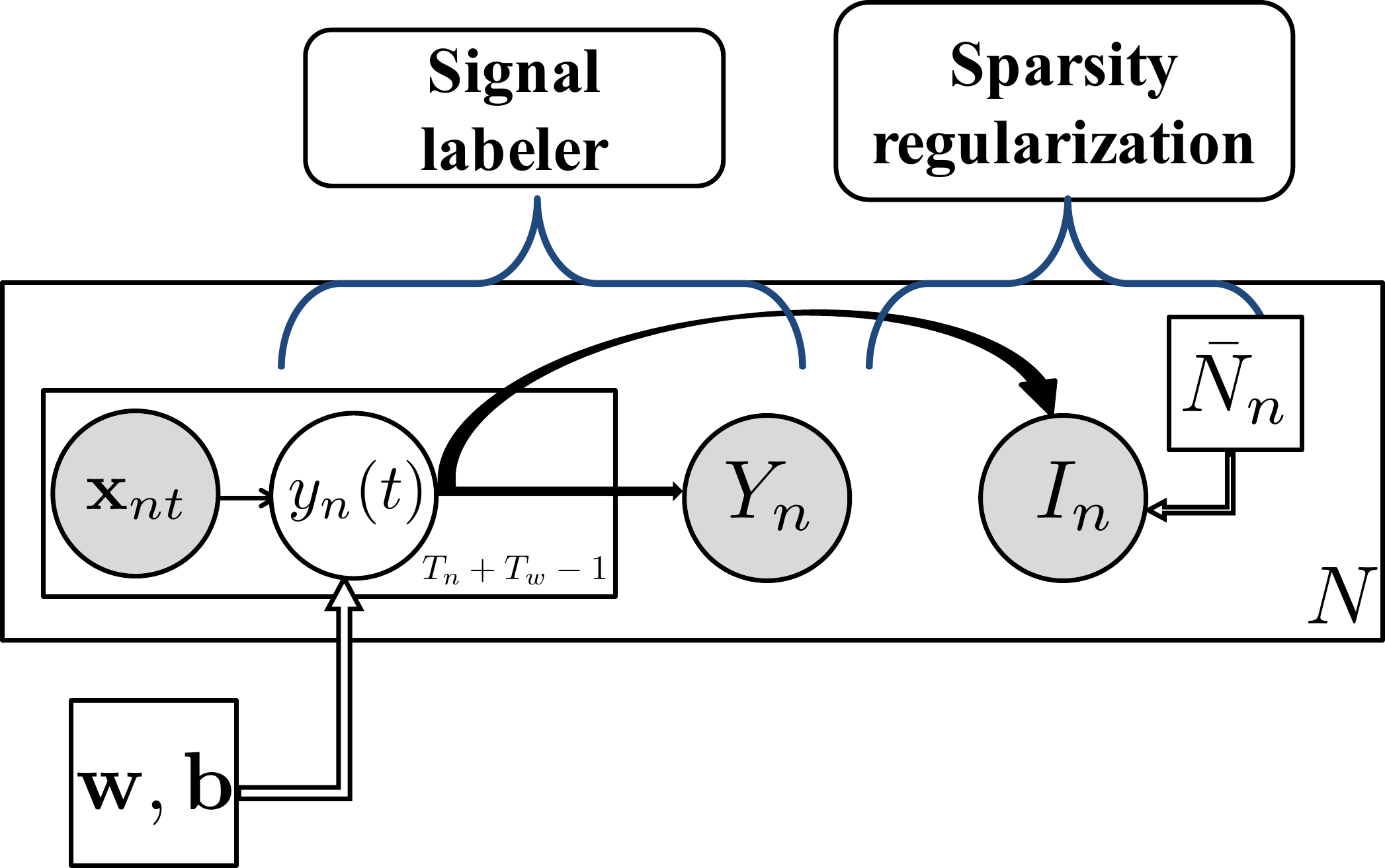}
  \caption{The proposed graphical model for WSCADL}
  \label{fig:gm}
\end{figure}

\subsubsection{Model assumptions}
To develop our model, we introduce additional assumptions to the aforementioned setting to explain the link between the analysis result and the signal label. Specifically, we assume that 

\noindent{\bf {A}.1 Convolutive instance labeler:} 
for each signal $x_n$, a hidden discrete-value label signal $y_n (t)\in \{0,1,\ldots,C\}$ is produced given only the analysis result at time $t$, i.e., the probability $P(y_n(t)=c | x_n )$ depends on signal $x_n$ only through $[w_0*x_n~|_{t},w_1*x_n~|_{t},\ldots, w_C*x_n~|_{t}]$, the analysis result evaluated at time $t$:
\begin{eqnarray*}
P(y_n(t)=c | x_n )= f_c(w_0*x_n~|_t,\ldots,w_C*x_n~|_t)
\end{eqnarray*}
for $c=0,1,\ldots,C$, where $f_c$ is an arbitrary multivariate function such that $f_c: \mathds{R}^{C+1} \to [0,1]$ and $\sum_c f_c = 1$.

\noindent{\bf {A}.2 Sparse instance label signals:}
the instance label signal $y_n$ is sparse with the number of nonzero values at most $\bar N_n$:
\[ 
\sum_{t=-\Delta}^{T_n-1+\Delta}{\mathbb I}(y_n(t)\neq 0)\leq \bar N_n.
\] 
\noindent{\bf {A}.3 Signal label union assumption:}
the signal label $Y_n$ is produced by taking the union of all the nonzero values of $y_n$
\[
Y_n=\bigcup_{\stackrel{t=-\Delta,}{y_n(t)\neq 0}}^{T_n-1+\Delta} \{ y_n(t) \}.
\]
Note that this assumption makes it is possible to have an empty signal label set in the case that all instantaneous labels are zero and no positive class is associated with any time instance. For simplicity, we remove the braces of ${y_n(t)}$ and change to $y_n(t)$ to represent a set of union of time instances. 


\subsubsection{Model description}
The probabilistic graphical model for the weakly-supervised convolutive analysis dictionary learning (WSCADL) is shown in Figure~\ref{fig:gm}, in which, our target is to estimate the model parameters ${\bf w}=[{\bf w}_0^T,{\bf w}_1^T,\ldots,{\bf w}_C^T]^T$ and ${\bf b}=[b_0,b_1,\ldots, b_C]^T$. As explained earlier, the latent label signal at time $t$ given by $y_n(t)$ depends on the entire signal $x_n$ through the convolution $w_c*x_n$ for $c=0,1,\ldots,C$ evaluated at time $t$ and hence through signal window ${\bf x}_{nt}$. The probabilistic model for $y_{n}(t)$ follows the multinomial logistic regression model given by:
\begin{equation}
P(y_{n}(t) = c | x_n; {\bf w},{\bf b}) =  \frac{e^{{\bf w}_{c}^T{\bf x}_{nt}+b_c}}{\sum_{u=0}^C  e^{{\bf w}_{u}^T{\bf x}_{nt}+b_u}},
\label{eq:prior}
\end{equation}
for $c=0,1,\ldots,C$, where ${\bf w}_c$ is the $c$th analysis word and $b_c$ is a scalar bias term for the logistic regression model.

To encode the notion of sparsity in the instance label $y_n(t)$, we introduce class $0$ following the novel class concept of \cite{pham2015multi}. To integrate a constraint on the number of non-zero instances in the $n$th signal (i.e., the sparsity of $y_n(t)$) into our probability model, 
we introduce the latent random variable $I_n$, an indicator that takes the value $1$ if the number of nonzero $y_n(t)$ is less than or equal
a sparsity upper bound $\bar N_n$ and zero otherwise. We treat $\bar N_n$ as a tuning (or hyper-) parameter of 
the graphical model. The smaller the value of $\bar N_n$, the sparser the label signal $y_n(t)$ is.
The probability model for sparsity indicator $I_n$ of label signal $y_n$ is given by
\begin{eqnarray}\label{eq:model}
 P(I_n=1 | y_n; \bar N_n) = \mathbb{I}_{({\sum_{t=-\Delta}^{T_n-1+\Delta}} \mathbb{I}(y_{n}(t)\neq 0)\leq \bar N_n)}.
\end{eqnarray}
Using this notation, the sparsity constraint can be encoded as observing $I_n=1$. 

Since the class $0$ is not represented in the signal label set, to obtain the signal label $Y_n$ from $y_{n}(-\Delta),\ldots,y_{n}(T_n-1+\Delta)$, we consider two possibilities. First, if the label signal $y_n(t)$ does not contain zeros then we expect $Y_n=\cup_t \{y_n(t)\}$. 
Alternatively, if the label signal $y_n(t)$ contains zeros then we expect $Y_n \cup \{0\}=\cup_t \{y_n(t)\}$. Consequently, the corresponding probabilistic model for the signal label $Y_n$ is given by:
\begin{eqnarray}\label{eq:bmodel}
\nonumber P(Y_{n} | y_n) = &\mathbb{I}_{(Y_{n}  = \cup_{t=-\Delta}^{T_n-1+\Delta} \{y_n(t)\})} +\\
&\mathbb{I}_{(Y_{n} \cup \{0\} = \cup_{t=-\Delta}^{T_n-1+\Delta} \{y_n(t)\})}.
\end{eqnarray}

\vspace{-0.4cm}
\subsection{Model parameter estimation}
\label{sec:paraest}
Given the parametric representation of our proposed model it is natural to consider estimating the model parameter using  a maximum likelihood estimation (MLE).
Since the model contains hidden variables, we adopt an expectation-maximization (EM) framework to  facilitate the MLE estimator. We continue with the derivation of the complete and  incomplete data likelihood. 
\subsubsection{Complete and incomplete data likelihood}
\label{sec:datalikelihood}
Define the observed data as $\mathcal{D}=\{\mathcal{X},\mathcal{Y}, I_1=1,\ldots, I_N=1\}$, 
the hidden data as $\mathcal{H}=\{y_1,\ldots, y_N\}$, 
the unknown parameters as ${\mathbold \theta}=[{\bf w}^T,{\bf b}^T]^T$, and the tuning parameters as ${\mathbold \phi}=[\bar N_1,\ldots, 
\bar N_N]^T$. According to the graphical model shown in Figure~\ref{fig:gm}, 
the complete data likelihood $P({\mathcal D}, {\mathcal H} ;  {\mathbold \theta},{\mathbold \phi})$ is computed as 
\begin{eqnarray}\label{eq:clikelihood}
\nonumber && \hspace{-0.8cm}P(\mathcal{X}) \prod_{n=1}^N  \overbrace{[\mathbb{I}_{(Y_{n}=\cup_{t=-\Delta}^{T_n-1+\Delta} \{y_n(t)\})} + \mathbb{I}_{(Y_{n}\cup \{0\} =\cup_{t=-\Delta}^{T_n-1+\Delta} \{y_n(t)\})}]}^{P(Y_n|y_n)} \\
&& \hspace{-0.5cm} \overbrace{\mathbb{I}_{(\sum_{t=-\Delta}^{T_n-1+\Delta} \mathbb{I}(y_{n}(t)\neq 0) \leq \bar N_n)}}^{P(I_n=1|y_n;\bar N_n)} \overbrace{\prod_{t=-\Delta}^{T_n-1+\Delta} P (y_n(t) | x_{n};{\bf w},{\bf b})}^{P(y_n|x_n;{\bf w},{\bf b})}.
\end{eqnarray}
The incomplete data likelihood is calculated by marginalizing out the hidden variables as
\begin{eqnarray}\label{eq:likelihood}
 \nonumber L({\mathbold \theta}) \hspace{-0.3cm}&=&\hspace{-0.5cm} \sum_{ y_{1}(-\Delta)=0}^C \hspace{-0.2cm} \ldots \hspace{-0.2cm} \sum_{ y_{1}(T_1+\Delta)=0}^C \sum_{y_{2}(-\Delta)=0}^C \hspace{-0.2cm}\ldots\hspace{-0.2cm} \sum_{ y_{2}(T_2+\Delta)=0}^C \\
  && \ldots \sum_{ y_{N}(-\Delta)=0}^C \hspace{-0.2cm}\ldots\hspace{-0.2cm} \sum_{y_{N}(T_N+\Delta)=0}^C P({\mathcal D}, {\mathcal H} ; {\mathbold \theta},{\mathbold \phi}) 
\end{eqnarray}
Taking the natural logarithm of (\ref{eq:likelihood}) and plugging in the probability with (\ref{eq:clikelihood}) produces the incomplete log-likelihood:
\begin{eqnarray}
  \nonumber   l({\mathbold \theta}) \hspace{-0.3cm}&=&\hspace{-0.3cm}\log P({\mathcal X})+\sum_{n=1}^N \log \bigl (\sum_{ y_{n}(-\Delta)=0}^C \ldots \sum_{y_{n}(T_n-1+\Delta)=0}^C  \\
  \nonumber & &\hspace{-0.6cm}[\mathbb{I}_{(Y_{n}=\cup_{t=-\Delta}^{T_n-1+\Delta} \{y_n(t)\})} + \mathbb{I}_{(Y_{n}\cup \{0\} =\cup_{t=-\Delta}^{T_n-1+\Delta} \{y_n(t)\})}]\\
& &\hspace{-0.6cm}  \mathbb{I}_{(\sum_{t=-\Delta}^{T_n-1+\Delta}\mathbb{I}(y_{n}(t)\neq 0) \leq \bar N_n)} \hspace{-0.2cm}\prod_{t=-\Delta}^{T_n-1+\Delta} \hspace{-0.2cm} P (y_n(t) | x_{n};{\bf w},{\bf b})\bigr ).\hspace{0.2cm}
\end{eqnarray}
Calculating the incomplete data likelihood in (\ref{eq:likelihood}) involves enumerating all possible instance labels, which is computationally intractable especially when the number of instance per signal is large.

\begin{figure*}[t]
\centering
\begin{subfigure}[b]{0.9\columnwidth}
\centering
\resizebox{0.8\textwidth}{!}{
\includegraphics[width=4cm]{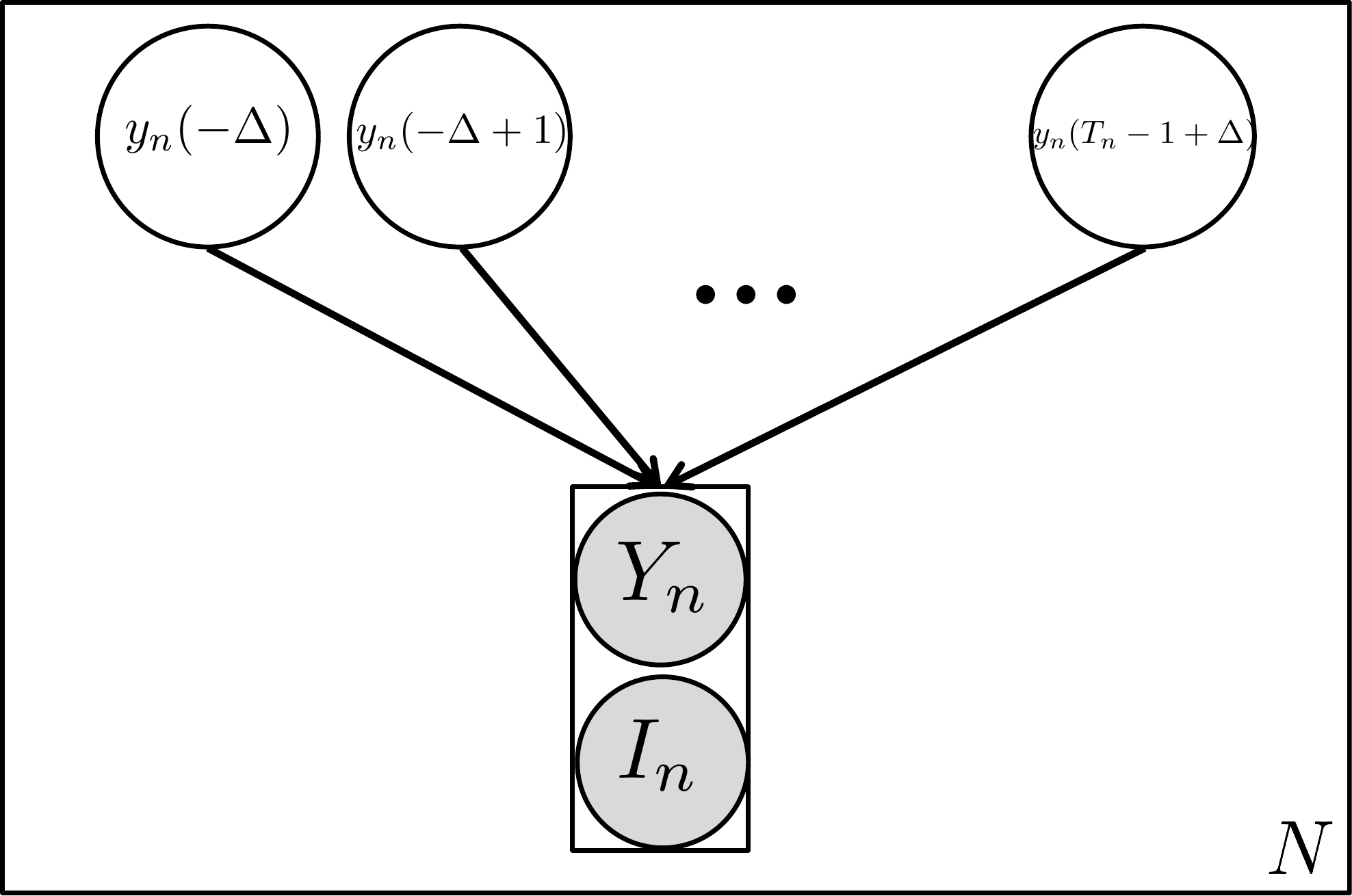}}
\caption{Original model}\medskip
\end{subfigure}
\begin{subfigure}[b]{0.96\columnwidth}
\centering
\resizebox{0.8\textwidth}{!}{
\includegraphics[width=4cm]{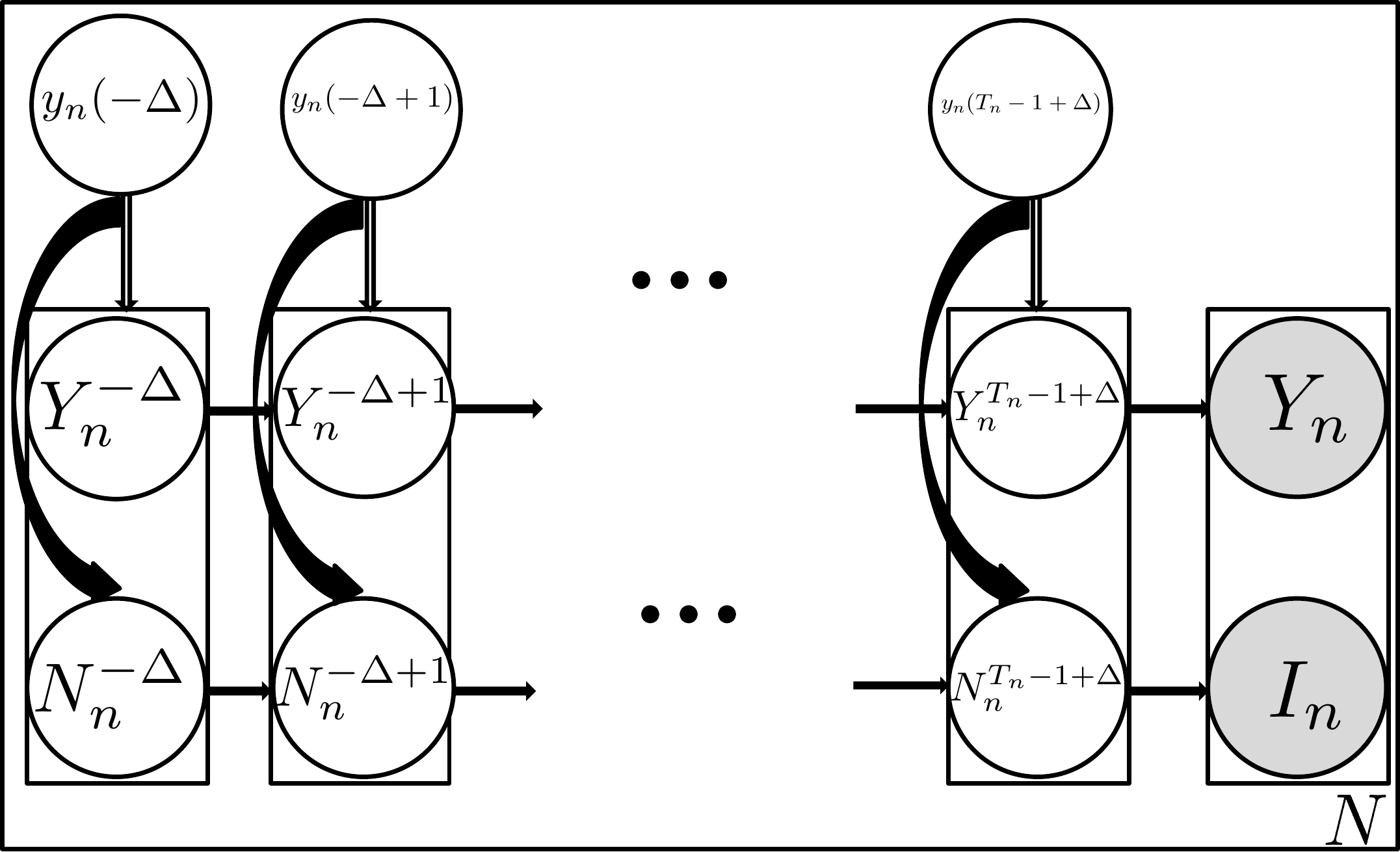}}
\caption{Graphical chain model}\medskip
\end{subfigure}
\caption{The label portion of the proposed graphical model (a) and its reformulation as a chain (b).}
\label{fig:chainb}
\end{figure*}

\subsubsection{Expectation maximization}
\label{sec:solution}
Exact inference which computes the likelihood in a brute force manner by marginalizing over all instance value combinations is intractable. To resolve this, we consider an expectation maximization (EM) approach \cite{moon1996expectation}, where the proposed approach is very similar to the one in \cite{pham2015multi}. Specifically, the EM algorithm alternated between the expectation over the hidden variable and maximization of the auxiliary function as the following two steps:
\begin{itemize}
 \item {\bf E-step}: Compute $Q({\mathbold \theta}, {\mathbold \theta}^{i})= E_{{\mathcal H}|{\mathcal D};{\mathbold \theta}^i}[\log P({\mathcal H},{\mathcal D};{\mathbold \theta})]$.
\item  {\bf M-step}: ${\mathbold \theta}^{i+1} = \arg\max_{\mathbold \theta} Q({\mathbold \theta}, {\mathbold \theta}^{i})$
\end{itemize}
The auxiliary function for the proposed model is given by 
\begin{eqnarray*}\label{eq:auxiliary}
 &&\hspace{-0.8cm} Q({\mathbold \theta}, {\mathbold \theta}^{i})= \sum_{n=1}^N\sum_{t=-\Delta}^{T_n-1+\Delta} [\sum_{c=0}^C P(y_n(t)=c |\mathcal{D}; \bar N_n, {\mathbold \theta}^{i}) \cdot\\
&& ({\bf w}_{c}^T{\bf x}_{nt}+b_c)- \log (\sum_{u=0}^C e^{{\bf w}_{u}^T{\bf x}_{n t}+b_u})] + const.
\end{eqnarray*}
The derivation of the auxiliary function for our model is explained in Appendix~\ref{append:aux}.
The maximization of the auxiliary function $Q({\mathbold \theta}, {\mathbold \theta}^{i})$ provides an update rule for both dictionary words ${\bf w}$ and bias terms ${\bf b}$ with a learning rate $\gamma$: 
\begin{eqnarray*}\label{eq:mstep}
\nonumber {\bf w}_{c}^{i+1} &=& {\bf w}_{c}^{i} + \gamma \frac{\partial {Q}({\mathbold \theta}, {\mathbold \theta}^{i})}{\partial {\bf w}_{c}} \mid_{{\mathbold \theta}={\mathbold \theta}^{i}}, \\
b_{c}^{i+1} &=& b_{c}^{i} + \gamma \frac{\partial {Q}({\mathbold \theta}, {\mathbold \theta}^{i})}{\partial b_{c}} \mid_{{\mathbold \theta}={\mathbold \theta}^{i}},
\end{eqnarray*}
for $c=0,1,\ldots,C$, where
\begin{eqnarray}\label{eq:Q:deriv}
 \nonumber \frac{\partial Q({\mathbold \theta}, {\mathbold \theta}^{i})}{\partial {\bf w}_{c}} \hspace{-0.3cm}&=& \hspace{-0.3cm} \sum_{n=1}^N\sum_{t=-\Delta}^{T_n-1+\Delta}[P(y_n(t)=c |Y_n, x_n,I_n; \bar N_n,{\mathbold \theta}^{i})\\
 && ~~~~- 
P(y_n(t)=c |x_n; {\bf w},{\bf b}) 
 ]{\bf x}_{nt},
\end{eqnarray}
\begin{eqnarray}\label{eq:Q:derivb}
\text{and}~\nonumber \frac{\partial Q({\mathbold \theta}, {\mathbold \theta}^{i})}{\partial b_{c}} \hspace{-0.3cm}&=& \hspace{-0.3cm} \sum_{n=1}^N\sum_{t=-\Delta}^{T_n-1+\Delta}[P(y_n(t)=c |Y_n, x_n,I_n; \bar N_n,{\mathbold \theta}^{i})\\
 && ~~~~-
P(y_n(t)=c |x_n; {\bf w},{\bf b})].
\end{eqnarray}
The term $P(y_n(t)=c |x_n; {\bf w},{\bf b})$ in (\ref{eq:Q:deriv}) and (\ref{eq:Q:derivb}), which is calculated using (\ref{eq:prior}), is regarded as a prior probability of the instance label, i.e., without any information about the signal label or a sparsity constraint. The term $P(y_n(t)=c |Y_n, x_n,I_n; \bar N_n,{\mathbold \theta}^{i})$ can be viewed as a posterior instance label probability that takes into account the signal label and the sparsity constraint.
Denote the difference between the posterior probability of $y_n(t)$ and its prior by $a_{nc}(t)=P(y_n(t)=c |Y_n,x_n,I_n; \bar N_n,{\mathbold \theta}^{i})-P(y_n(t)=c |x_n; {\bf w},{\bf b})$.

The gradient calculation w.r.t. ${\bf w}_c$ in (\ref{eq:Q:deriv}) 
is performed as a convolution between $a_{nc}(t)$ and the time-reversed signal $x_n(-t)$ such that 
\begin{eqnarray*}
\frac{\partial Q({\mathbold \theta}, {\mathbold \theta}^{i})}{\partial w_{c}(t)}&=&\sum_{n=1}^N\sum_{\tau=0}^{T_n-1}a_{nc}(t+\tau)x_n(\tau)\\
&=&\sum_{n=1}^Na_{nc}(t)*x_n(-t)
\end{eqnarray*}
for $t=-\Delta,-\Delta+1,\ldots,\Delta$. 
When both signal ${\bf x}_n$ and kernel $w_c$ are 2-D, the gradient step in (\ref{eq:Q:deriv}) is
\begin{eqnarray*}
\frac{\partial Q({\mathbold \theta}, {\mathbold \theta}^{i})}{\partial w_{c}(f,t)}&=&\sum_{n=1}^N\sum_{\tau=0}^{T_n-1}a_{nc}(t+\tau)x_n(f,\tau)\\
&=&\sum_{n=1}^Na_{nc}(t)*x_n(f,-t)
\end{eqnarray*}
for $f=1,2,\ldots, F$ and $t=-\Delta,-\Delta+1,\ldots,\Delta$.

\vspace{0.2cm}
\noindent {\bf Regularization:}
To guarantee the boundedness of the solution, we add an $L_2$-regularization term $-\lambda_r/2 \sum_{c=0}^C \|{\bf w}_c\|^2$ in the M-step.

\subsubsection{Key challenge}
\label{sec:challenge}
The computation of the gradient in (\ref{eq:Q:deriv}) and (\ref{eq:Q:derivb}) involves the computation of the posterior probability $P(y_n(t)=c |Y_n, x_n,I_n; \bar N_n,{\mathbold \theta}^{i})$ for each $y_{n}$ signal. This term presents one of the challenges of EM inference for the proposed model. Brute-force calculation requires marginalization over all other instance level labels, i.e., $y_{n}(s)$ for $s \neq t$. This marginalization is exponential in the number of instances per signal and hence makes the brute-form calculation prohibitive. In the following, we present the proposed efficient approach for calculating the posterior instance level label probability.

\section{Graphical model reformulation for the E-step}
\label{sec:reform}
The goal of the E-step is to obtain the posterior probability $P(y_n(t)=c |Y_n,x_n, I_n; \bar N_n, {\mathbold \theta}^i)$, which first requires the calculation of the prior probability. To efficiently compute the prior probability $P(y_n(t)|x_n;{\bf w},{\bf b})$ as a function of $t$ for each signal $x_n$, $C+1$ convolutions of the form $w_{c} * x_n$ are performed to obtain the values of ${\bf w}_{c}^T {\bf x}_{nt}$ in (\ref{eq:prior}) for $t=-\Delta,-\Delta+1,\ldots, T_n-1+\Delta$.  Under some settings, the fast Fourier transform (FFT) and its inverse can be used as a computationally efficient implementation of the convolution. We proceed with two efficient procedures for calculating the posterior probability of  $y_n(t)$ given the prior probabilities.

\begin{figure*}[ht]
\centering
\begin{subfigure}[b]{0.96\columnwidth}
\centering
\resizebox{!}{0.18\textheight}{
\includegraphics{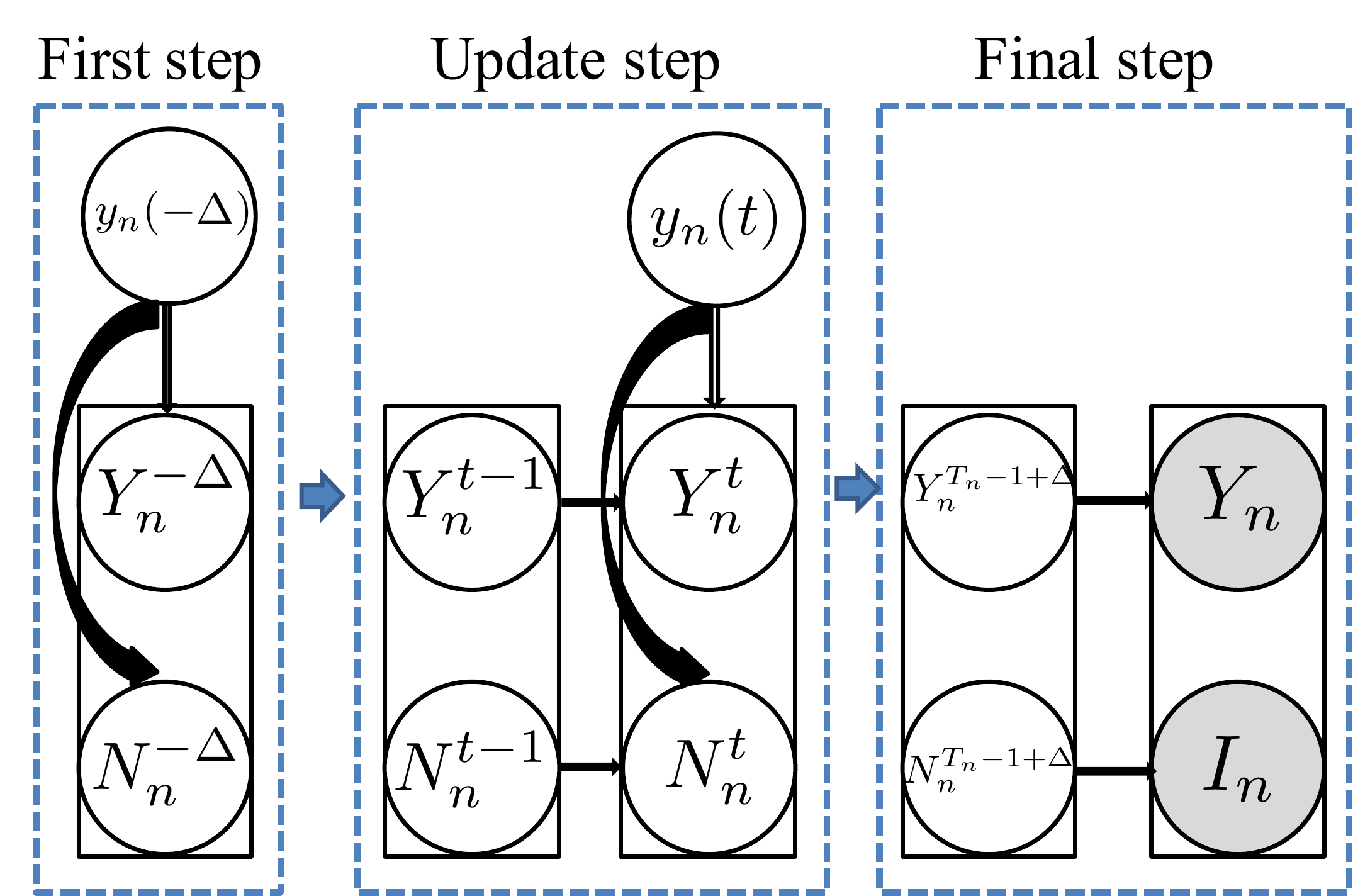}}
\caption{Forward algorithm}\medskip
\end{subfigure}
%
\begin{subfigure}[b]{0.96\columnwidth}
\centering
\resizebox{!}{0.18\textheight}{
\includegraphics{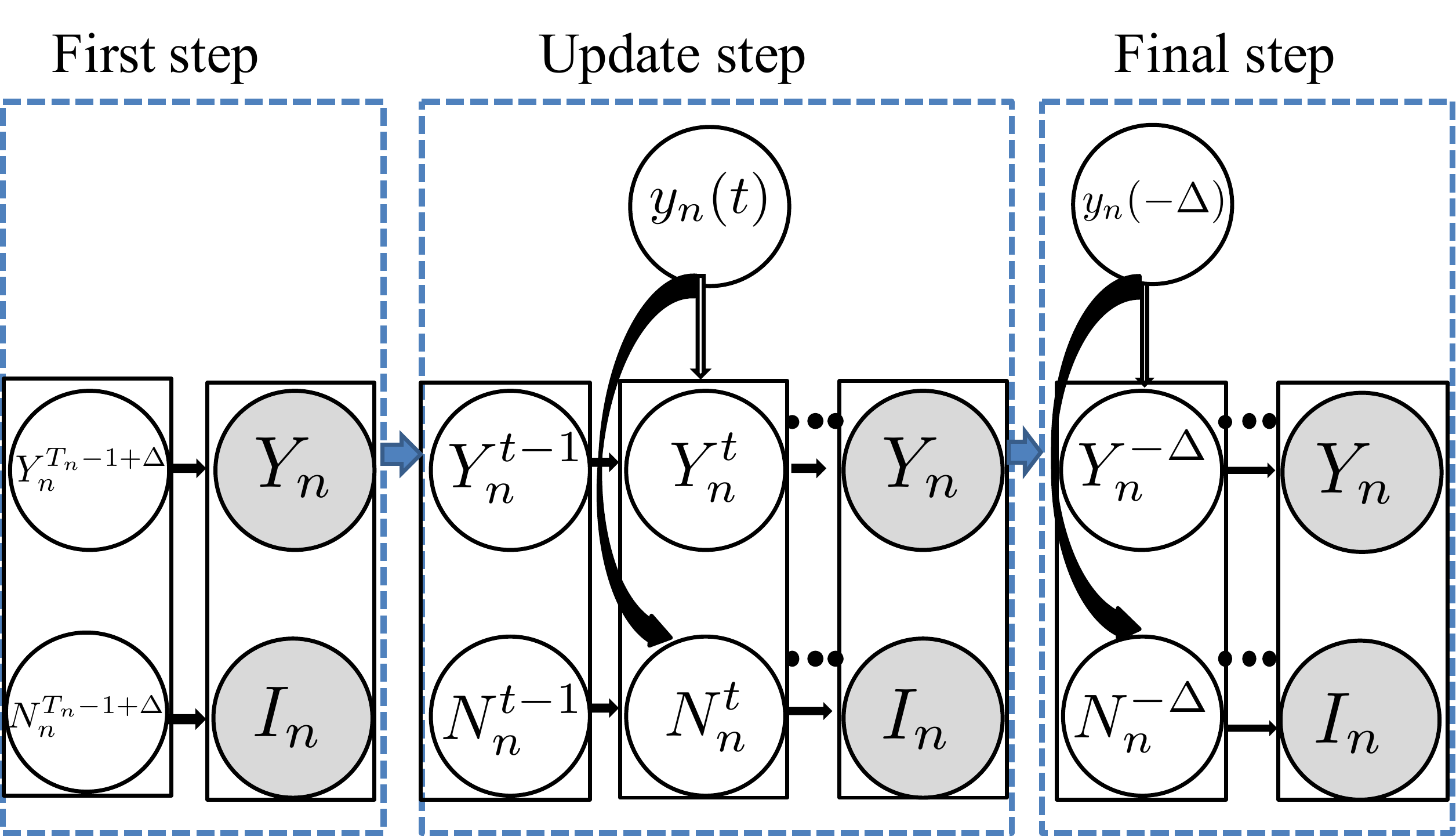}}
\caption{Backward algorithm}\medskip
\end{subfigure}
\caption{Graphical illustration of the chain forward and backward message passing routines}
\label{fig:chain1inf}
\end{figure*}

\subsection{Chain model reformulation}
Consider the label portion of the original graphical model in Figure~\ref{fig:chainb}(a). Enumerating over the set of all possible values of $(y_{n}(-\Delta),\ldots,y_{n}(T_n-1+\Delta))$ to compute the posterior is exponential with respect to the number of time instances. The v-structure graph in Figure~\ref{fig:chainb}(a) does not offer an immediate efficient approach for computing the posterior. Hence, we propose a reformulation of the model as follows. We denote
\[Y_n^{t}=\bigcup_{\stackrel{i=-\Delta,}{y_n(i)\neq 0}}^{t}  \{y_{n}(i)\},~~ N_n^t=\sum_{i=-\Delta}^{t} \mathbb{I}(y_{n}(i)\neq 0)\] 
as the label set and the number of non-zero class instances of the first $t$ instances in $n$th signal.
Both of the aforementioned newly introduced variables follow a recursive rule 
\begin{eqnarray}
Y_n^{t+1} & = & Y_n^{t} \cup \{y_{n}(t+1)~|~y_{n}(t+1)\neq 0\} \label{eq:chainY} \\
N_n^{t+1} & = & N_n^{t} + {\mathbb I}( y_{n}(t+1) \neq 0)\label{eq:chainN} 
\end{eqnarray}
for $t=-\Delta,\ldots,T_n-1+\Delta-1$. 
This reformulation gives rise to the chain model in Figure~\ref{fig:chainb}(b). We proceed with a linear time procedure for calculating the posterior probabilities that takes advantage of the reformulation of our model as a chain. 

\vspace{0.2cm}
\noindent {\bf Forward and backward message passing on the chain:}

Given the prior probability of $y_n(t)=c$ for $c \in \{0,1,\ldots,C\}$ and $t=-\Delta,\ldots,T_n-1+\Delta$, our goal is to obtain the posterior probability of $y_n(t)=c$. This can be done by first computing the joint probability defined by $P_{ntc}=P(y_n(t)=c,Y_n ,I_n|x_{n};\bar N_n,{\mathbold \theta}^i)$ and then applying Bayes rule as
\begin{equation}\label{eq:bposterior}
 P(y_n(t)=c|y_n,x_n, I_n;\bar N_n,{\mathbold \theta}^i) = P_{ntc} / \textstyle \sum_{c=0}^C P_{ntc}.
\end{equation} 
We compute the joint probabilities $P_{ntc}$ 
using a dynamic programming approach that is summarized in the following three steps:

\vspace{0.2cm}
\noindent {\bf Step 1: Forward message passing.} In this step, the goal is to compute the joint state probability $P(Y^{t}_n,N^t_n|x_n;{\mathbold \theta}^i)$ for $t=-\Delta,\ldots,T_n-1+\Delta$. Denote an element in the power set of all class labels in the $n$th signal by ${\mathbb L}\in 2^{Y_n}$. A forward message is defined as 
\[
\alpha_t({\mathbb L},l)\triangleq P(Y_n^{t}={\mathbb L},N_n^t=l|x_n; {\mathbold \theta}^i).
\]
The first message is initialized as Figure~\ref{fig:chain1inf}(a) first step shows

\hspace{-0.3cm}$\alpha_1({\mathbb L},l)=$
\begin{equation*}
  \begin{cases}
    P(y_{n}(-\Delta)=0|x_{n};{\bf w}^i,{\bf b}^i), & l=0, {\mathbb L}=\{0\};\\
    P(y_{n}(-\Delta)=c|x_{n};{\bf w}^i,{\bf b}^i), & l=1, {\mathbb L}=\{c\},c\in Y_n;\\
    0, & Otherwise,
  \end{cases}
\end{equation*} 
The update equation for the forward message of the $t$th instance is calculated by marginalizing over the ($t-1$)th state and the $t$th instance label as Figure~\ref{fig:chain1inf}(a) update step shows: 
\begin{eqnarray}\label{eq:chainstep1}
\nonumber  \alpha_t({\mathbb L},l)\hspace{-0.3cm}&=&\hspace{-0.3cm}\alpha_{t-1}({\mathbb L},l)P(y_n(t)=0|x_n;{\bf w}^i,{\bf b}^i)\\
 \nonumber &+&\hspace{-0.3cm}{\mathbb I}_{(l \neq 0)} \sum_{c=1}^C P(y_n(t)=c|x_n;{\bf w}^i,{\bf b}^i) \\
 &\cdot&\hspace{-0.3cm} [\alpha_{t-1}({\mathbb L},l-1) + {\mathbb I}_{(c\in {\mathbb L})}\alpha_{t-1}({\mathbb L}_{\setminus c},l)].
\end{eqnarray}
In the final step, $P(Y_n={\mathbb L}, I_n=1|x_n;{\mathbold \theta}^i)=\sum_{l=1}^{\bar N_n}\alpha_{T_n-1+\Delta}({\mathbb L},l)$.

\vspace{0.2cm}
\noindent {\bf Step 2: Backward message passing.} In this step, the goal is to compute the conditional joint state probability defined as $P(Y_n, I_n=1 |Y^{t}_n,N^t_n, x_n;{\mathbold \theta}^i,\bar N_n)$. We denote a backward message as 
\[
\beta_t({\mathbb L}, l)\triangleq P(Y_n, I_n=1 | Y_n^{t}={\mathbb L},N_n^t=l, x_n; {\mathbold \theta}^i,\bar N_n).
\]
According to the graphical model in Figure~\ref{fig:chain1inf}(b) such that $Y_n, I_n$ is only dependent on $Y_n^{T_n-1+\Delta}, N_n^{T_n-1+\Delta}$,
the first backward message is initialized as 
\[\beta_{T_n-1+\Delta}({\mathbb L}, l)={\mathbb I}_{(l \leq \bar N_n)} {\mathbb I}_{({\mathbb L} = Y_n)}.\] 
The update equation for the $t-1$th backward message is calculated by marginalizing over the  $t$th backward message as Figure~\ref{fig:chain1inf}(b) update step shows $ \beta_{t-1}({\mathbb L},l)=$
\begin{equation}\label{eq:chainback1}
 \hspace{-0.2cm}\sum_{c=0}^C \beta_t({\mathbb L} \cup \{c\neq 0\}, l+{\mathbb I}_{(c\neq 0)})P(y_n(t)=c | x_{n},{\bf w}^i,{\bf b}^i).
\end{equation}
\noindent {\bf Properties:} To understand the range that should be used in computing the joint probability, we examine the values for which the forward and the backward messages are non-zero. The forward and backward messages for $t=-\Delta,\ldots,T_n-1+\Delta$ have the following properties: (i) $\alpha_t({\mathbb L},l)=0$ for $l\geq t+1$,  
(ii) $\beta_t({\mathbb L},l)=0$ for $l > \bar N_n$, ${\mathbb L}\notin 2^{Y_n}$. Where (i) is from the definition of $N_n^t$ in (\ref{eq:chainN}) and (ii) is from the sparsity constraint in (\ref{eq:model}) and the definition of $Y_n^t$ in (\ref{eq:chainY}) such that each $Y_n^{t-1}\subseteq Y_n^{t}$, and $Y_n^{T_n-1+\Delta}={\mathbb L}$.

\vspace{0.2cm}
\noindent {\bf Step 3: Joint probability.}
Finally, the numerator of (\ref{eq:bposterior}) for $t=-\Delta,\ldots,T_n-1+\Delta$ is computed using all of the forward messages and the backward messages as
\begin{eqnarray}\label{eq:joint1}
  \nonumber && \hspace{-0.8cm}P( y_{n}(t)=c, Y_n, I_n =1| x_n; {\mathbold \theta}^i,\bar N_n)=p(y_n(t)=c|x_n;{\bf w}^i,{\bf b}^i)\\
&&~~~\cdot \sum_{\mathbb L\in 2^{Y_n}} \sum_{l=0}^{\bar N_n^*} \beta_t({\mathbb L} \cup \{c\neq 0\}, l+{\mathbb I}_{(c\neq 0)}) \alpha_{t-1}({\mathbb L}, l)),\hspace{0.6cm}
\end{eqnarray}
where $\bar N_n^*=\min (\bar N_n-{\mathbb I}_{(c\neq 0)}, t)$.
Since $Y_n^{-\Delta}, N_n^{-\Delta}$ is only dependent on the first instance $y_{n}(-\Delta)$ as Figure~\ref{fig:chain1inf}(b) shows, 
\begin{eqnarray*}
&\hspace{-2.5cm}P(y_{n}(-\Delta)=c,Y_n, I_n=1| x_n, \bar N_n,{\mathbold \theta}^i)=\\
&\beta_{-\Delta}(\{c\neq 0\},{\mathbb I}_{(c\neq 0)})p(y_{n}(1)=c|x_{n};{\bf w}^i,{\bf b}^i).
\end{eqnarray*}

\vspace{0.1cm}
\noindent {\bf Note:} Based on property (i) that $\alpha_{t-1}({\mathbb L}, l)=0$ when $l>t$, and property (ii) that $\beta_t({\mathbb L}, l)=0$ when $l>\bar N_n$, the effective calculation and actual need of storing both forward and backward message is for $0\leq l \leq \min(\bar N_n, t)$.

\begin{figure}[ht]
\centering
%
%
\begin{subfigure}[b]{0.96\columnwidth}
\centering
\resizebox{0.8\textwidth}{!}{
\includegraphics{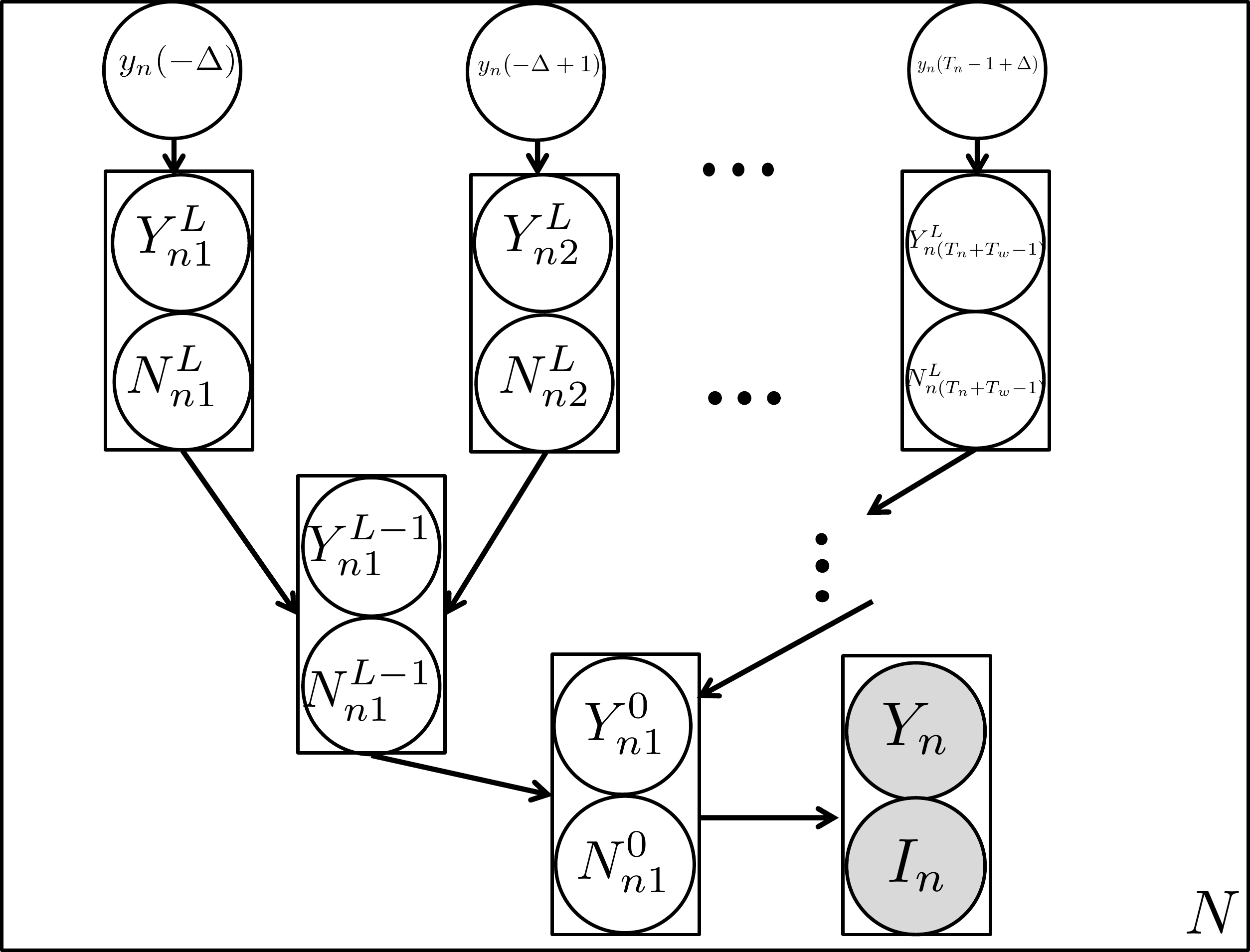}}
\end{subfigure}
\caption{Graphical model reformulation as a tree}
\label{fig:treeb}
\end{figure}

\begin{figure*}[t]
\centering
\begin{subfigure}[b]{0.96\columnwidth}
\centering
\resizebox{!}{0.18\textheight}{
\includegraphics[width=4cm]{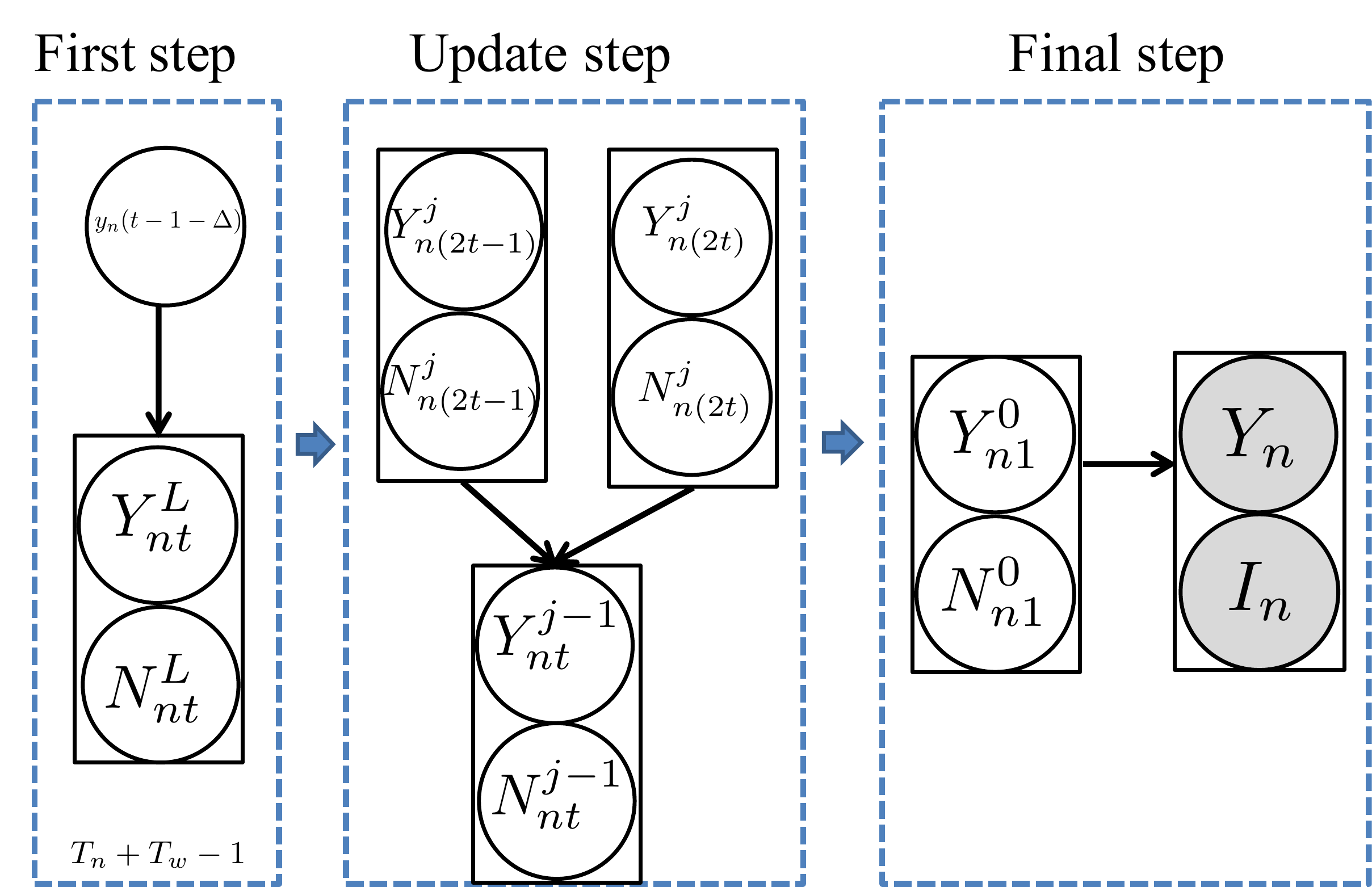}}
\caption{Forward algorithm}\medskip
\end{subfigure}
%
\begin{subfigure}[b]{0.96\columnwidth}
\centering
\resizebox{!}{0.18\textheight}{
\includegraphics[width=4cm]{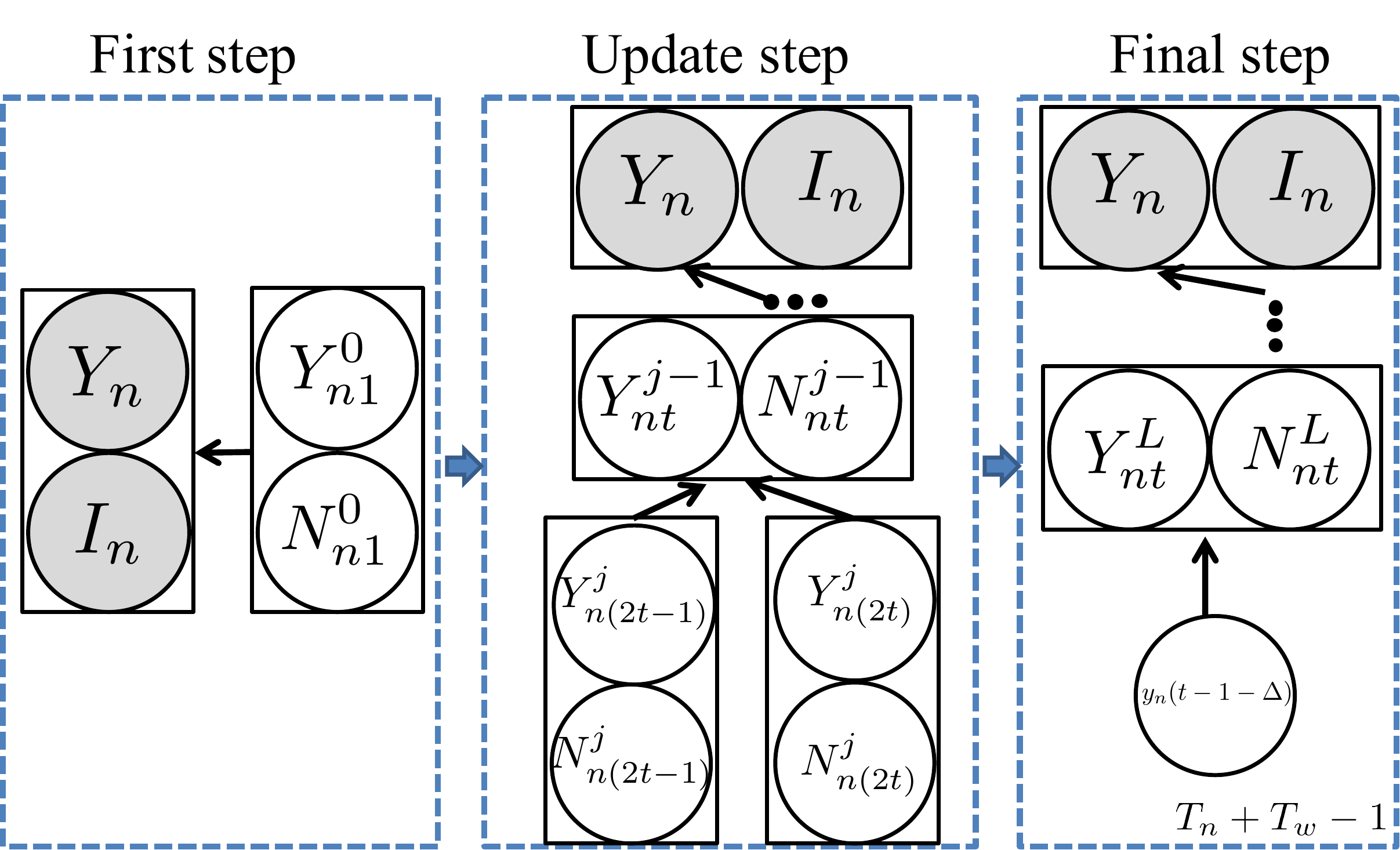}}
\caption{Backward algorithm}\medskip
\end{subfigure}
\caption{Graphical illustration of the tree forward and backward message passing routines}
\label{fig:treefb}
\end{figure*}

\subsection{Tree model reformulation}
When both the cardinality constraints $\bar N_n$ and size of signal $T_n$ are large, chain model reformulation become computational-wise inefficient, since the complexity for both time and space grows as $\bar N_n \times T_n$ increases. Instead, we propose a complete full binary tree (denoted as ${\mathcal T}(S^j_{nt},j)$ with depth $L+1$, where $j$ indicates the tree level, $S^j_{nt}$ is the node variable at index $t$ in $j$th level and $L=\ceil {\log_2  (T_n+T_w-1)}$) graph structure reformulation of the original graphical model in Figure~\ref{fig:chainb}(a) to make the E-step calculation more efficient. 

In the complete full binary tree structure, each node of the tree $S^j_t$ is considered as the joint state node ($Y_{nt}^{j}$ and $N_{nt}^{j}$). We denote $Y_{nt}^{j}$ as the label set of all ancestors of node $t$ in level $j$ of the $n$th tree and $N_{nt}^j$ as the number of non-zero class instances of all ancestors of node $t$ in level $j$ of the $n$th tree. We present the recursive relation  At the leaf's level, we assign the values as:
\begin{eqnarray*}
Y_{nt}^{L}&=&\{y_n(t-\Delta-1)|y_n(t-\Delta-1)\neq 0\},\\ N_{nt}^{L}&=&{\mathbb I}_{(y_n(t-\Delta-1)\neq 0)}
\end{eqnarray*}
for $t=1,2,\ldots,T_n+T_w-1$, and $Y_{nt}^{L}=\emptyset, ~N_{nt}^{L}=0$
for $t=T_n+T_w,T_n+T_w+1,\ldots,2^L$,
The relationship between the child node and its left parent node and its right parent node is using the following recursive formula:
\begin{eqnarray}
Y_{nt}^{j-1} & = & Y_{n(2t-1)}^{j} \cup Y_{n(2t)}^{j} \label{eq:treeY}\\
N_{nt}^{j-1} & = & N_{n(2t-1)}^{j}+N_{n(2t)}^{j}\label{eq:treeN}
\end{eqnarray}
for $t=1,2,\ldots, 2^{j-1}$. This reformulation gives rise to the tree model in Figure~\ref{fig:treeb}. Note that $Y_{n1}^0=\bigcup_t\{y_n(t)|y_n(t)\neq 0\}=Y_n$ and $N_{n1}^0=\sum_t I_{(y_n(t) \neq 0)}$ which is used to determine $I_n$.

\vspace{0.2cm}
\noindent {\bf Forward and backward message passing on the tree:}

In the tree inference, the target is the same as the chain inference as to compute the posterior probability of $y_n(t)=c$ for $c \in \{0,1,\ldots,C\}$ and $t=-\Delta,\ldots,T_n-1+\Delta$.
Using a dynamic programming approach, the joint probabilities $P(y_n(t),Y_n ,I_n|x_{n};\bar N_n,{\bf w}^i)$ can be computed efficiently with the following three steps:

\vspace{0.2cm}
\noindent {\bf Step 1: Forward message passing.} In this step, the goal is to compute the joint state probabilities $P(Y^j_{nt},N^j_{nt}|x_n;{\mathbold \theta}^i)$ for all $t=1,2,\ldots,2^j$ and $0\leq j\leq L$.  
Denote an element in the power set of all class labels in $n$th signal by  ${\mathbb L}\in 2^{Y_n}$. The forward message is defined as 
\[
\alpha^j_t({\mathbb L},l)\triangleq P(Y^j_{nt}={\mathbb L},N^j_{nt}=l|x_n; {\mathbold \theta}^i).
\]
At the leaf level, the forward messages are initialized as
$\alpha^L_t({\mathbb L},l)=$
\begin{equation*}
  \begin{cases}
    P(y_n(t-\Delta-1)=0|x_{n};{\mathbold \theta}^i), & l=0, {\mathbb L}=\emptyset;\\
    P(y_n(t-\Delta-1)=c|x_{n};{\mathbold \theta}^i), & l=1, {\mathbb L}=\{c\},c\in Y_n;\\
    0, & Otherwise,
  \end{cases}
\end{equation*}
for $t=1,2,\ldots,T_n+T_w-1$, and 
\begin{equation*}
\alpha^L_t({\mathbb L},l)=
  \begin{cases}
    1, & l=0, {\mathbb L}=\emptyset;\\
    0, & Otherwise,
  \end{cases}
\end{equation*}
for $t=T_n+T_w,T_n+T_w+1,\ldots,2^L$.

The update for the forward message of the $t$th node in $j$-$1$th level is calculated by marginalizing over its left parent (the ($2t-1$)th message in $j$th level) and the right parent (($2t$)th message in $j$th level) as 
\begin{equation}\label{eq:treestep1}
\alpha^{j-1}_t({\mathbb L},l) =\sum_{\mathbb A \subseteq \mathbb L}\sum_{\mathbb B\subseteq \mathbb L}\sum_{a=0}^l {\mathbb I}_{(\mathbb A+\mathbb B=\mathbb L)}\alpha^j_{2t-1}({\mathbb A},a)\alpha^j_{2t}({\mathbb B},l-a).
\end{equation}
We summarize the forward message step in Figure~\ref{fig:treefb}(a).

\vspace{0.2cm}
\noindent {\bf Step 2: Backward message passing.} In this step, the goal is to compute the joint state posterior probability $P(Y_n, I_n=1 |Y^{j}_{nt},N^{j}_{nt}, x_n; {\mathbold \theta}^i,\bar N_n)$. We denote a backward message as 
\[
\beta^j_t({\mathbb L}, l)\triangleq P(Y_n, I_n=1 | Y_{nt}^{j}={\mathbb L},N_{nt}^{j}=l, x_n; {\mathbold \theta}^i, \bar N_n).
\]
The first backward message is initialized as 
\[\beta^0_1({\mathbb L}, l)={\mathbb I}_{(l \leq \bar N_n)} {\mathbb I}_{({\mathbb L} = Y_n)} .\] 
The update equation for the backward messages are calculated as follows:
\begin{equation}\label{eq:treeback1}
\beta^j_{2t-1}({\mathbb A},a) = \sum_{\mathbb E \in 2^{Y_n}}\sum_{e=0}^{\bar N_n-a} \beta^{j-1}_t({\mathbb A}\cup {\mathbb E},a+e)\alpha^j_{2t}({\mathbb E},e).
\end{equation}
\begin{equation}\label{eq:treeback2}
\beta^j_{2t}({\mathbb E},e) = \sum_{\mathbb A\in 2^{Y_n}}\sum_{a=0}^{\bar N_n-e} \beta^{j-1}_t({\mathbb A}\cup {\mathbb E},a+e)\alpha^j_{2t-1}({\mathbb A},a).
\end{equation}
We summarize the backward message step in Figure~\ref{fig:treefb}(b).
\vspace{0.1cm}
\noindent {\bf Note:} To efficiently calculate and store the forward and backward messages, we consider the following results: (i) $\alpha^j_t({\mathbb L},l)=0$ for $l > 2^{L-j}+1$.  
(ii) $\beta^j_t({\mathbb L},l)=0$ for $l > \bar N_n$ or ${\mathbb L}\notin 2^{Y_n}$ for $j=0,1,\ldots,L$ and $t=1,\ldots,2^{j}$. 
Where (i) is obtained from the recursive formula of $N^{j}_{nt}$ in (\ref{eq:treeN}) with the initialization of $N^L_{nt}$ and (ii) is obtained from the sparsity constraint in (\ref{eq:model}) and the definition of $Y_{nt}^j$ in (\ref{eq:treeY}) such that each $Y_{nt}^j\subseteq Y_{nt/2}^{j-1}$, and $Y_{n1}^{0}={\mathbb L}$.
Based on summary (i) that $\alpha^j_t({\mathbb L},l)=0$ when $l > 2^{L-j}+1$, and (ii) that $\beta^j_t({\mathbb L},l)=0$ when $l>\bar N_n$, the effective calculation and actual storing of both forward and backward message is for $0\leq l \leq \min(\bar N_n, 2^{L-j}+1)$. 

\vspace{0.2cm}
\noindent {\bf Step 3: Joint probability.}
Finally, the numerator on the RHS of (\ref{eq:bposterior}) for $t=1,\ldots,T_n$ is computed using the backward message $\beta^L_t({\mathbb L},l)$ 
such that
\begin{eqnarray}\label{eq:joint2}
  \nonumber P( y_{n}(t)=c, Y_n, I_n =1| x_n;{\mathbold \theta}^i, \bar N_n)=\beta^L_t(\{c\},{\mathbb I}_{(c\neq 0)})\cdot\\
  p(y_n(t)=c|x_n;{\bf w}^i,{\bf b}^i).
\end{eqnarray}

\noindent {\bf Convolutive model on tree:} Based on update equation of the forward message in (\ref{eq:treestep1}), if we treat each $\alpha^{j-1}_t$ message of a particular set value $\mathbb L$ as a discrete signal $\alpha^{j-1}_t(\mathbb L,t)$, then the update of each forward message is performing a convolution between $\alpha^j_{2t-1}(\mathbb A,t)$ and $\alpha^j_{2t}(\mathbb B,t)$. For the update on the backward message in (\ref{eq:treeback1}) and (\ref{eq:treeback2}), the update of backward message signal $\beta^j_{2t-1}({\mathbb A},t)$ is a convolution between $\beta^{j-1}_t({\mathbb A}\cup {\mathbb E},-t)$ and $\alpha^j_{2t}({\mathbb E},t)$ and the update of backward message signal $\beta^j_{2t}({\mathbb E},t)$ is a convolution between $\beta^{j-1}_t({\mathbb A}\cup {\mathbb E},-t)$ and $\alpha^j_{2t-1}({\mathbb A},t)$. 

\subsection{Complexity analysis} 
The complexity analysis can be divided into three parts (i) prior calculation, (ii) posterior calculation in E-step, and (iii) gradient calculation in M-step.
We evaluate both prior probability and gradient update by forming $(C+1)\times F$ number of convolutions in the time domain for the $n$th signal. Therefore, the time complexity for both (i) and (iii) is ${\cal O}(\sum_{n=1}^N(C+1)FT_nT_w)$. The space complexity is ${\cal O}((C+1)F(T_n+T_w-1))$ and ${\cal O}((C+1)FT_w)$ respectively.

On the posterior calculation of the E-step, 
the chain forward and backward messages require to run over all possible values of $y_n(t)$ and the previous state values of $Y^{t-1}_n\in 2^{Y_n}$ and $0\leq N^{t-1}_{n}\leq \bar N_n$.
Therefore the overall time and space complexity is ${\cal O}(\sum_{n=1}^N|Y_n|2^{|Y_n|}(T_n+T_w)\bar N_n)$ and ${\cal O}(2^{|Y_n|} (T_n+T_w) \bar N_n)$ respectively. To formulate the tree forward and backward messages, we need to run over all possible values of the previous two parents' states. Therefore the resulting time and space complexity is ${\cal O}(\sum_{n=1}^N4^{|Y_n|}(T_n+T_w) (\log_2 \bar N_n)^2)$ and ${\cal O}(2^{|Y_n|} (T_n+T_w) \log_2 \bar N_n)$ respectively. 

\subsection{Prediction}
In addition to identifying the analysis words ${\bf w}_{c}$, the discriminative model allows for the prediction of time instance labels $y_n(t)$ for both labeled and unlabeled signals as well for the prediction of the signal label. Given a test signal $x^{\text{test}}_n(t)$ for $t=1,2,\ldots, T_n$, the goal is to predict the time instance label signal $\hat y_n(t)$ and the signal label $Y_n$.
\subsubsection{Time instance prediction}
\[
\hat y^{\text{test}}_n(t)=\arg\underset{0\leq c\leq C}{\max} P(y_n(t)=c|x^{\text{test}}_n; {\bf w},{\bf b}).
\]
\subsubsection{Signal label prediction}
For an unlabeled test signal $x^{\text{test}}_n$, the predicted signal label set using the {\bf union rule} is
\[
\hat Y^U_n = \cup_{t=-\Delta}^{T_n-1+\Delta} \{\hat y^{\text{test}}_{n}(t)~|~\hat y^{\text{test}}_{n}(t)\neq 0\}.
\]
Alternatively, the signal label can be predicted by {\bf maximizing a posterior probability (MAP) rule}:
\[
\hat Y^P_n = \arg\underset{\mathbb A \in \{0,1\}^C}{\max}P(Y_n={\mathbb A},I_n=1|x^{\text{test}}_n; \bar N_n, {\bf w},{\bf b}), \text{where}
\]
$P(Y_n={\mathbb A},I_n|x^{\text{test}}_n; \bar N_n, {\bf w},{\bf b}) = \textstyle \sum_{l=0}^{\bar N_n}\alpha_{T_n-1+\Delta}({\mathbb A},l)$.

\begin{figure*}[ht]
\vspace{-40pt}
\centering
\captionsetup[subfigure]{aboveskip=-40pt,belowskip=-5pt}
\begin{subfigure}[b]{0.6\columnwidth}
\centering
\resizebox{1\textwidth}{!}{
\includegraphics{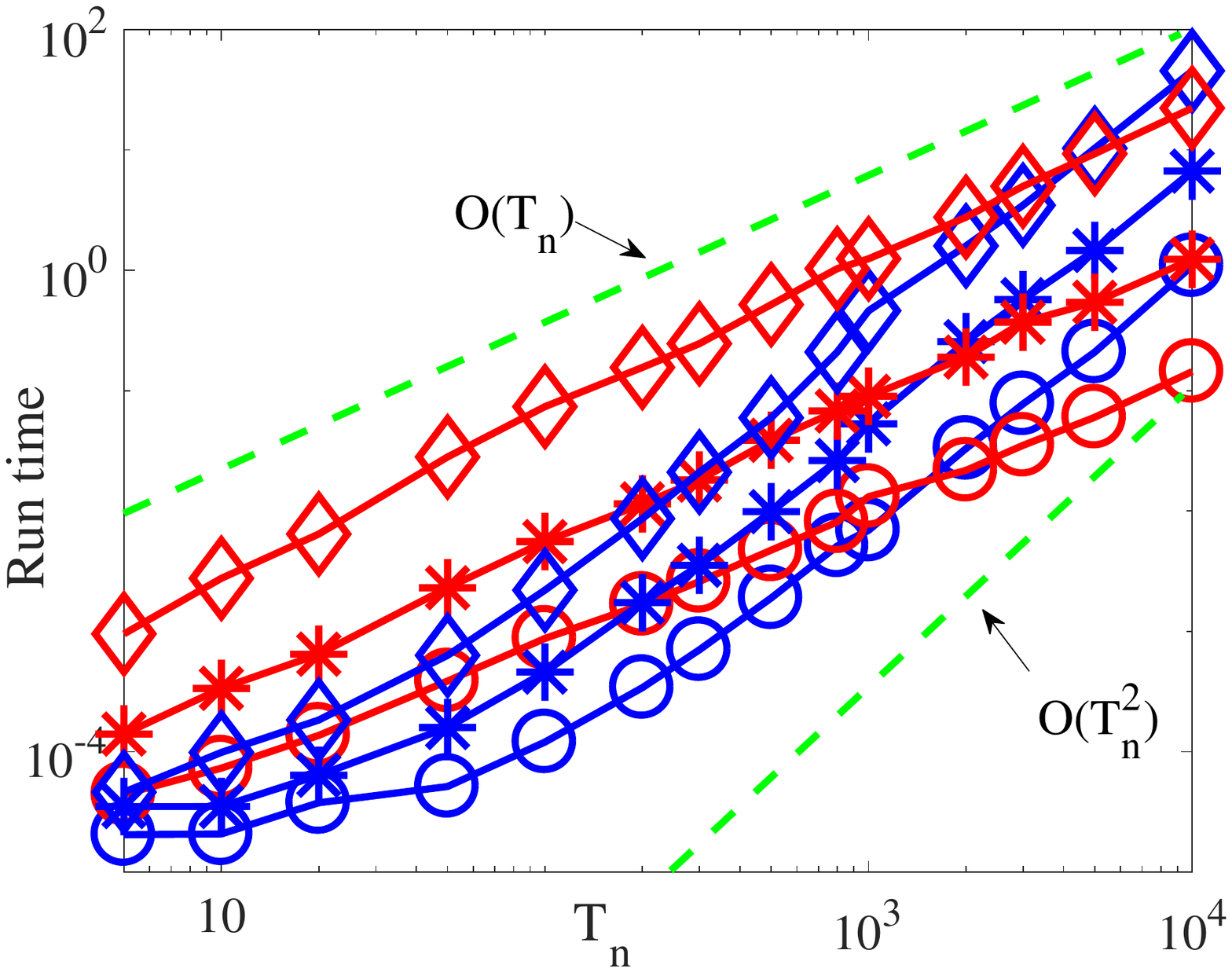}}
\caption{$\bar N_n/T_n=0.2$}\medskip
\end{subfigure}
\begin{subfigure}[b]{0.6\columnwidth}
\centering
\resizebox{1\textwidth}{!}{
\includegraphics{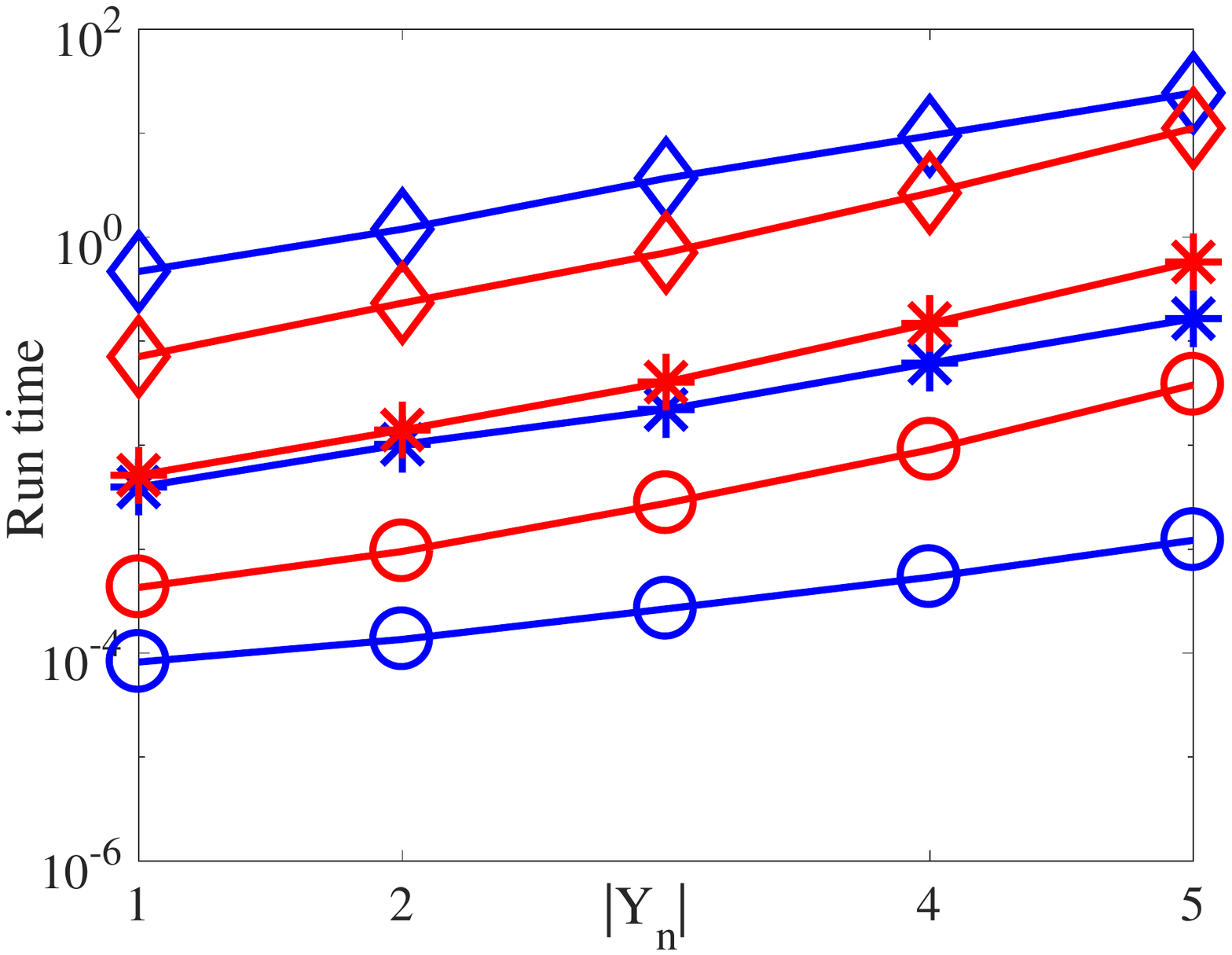}}
\caption{$\bar N_n/T_n=0.5$}\medskip
\end{subfigure}
\begin{subfigure}[b]{0.6\columnwidth}
\centering
\resizebox{1\textwidth}{!}{
\includegraphics{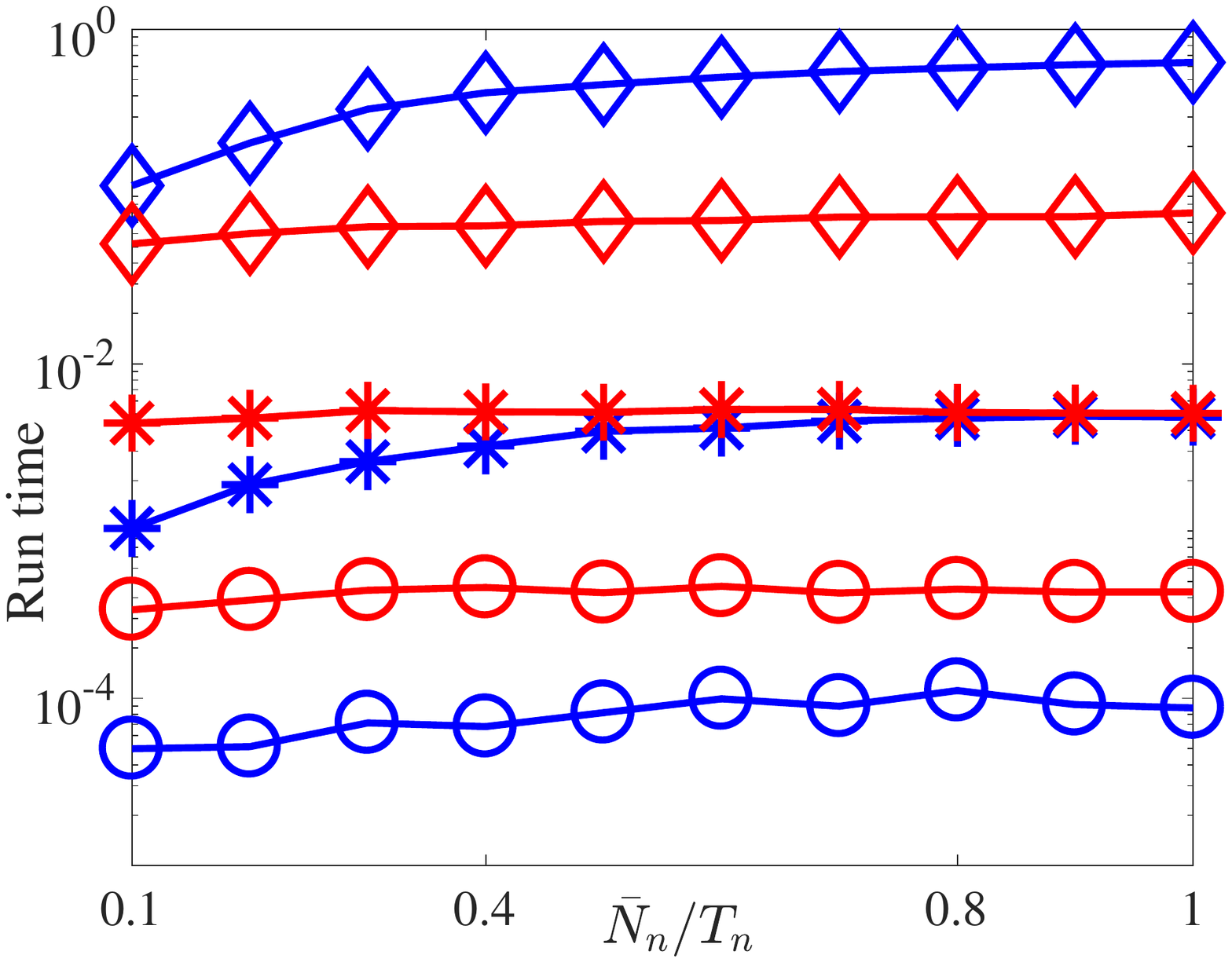}}
\caption{$|Y_n|=2$}\medskip
\end{subfigure}
\caption{Running time versus $T_n,|Y_n|,\bar N_n$. (Blue color for chain and red color for tree algorithm. (a) $\circ: |Y_n|=1$, $\star: |Y_n|=3$, $\diamond: |Y_n|=5$. (b)-(c) $\circ: T_n=50$, $\star: T_n=500$, $\diamond: T_n=5000$.)}
\label{fig:runtime}
\end{figure*}

\section{Experiments}
\label{sec:exp}
In this section, we first present a runtime comparison between chain and tree model reformulations of the E-step inference. We continue by evaluating the performance of the proposed approach 
using synthetic datasets and real world datasets.

\vspace{-0.1cm}
\subsection{Run time analysis}
\label{sec:syn-runtime}
The computational complexity is due to three main calculations namely the prior calculation, the posterior calculation and the gradient calculation in the M-step. Since the posterior calculation dominates the computational complexity and since we have focused on developing an efficient computation for this step, the following results are on the runtime analysis of the posterior calculation during the E-step based on the chain and the tree reformulations of our model. We used a randomly generated prior probability as an input to the posterior calculation. We illustrate the relationships between the E-step posterior calculation time and the number of classes per signal $|Y_n|$, the number of time instances per signal $T_n$, the sparsity regularization per signal $\bar N_n$, we vary $|Y_n|\in\{1,2,3,4,5\}$, $T_n\in\{5,10,20,50,100,\ldots,10000\}$ and $\bar N_n/T_n \in\{0.1,0.2,\ldots,1.0\}$.

\begin{table}[ht]
\fontsize{8}{10}\selectfont
\centering
\begin{tabular}{|l||c|c|c|c|c|}\hline
$T_n$&5&50&500&5000&10000\\ \hline
    \multicolumn{6}{|c|}{$\bar N_n/T_n=0.2$, $|Y_n|=2$} \\\hline
   chain        &{\bf 0.024ms} &{\bf 0.08ms} & {\bf 4.27ms}   &0.55s &2.32s\\ \hline
   tree        &0.074ms &0.84ms & 11.80ms   &{\bf 0.21s} &{\bf 0.39s}\\ \hline
       \multicolumn{6}{|c|}{$\bar N_n/T_n=0.5$, $|Y_n|=2$} \\\hline
   chain        &{\bf 0.028ms}&{\bf 0.13ms} & {\bf 10.20ms}   &1.19s&5.52s\\ \hline
   tree        &0.071ms&0.95ms & 13.97ms   &{\bf 0.23s}&{\bf 0.46s}\\ \hline    \multicolumn{6}{|c|}{$\bar N_n/T_n=0.2$, $|Y_n|=5$} \\\hline
   chain     &{\bf 0.046ms}   &{\bf 0.63ms} & {\bf 0.06s}   &10.32s & 45.34s\\ \hline
   tree      &0.95ms  &28.11ms & 0.52s   &{\bf 9.29s} & {\bf 22.25s}\\ \hline    
   \multicolumn{6}{|c|}{$\bar N_n/T_n=0.5$, $|Y_n|=5$} \\\hline
   chain     &{\bf 0.074ms}   &{\bf 1.21ms} & {\bf 0.16s}   &24.39s &198.78s\\ \hline
   tree      &0.90ms  &37.91ms & 0.57s   &{\bf 11.15s} &{\bf 27.00s}\\ \hline
   \end{tabular}
   \caption{Runtime values for the chain-based and the tree-based E-step calculation as a function of $T_n$ for four scenarios. 
   }    \label{table:runtime}
\end{table}
Figure~\ref{fig:runtime}(a) shows the posterior calculation time per signal based on the tree reformulation grows in a nearly-linear rate with respect to $T_n$ when setting the sparsity level to $0.2T_n$. In addition, it shows the chain based inference time grows quadratically in $T_n$ when $T_n>100$. However, the chain reformulation is more efficient than the tree approach when $T_n$ is small or when $|Y_n|$ is large. Even though Figure~\ref{fig:runtime}(b) shows the posterior calculation time is exponential with respect to $|Y_n|$, the number of classes per signal is usually a small number in practice (see \cite{pham2017dynamic}).
Figure~\ref{fig:runtime}(c) exhibits a near-constant runtime with respect to the sparsity factor $\bar N_n/T_n$ for the tree inference. 
Runtime values for both models are shown in Table~\ref{table:runtime}. 


\subsection{Synthetic datasets and settings}
\label{sec:syn-data}
\mycolor{In designing the synthetic datasets, our goal is to test the performance of the proposed algorithms on different types of data both in terms of the dimension of the data and whether a  generative or discriminative approach is taken for the data generation mechanism. }

\subsubsection{Data generation}
\label{sec:syn-data1}

Below we describe the two synthetic datasets.\\
{\bf Gabor basis dataset:}
\begin{figure}[ht]
\vspace{-1.8cm}
\captionsetup[subfigure]{aboveskip=-20pt,labelformat=empty,belowskip=-20pt}
\centering
\begin{subfigure}[b]{0.98\columnwidth}
\centering
\resizebox{!}{0.35\textheight}{
\includegraphics{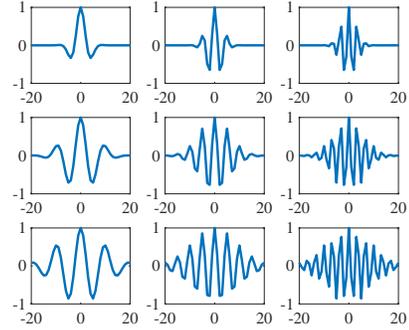}}
\end{subfigure}
%
%
\vspace{-1.8cm}
\caption{\mycolor{Nine Gabor basis used in the experiment.}}
\label{fig:Gaborbases}
\end{figure}
This dataset is constructed with nine different Gabor filters as 1-$D$ signal templates as shown in \ref{fig:Gaborbases}(a):
\[
s_{(a,f)}(t)=\cos{(2\pi f t)}~e^{-\frac{t^2}{2a^2}},~\textrm{for}~t=-20,-19,\ldots, 20
\]
by setting $a=1,2,3$ and $f=0.1,0.2,0.3$. 
Each of the nine templates is used to generate signal of a particular class. 

The data is generated as follows. 1). First, we generate the label sets using a fixed proportion such that $50\%$ contains only a single label, $20\%$ contain two, $20\%$ contain three, and $10\%$ contain no label and are pure noise. The labels in the label set are generated by sampling uniformly with replacement from the nine classes until the target size is reached. 2). Given a non-empty label set $Y_n$, to generate its signal $x_n$, we first decide $m_n$, the number of active time instances in the signal that contain true templates, by randomly choosing between $|Y_n|$ and $|Y_n|+1$ for $|Y_n| \leq 2$ and sampling uniformly from $|Y_n|$ to $10$ for $|Y_n|>2$. 3) For each active time instance $k \in \{1,...,m_n\}$, its exact location $t_k$ is sampled uniformly without replacement from $1$ to $T_n=200$, its class label $c_k$ is uniformly sampled from $Y_n$, and its scaling factor $A_k$ is sampled from ${\mathbb U}[1,2]$. 4) We then generate $y^n_c(t)=\sum_{k=1}^{m_n}A_k\mathbb{I}_{(c-c_k)}\mathbb{I}_{(t-t_k)}$ for each class $c$ and generate the signal $\tilde{x}_n=\sum_{c=1}^9 y_c^n*s^c$, where $s^1(t)=s_{(1,0.1)}(t),s^2(t)=s_{(1,0.2)}(t),\ldots, s^9(t)=s_{(3,0.3)}(t)$. 5) Lastly, we generate the final ${x}_n$ by adding to $\tilde{x}_n$ the white Gaussian noise, whose variance is set to $\sigma^2=E/(T_n\textrm{SNR})=E/(T_n 10^{\textrm{SNR}_{dB}/10})$. Here $E$ is the average signal energy of all $\tilde{x}_n$s and we use $\textrm{SNR}_{dB}$ to control the signal to noise ratio for our final data. 

\noindent {\bf Binary patterns dataset:}
For this dataset, we work with 2-D signals and generate the data following the discriminative assumption.  First we randomly generated $200$ binary sequences of size $3\times 200$ as synthetic 2-D signals ($N=200$, $F=3$, $T=200$). For each generated 2-D signal, we then determined the label for each of its time instance $t$ by matching the $3\times 3$ sub-window starting at $t$ to three pre-defined class-specific templates\footnote{These class templates are defined by selecting the most frequent $3$ patterns in the generated 2-D signals.} shown in Figure~\ref{fig:syndicr}(a). The label was set to $1,2$ or $3$ if the sub-window matched the template of class $1, 2$ or $3$ respectively, and $0$ otherwise.
After all time instance labels were created, the signal label was set to the union of its corresponding instance labels. 


\subsubsection{Experimental setting}
\label{sec:syn-data2}
To demonstrate the performance of the proposed approach, we used $10$ random splits of $100$ signals and trained on each split of $80\%$ data and tested on the rest $20\%$ data. For each random split, we denote it as one Monte-Carlo (MC) run. We evaluated the performance on the test data with all 10 MC runs to find kernel size $T_w$, regularization term $\lambda_r$, and the cardinality constraints $\bar N_n$. 
For the Gabor basis dataset, we first tuned the model parameters by evaluating the average signal label prediction accuracy with $\lambda_r \in \{10^{-8},10^{-6},10^{-4},10^{-2},10^0,10^2\}$ and 
$T_w \in \{5,10,20,40,60,80\}$
. The iteration number was set to $10,000$. 
Using cross-validation for prediction accuracy, we found the optimal $\lambda_r$ and $T_w$ and used those to present the prediction performance as a function of the cardinality constraint parameter 
$N \in \{5, 10, 20, 50, 100, 200\}$ 
(setting $\bar N_1,\ldots,\bar N_N = N$) and the $\textrm{SNR}_{dB} \in \{-10,-5,0,5,10,15,20,25\}$ (see Figure~\ref{fig:tunepara}(a)).

For the binary patterns dataset, we tune the kernel size $T_w \in \{1,3,5,10\}$, regularization term $\lambda_r \in \{10^{-6},10^{-4},10^{-2},10^0\}$ and the cardinality constraint $\bar N_n\in \{3,5,10,50,100\}$. 

\noindent \textbf{Benchmark competing algorithm - A generative dictionary learning followed by logistic regression (GDL-LR) approach:} To the best of our knowledge, we are unaware of other weak-supervision methods for convolutive dictionary learning. In order to provide a benchmark, we considered a two-step approach: a generative convolutive dictionary learning method followed by a classifier.\footnote{Although the two steps can be combined to yield improved performance, the combination of the two steps requires further research beyond the scope of this paper.} For the implementation of the generative dictionary learning method, we chose \cite{ruiz2015dictionary} (used previously on the HJA dataset) and constructed a generative dictionary $D=\{d_1, d_2,\ldots, d_K\}$. We used a matched filter approach to compute a test statistic for each of the $K$ dictionary works as $\max_{t} \tilde{d_k}*x^{\text{train}}_n~|_t$, where $\tilde{d_k}$ is a time reversed version of $d_k$ ($\tilde{d_k}(t) = d_k(-t)$). We combined the $K$ test statistics into one feature vector and trained $C$ logistic regression classifiers based on the feature vectors and their corresponding binary labels indicating the presence and absence of a class $c \in \{1,2,\ldots,C\}$. We use the resulting $C$ classifiers in our performance evaluation for instance level classification and for signal classification.

Using the 10 MC runs, we evaluated the proposed GDL-LR approach by trained on a fixed number of $5000$ outer iterations as in \cite{ruiz2015dictionary}. We vary the dictionary window size $T_d \in\{5, 10,20,40,60,80\}$, sparsity regularization $\lambda_s \in \{10^{-8},10^{-6},10^{-4},10^{-2},10^0,10^2\}$ and the number of dictionary words $K \in \{3,5,7,9,15,18\}$ for the Gabor basis dataset. For the binary patterns dataset, we vary $T_d \in\{1,3,5,10\}$, sparsity regularization $\lambda_s \in \{10^{-4},10^{-2},1,5\}$ and $K \in \{3,7,9,15\}$.

\noindent \mycolor{{\bf Evaluation metric:} In computing the instance level detection area-under-the-curve (AUC), we calculate an AUC for each class $c$ and obtain $\textrm{AUC}=1/C\sum_c \textrm{AUC}_c$. For each class $c$, we obtain the ground truth based on the presence and absence of a given class $c$ at each time stamp $t=0,\ldots,T_n-1$ and use $P_t=P(y_n(t)=c|x^{\text{test}}_n; {\bf w},{\bf b})$ as a test score~\cite{mann1947test}.}

\mycolor{The signal level detection AUC is obtained based on $\textrm{AUC}_c$ for all $c=1,2,\ldots,C$. For each class $c$, the $\textrm{AUC}_c$ is obtained based on the signal level ground truth and the corresponding test score defined as $1-\prod_{t=-\Delta}^{T_n-1+\Delta}(1-P_t)$.}
\subsection{Results on synthetic datasets}
\label{sec:syn-result}
\subsubsection{Gabor basis dataset}
\label{sec:syn-result1}
\begin{figure}[th]
\vspace{-1cm}
\captionsetup[subfigure]{aboveskip=-15pt,labelformat=empty}

\centering
\begin{subfigure}[b]{0.48\columnwidth}
\centering
\resizebox{1.2\textwidth}{!}{
\includegraphics{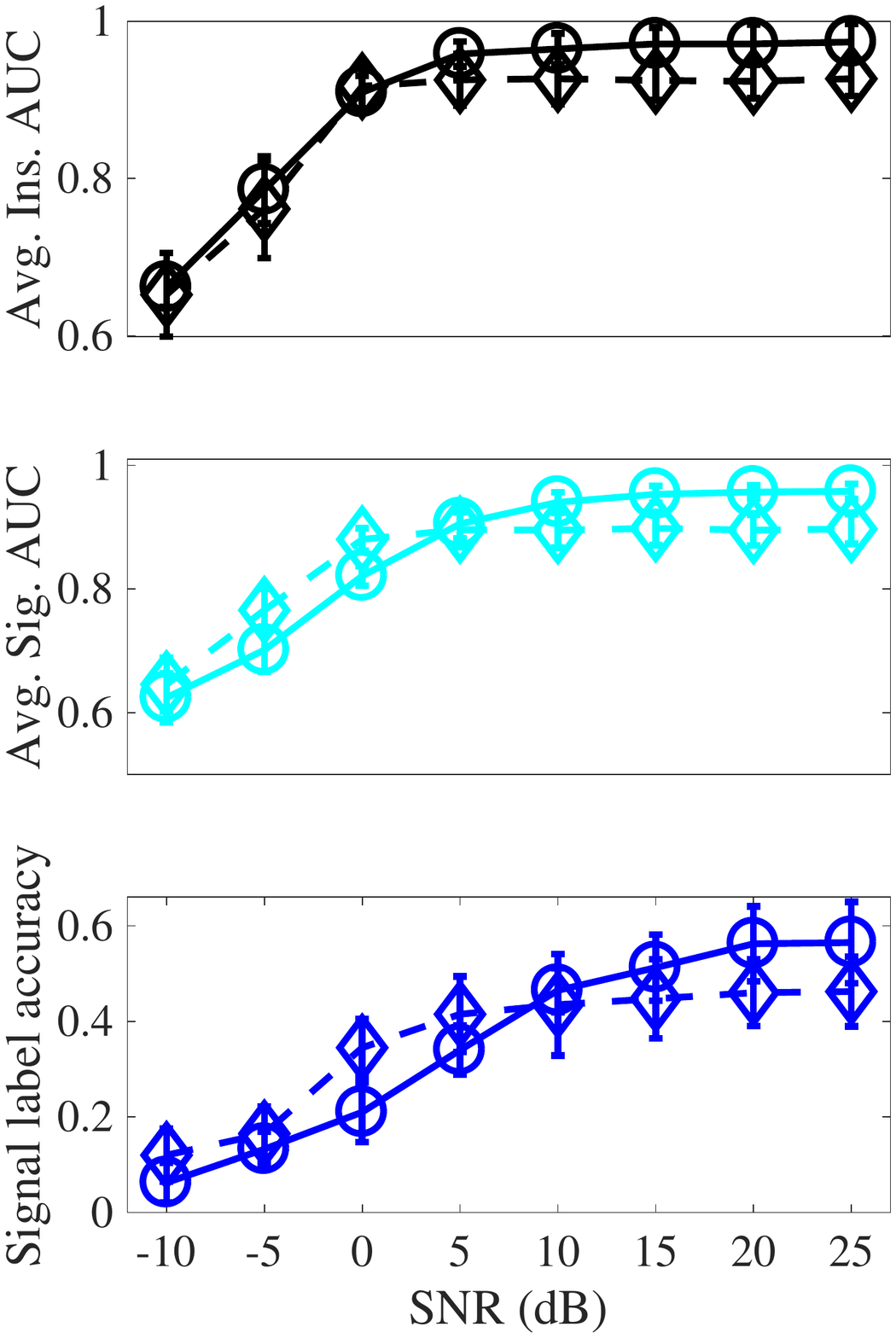}}
\caption{(a) $\bar N_n=20$}\medskip
\end{subfigure}
\begin{subfigure}[b]{0.48\columnwidth}
\centering
\resizebox{1.2\textwidth}{!}{
\includegraphics{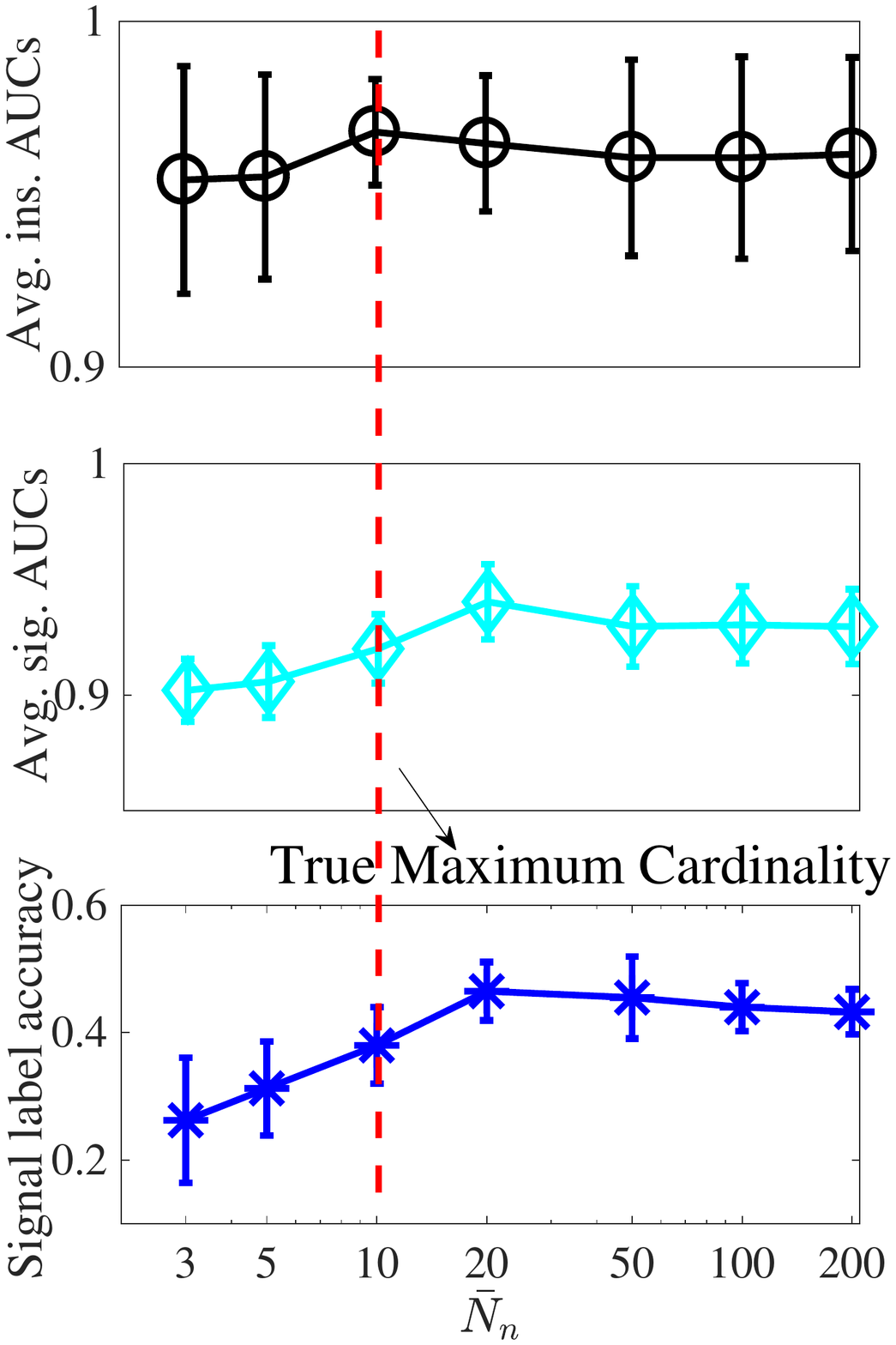}}
\caption{(b) $\textrm{SNR}_{dB}=10$}\medskip
\end{subfigure}
%
%
\caption{\mycolor{Gabor basis dataset performance metrics for the WSCADL approach (solid $\circ$) and the GDL-LR approach (dashed $\diamond$) as a function of $\textrm{SNR}_{dB}$ in (a), and for the WSCADL approach as a function of $\bar N_n$ in (b).}}
\label{fig:tunepara}
\end{figure}

Based on the highest prediction accuracy, the hyper-parameters of our WSCADL approach are set to be $\lambda_r=10^{-4}$ and $T_w=40$ via the aforementioned cross-validation. The hyper-parameters on the GDL-LR approach are set to be $\lambda_s=1,K=9$ and $T_d=40$. \mycolor{The optimal window size is learned to be $40$, which is close to the ground truth Gabor basis length. We believe the kernel size should at least cover the length of the signal patterns to obtain a good performance. If the kernel size is set to be too large, over-fitting may occur.} For the proposed WSCADL approach and the competing GDL-LR framework, we observe that the prediction performance increases when $\textrm{SNR}_{dB}$ increases in Figure~\ref{fig:tunepara}(a). While the two methods perform similarly at low $\textrm{SNR}_{dB}$ values, but for medium and high values the proposed WSCADL approach outperforms the competing GDL-LR approach.

\mycolor{To show the importance of the cardinality constraints, we present the signal label accuracy of the proposed approach, average instance-level and signal-level detection AUC as a function of the cardinality parameter $\bar N_n$ in Figure~\ref{fig:tunepara}(b). As Figure~\ref{fig:tunepara}(b) shows, when $\bar N_n<20$, the signal label accuracy and signal-level AUC drops significantly. We suspect that this setting forces some of the non-zero instance labels to be predicted as zero both in the training and test. When $\bar N_n>20$, the performance drops gracefully. In this setting, we allow some of the zero instance labels to be predicted as non-zero. From Figure~\ref{fig:tunepara}(b), the optimal $\bar N_n$ is $20$ in terms of signal label accuracy. We also observe the average instance-level AUC reaches the peak when $\bar N_n=10$, which is the ground truth maximum cardinality in the data. However, the average signal-level AUC and signal label accuracy reaches peak when $\bar N_n=20$, which is slightly higher than the ground truth.}

To evaluate the performance in terms of AUC, we fixed $\textrm{SNR}_{dB}$ to be $20$. We present the detection AUC performance for both methods in Table~\ref{table:aucs}. Comparing the instance level and the signal detection performance from class $1$ to $7$, we observe that our proposed WSCADL approach outperforms the GDL-LR approach. For class $8$ and $9$, the GDL-LR approach detection AUCs is comparable to our WSCADL and sometimes, the AUCs for the GDL-LR approach is slightly higher than our approach. The variance of the detection AUCs for the GDL-LR approach is mostly higher than our WSCADL approach. We suspect that since the GDL-LR approach performs an unsupervised dictionary learning followed by a classifier training in a separate fashion, the resulting words may have large variability. We believe that this can be fixed by combining the two steps into one. However, due to the weak-supervision setting, the combined approach  is a non-trivial extension, which to the best of our knowledge is unavailable. Hence, we provide the results for the two-step approach only.

\begin{table}[ht]
\fontsize{7.5}{9.5}\selectfont
\centering
\begin{tabular}{|p{5mm}||c|c|c|c|}\hline
Class&WSCADL-ins.&WSCADL-sig.&GDL-LR-ins.&GDL-LR-sig.\\ \hline  
c=1&\bf99.09$\pm$1.94&\bf99.89$\pm$0.36&92.68$\pm$4.33&91.77$\pm$5.18\\ \hline
c=2&\bf99.95$\pm$0.02&\bf96.74$\pm$2.40&90.74$\pm$12.78&81.92$\pm$7.83\\ \hline
c=3&\bf99.26$\pm$1.97&\bf99.67$\pm$0.72&95.45$\pm$10.33&90.00$\pm$5.84\\ \hline
c=4&\bf96.80$\pm$7.34&\bf97.65$\pm$1.72&90.40$\pm$9.67&87.85$\pm$5.46\\ \hline
c=5&\bf99.75$\pm$0.10&\bf92.84$\pm$2.15&97.27$\pm$2.65&86.56$\pm$7.60\\ \hline
c=6&\bf97.96$\pm$2.92&\bf95.63$\pm$5.30&93.24$\pm$6.77&89.58$\pm$4.85\\ \hline
c=7&\bf87.83$\pm$17.19&\bf94.32$\pm$9.89&81.26$\pm$15.95&83.73$\pm$15.15\\ \hline
c=8&\bf98.40$\pm$2.08&94.84$\pm$4.12&93.93$\pm$4.86&\bf96.54$\pm$3.49\\ \hline
c=9&94.96$\pm$5.18&85.59$\pm$5.29&\bf96.22$\pm$5.16&\bf97.95$\pm$1.21\\ \hline
   \end{tabular}
   \caption{\mycolor{Gabor basis dataset: Detection AUCs (\%) for the WSCADL and the GDL-LR approaches with optimal tuning parameters}}\label{table:aucs}
\end{table}

\subsubsection{Binary patterns dataset}
\label{sec:syn-result2}
The hyper-parameters are set to be $\lambda_r=10^{-2}$, $\bar N_n=3$ and $T_w=5$ via the aforementioned cross validation. \mycolor{The optimal kernel size $3\times 5$ is slightly higher than the ground truth window size $3\times 3$.}
For the GDL-LR approach, the optimal dictionary window size $T_d$ is $5$, sparsity constraint $\lambda_s$ is $1$ and the number of dictionary words $K$ is $15$.
\begin{figure}[th]
\vspace{-45pt}
\captionsetup[subfigure]{aboveskip=-15pt,labelformat=empty}
\centering
\begin{subfigure}[b]{0.48\columnwidth}
\centering
\resizebox{!}{0.27\textheight}{
\includegraphics{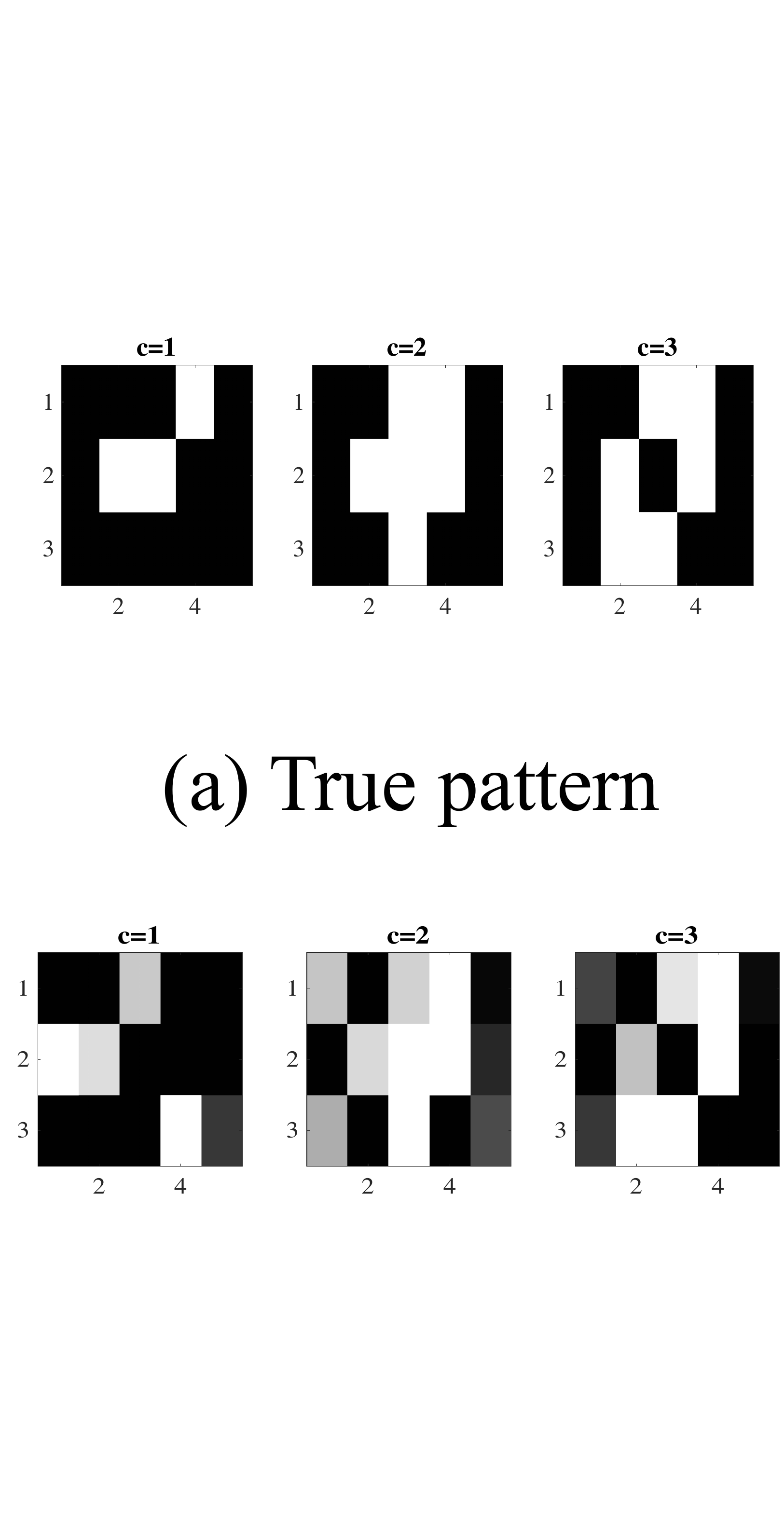}}
\caption{(b) Learned WSCADL words}\medskip
\end{subfigure}
\begin{subfigure}[b]{0.48\columnwidth}
\centering
\resizebox{1.1\textwidth}{!}{
\includegraphics{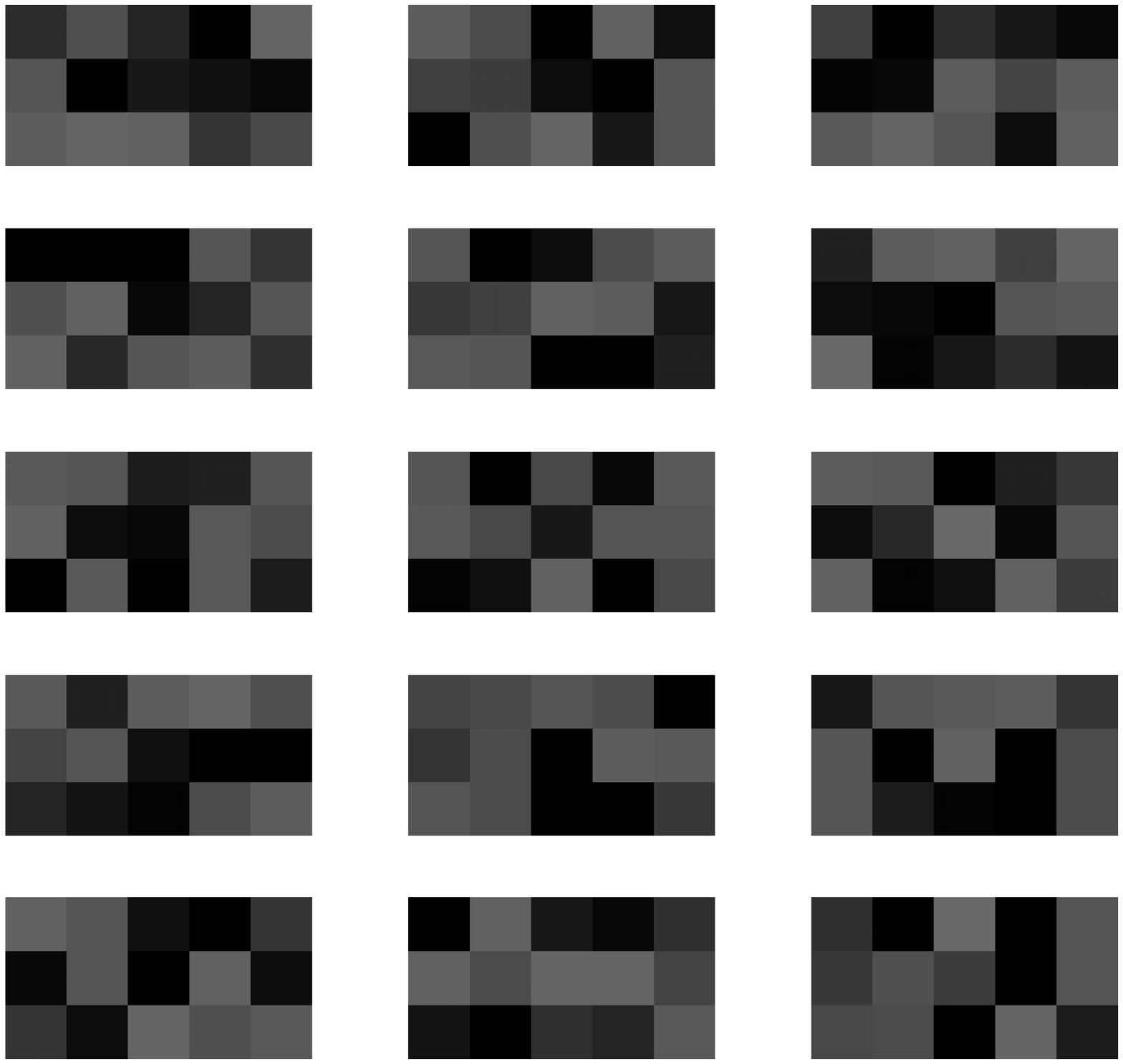}}
\caption{(c) Learned GDL-LR words}\medskip
\end{subfigure}
\begin{subfigure}[b]{0.48\columnwidth}
\centering
\resizebox{1.2\textwidth}{!}{
\includegraphics{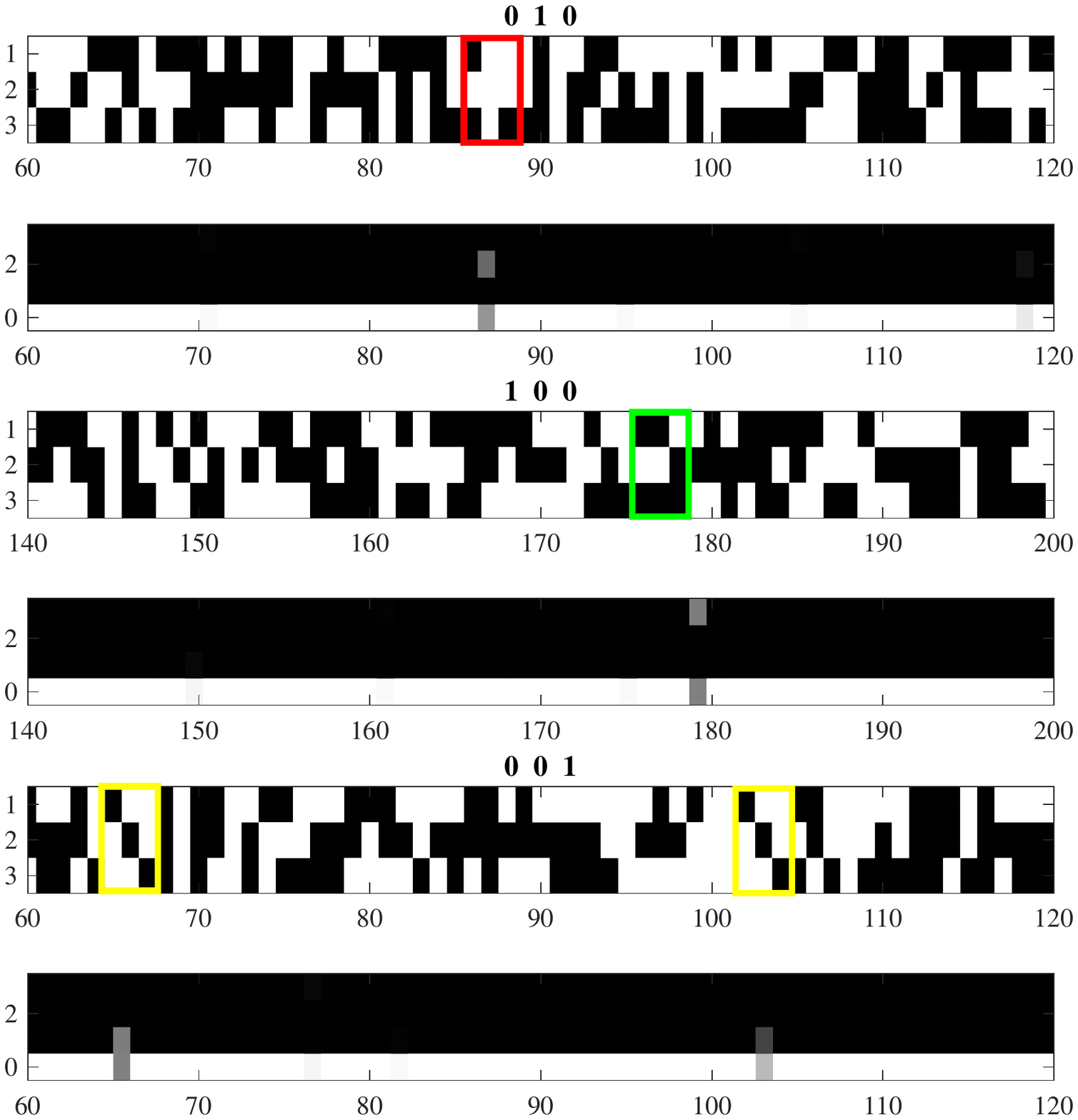}}
\caption{(d) Localization for WSCADL}\medskip
\end{subfigure}
\begin{subfigure}[b]{0.48\columnwidth}
\centering
\resizebox{1.2\textwidth}{!}{
\includegraphics{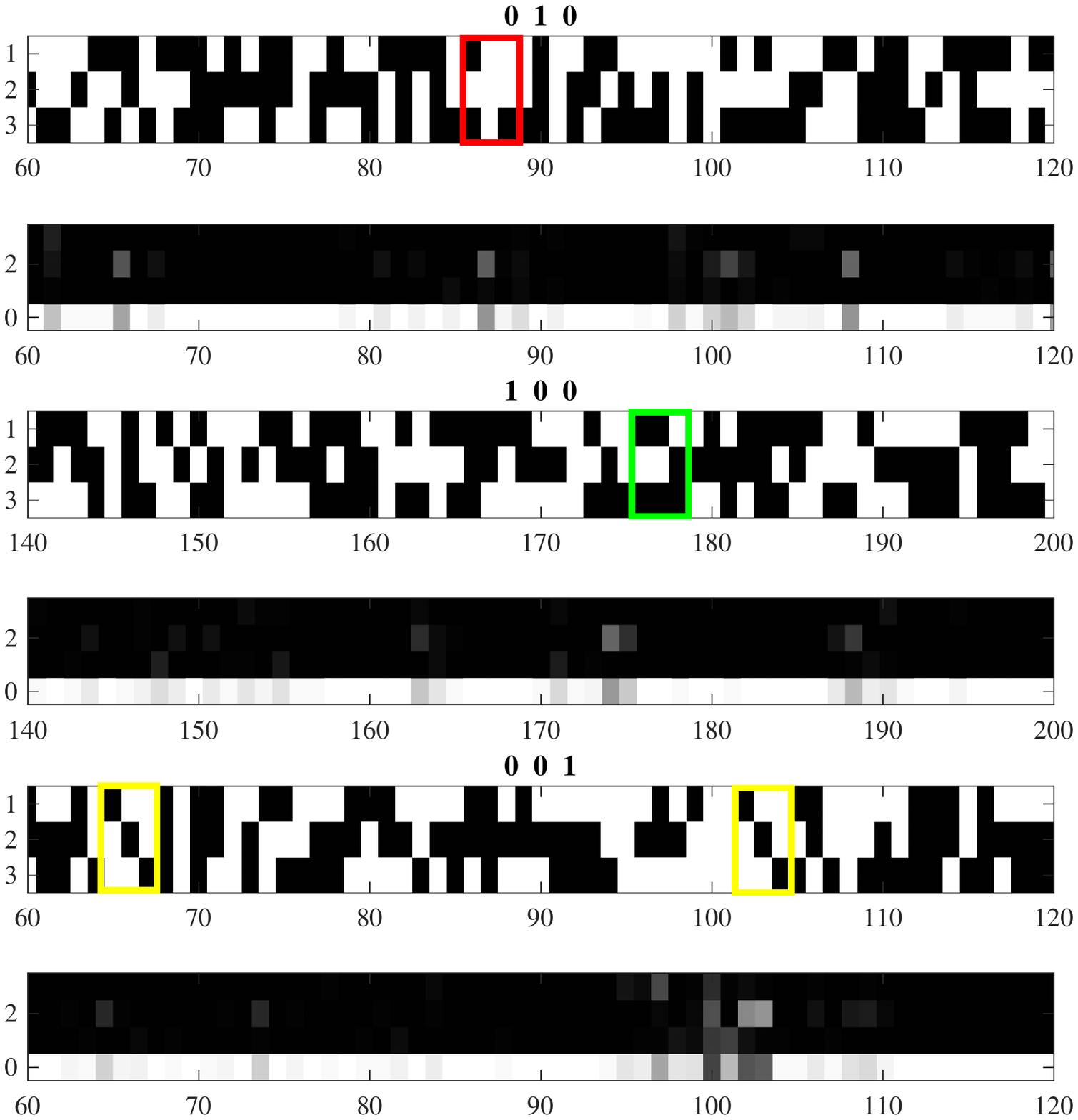}}
\caption{(e) Localization for GDL-LR}\medskip
\end{subfigure}
\caption{Binary patterns dataset setting and results}
\label{fig:syndicr}
\end{figure}

The learned WSCADL words in Figure~\ref{fig:syndicr}(b) and the learned GDL-LR words in \ref{fig:syndicr}(c) show that WSCADL is able to recover the true patterns while the GDL-LR approach fails. Figure~\ref{fig:syndicr}(d) and (e) also shows that WSCADL can localize the corresponding class patterns ideally while the GDL-LR approach is failing in this task. The resulting detection AUCs in both WSCADL and GDL-LR approaches are shown in Table~\ref{table:baucs}.
\begin{table}[ht]
\fontsize{7.5}{9.5}\selectfont
\centering
\begin{tabular}{|p{5mm}||c|c|c|c|}\hline
Class&WSCADL-ins.&WSCADL-sig.&GDL-LR-ins.&GDL-LR-sig.\\ \hline  
c=1&\bf100.00$\pm$0.00&\bf99.59$\pm$1.08&57.92$\pm$17.20&48.11$\pm$5.18\\ \hline
c=2&\bf100.00$\pm$0.00&\bf99.34$\pm$0.62&60.16$\pm$18.93&56.64$\pm$8.63\\ \hline
c=3&\bf100.00$\pm$0.00&\bf99.78$\pm$0.45&56.30$\pm$8.77&51.89$\pm$10.97\\ \hline
   \end{tabular}
\caption{\mycolor{Binary patterns dataset: Detection AUCs (\%) for the WSCADL and the GDL-LR approaches with optimal tuning parameters}}\label{table:baucs}
\end{table}

Due to the discriminative nature of the data, our proposed WSCADL model outperforms the GDL-LR approaches significantly. Since the data was not constructed as a linear combination of dictionary words, the GDL-LR approach was not able to recover a dictionary that could reconstruct the data accurately. Under the discriminative data generation setting, the GDL-LR approach can reproduce the original data only when the sparsity constraint is relaxed. However, regardless of sparsity the GDL-LR approach seems to perform poorly on classification. We suspect that this is due to the lack of discriminative power in the GDL-LR dictionary words obtained.

\subsection{Real-world datasets and results}
\label{sec:real-data}

\subsubsection{Dataset description}
\label{sec:real-data2}
Below we describe the two real-world datasets.\\
\noindent {\bf AASP challenge - office live scene dataset:} This dataset consists of audio recordings of sounds taken in an office environment  \cite{giannoulis2013detection}. The training dataset consists of $20$ to $22$ individual events (such as door slam, phone ringing, and pen drop) recording with varying time from $0.05$s to $20$s for $16$ various class. 
The test dataset contains seven roughly three minutes long office live sound recordings, where each single recording is multiple labeled. The task is to detect the presence and absence of events on the test set. 


\noindent {\bf HJA bioacoustic dataset:}
The HJA dataset contains $548$ labeled $10$-second recordings of $13$ different bird species. The audio recordings of bird song are collected at the H.~J.~Andrews (HJA) Experimental Forest, using unattended microphones~\cite{briggs2012rank}. Each recording may contain multiple species. 

\subsubsection{Data preprocessing}
\label{sec:data-prep}


For AASP challenge office live training dataset, we compared our proposed approach with the supervised dictionary learning approaches. Since the competing supervised dictionary learning algorithms use a fixed size feature vector, we created a fixed duration training signals from the various duration training data. For a fair comparison, we used this modified short duration training data for all algorithms. The fixed short duration training data is selected to be 1 sec duration because (i) most single occurrence of a sound event lasts less than 1 second and (ii) over 80\% of the recordings are around 1sec duration. Recordings longer than 1s were chunked into 1s duration signals. Recordings shorter than 1s were extended to 1s using the last sample value. Note that our proposed WSCADL algorithm does not require the aforementioned preprocessing as it can handle varying signal length. To perform a detection task on the test audio with 3 minutes long, we chunk the test recordings into 10s and apply the following procedures.


For both datasets, each audio recording was applied with (i). Spectrogram generation: FFT is applied to each windowed signal with 16ms window size of 0.9 overlap ratio and the number of FFT bins is twice of the window samples; (ii). Noise whitening: each column on the spectrogram was divided by the noise spectrum~\cite{briggs2012rank}. 
(iii). Spectrogram down-sampling: a Matlab built-in imresize function is applied (For office live dataset on experiment 1, training spectrogram is down-sampled from $\mathbb{R}^{707\times 612}$ to $\mathbb{R}^{256\times 200}$ and test spectrograms are from $\mathbb{R}^{707\times 6120}$ to $\mathbb{R}^{256\times 2000}$, on experiment 2, spectrograms are down-sampled from $\mathbb{R}^{707\times 6120}$ to $\mathbb{R}^{256\times 200}$. For HJA dataset, spectrograms are from $\mathbb{R}^{256\times 1249}$ to $\mathbb{R}^{256\times 200}$).

\subsubsection{Experimental setup}
\label{sec:real-setup}
Below we present two real-world experimental setting.

\noindent{\bf Office live experimental setting:}
we considered two experiments. In the first experiment, we trained on the training dataset, which consists of the 1s duration training examples, and tested on the 10sec-long recording test set. In the second experiment, we use the 10sec-long recordings for both training and test.

{\bf Experiment 1:} For cross-validated parameter tuning, we trained on $80\%$ of the original labeled data and validated on the independent $20\%$ of the data. Parameter tuning was performed for all dictionary learning approaches and the parameters that yielded the highest prediction accuracy were selected. For tuning our approach, we set the dictionary window size $T_w \in
\{10,20,30,40\}$, the cardinality constraint $\bar N_n \in
\{5,10,60,100,200\}$ along with a regularization term $\lambda_r \in \{10^{-6},10^{-4},10^{-2},10^{-1},1,10\}$. 
Using the learned WSCADL words for the optimal tuning parameter value, we evaluated both signal and instance level detection performance on the test set. For the other supervised dictionary learning approaches, it is not easy to perform the detection task since their approaches are non-convolutive.

{\bf Experiment 2:} We trained on $80\%$ of the sub-sampled test set along with the signal labels generated by the union of event ground truth labels. For choosing the optimal model parameters, we considered the same range as in experiment $1$. We evaluated the detection performance on the remaining $20\%$.

\noindent {\bf HJA bioacoustic experimental setting:}
For cross-validated parameter tuning, we trained on $80\%$ of the training data and evaluated the performance on the independent $20\%$ of the data. The tunning parameters considered were: the window size  $T_w \in \{10,20,50,100\}$, the cardinality constraint  $\bar N_n \in
\{10,20,40,60,100,160,200\}$ and the regularization term $\lambda_r \in \{10^{-6},10^{-4},10^{-2},10^{-1},10^0,10^1\}$. After we learned the analysis words for the optimal tuning parameter value, we used the dictionary to predict the signal label on the test set.


\subsubsection{Results}
\label{sec:real-result}
Below we present the results on two real-world datasets. 
\begin{table}[ht]
\fontsize{6}{10}\selectfont
\centering
\begin{tabular}{|l||c|c|c|c|}\hline
Experiment-ins./sig.&minimum class&mean over class&maximum class\\ \hline  
1-instance&{\bf 40.90$\pm$1.99}&53.29$\pm$1.00&65.66$\pm$0.14\\ \hline
2-instance&36.57$\pm$18.60&{\bf 54.75$\pm$3.32}&{\bf 77.65$\pm$16.74}\\ \hline
1-signal&45.57$\pm$4.62&64.23$\pm$0.39&95.12$\pm$4.62\\ \hline
2-signal&{\bf 45.58$\pm$9.18}&{\bf 70.19$\pm$3.54}&{\bf 99.88$\pm$9.18}\\ \hline
   \end{tabular}
   \caption{Instance level and signal detection AUCs (\%) for both experiments across five MC runs.}\label{table:aasp_aucs}
\end{table}

\noindent {\bf Office live event detection:}
We compared our WSCADL approach with discriminative dictionary learning approaches:
sparse representation-based classification (SRC)~\cite{wright2009robust}; label consistent K-SVD (LCKSVD1,LCKSVD2)~\cite{jiang2013label}; dictionary learning with structured incoherence and shared features (DLSI)~\cite{ramirez2010classification}; Fisher discrimination dictionary learning (FDDL)~\cite{yang2011fisher}; dictionary learning for separating the particularity and the commonality (COPAR)~\cite{kong2012dictionary}; fast low-rank shared dictionary learning for object classification (LRSDL)~\cite{vu2016learning}.

\begin{figure}[th]
\vspace{-1cm}
\captionsetup[subfigure]{aboveskip=-15pt,labelformat=empty}

\centering
\begin{subfigure}[b]{0.48\columnwidth}
\centering
\resizebox{1.2\textwidth}{!}{
\includegraphics{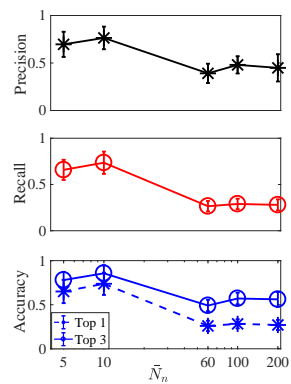}}
\end{subfigure}
%
%
\vspace{-0.8cm}
\caption{\mycolor{Prediction accuracy as a function of the cardinality parameter $\bar N_n$ on the AASP dataset.}}
\label{fig:realNn}
\end{figure}
\begin{table*}
\caption{\anothercolor{Signal evaluation metrics (\%) for various methods on HJA dataset. $\downarrow$ ($\uparrow$) next to a metric indicates that the performance improves when the metric is decreased (increased). The results from column MLR to M-NN are extracted from Table 4 in~\cite{pham2017dynamic}.}}
\label{tab:hja_acc}
\fontsize{8}{10}\selectfont
\centering
\begin{tabular}{|l||c|c||c|c|c|c||c|c|}
    \hline
        Method     &WSCADL         &GDL-LR    & MLR                  &SIM                    &Mfast                  &LSB               &M-SVM &M-NN                            \\ \hline
    $\downarrow$ Hamming loss       &{05.5$\pm$0.0}$^P$ &06.1$\pm$0.8    &09.6$\pm$1.0            &15.9$\pm$1.5           &05.5$\pm$1.1       &10.6$\pm$1.5  &{\bf 04.5$\pm$0.6} &04.7$\pm$1.1           \\ \hline
    $\downarrow$ rank loss        &{\bf 02.2$\pm$0.4}  &03.0$\pm$0.7      &02.7$\pm$0.6       &02.2$\pm$0.8     &02.5$\pm$0.7     &06.9$\pm$1.8  &02.7$\pm$1.1 &02.7$\pm$1.1            \\ \hline
    $\uparrow$ average precision           &{\bf 94.6$\pm$0.6} &92.4$\pm$0.8     &94.2$\pm$1.2       &94.1$\pm$1.8      &94.1$\pm$1.4      &89.7$\pm$2.6 &94.0$\pm$2.0 &93.9$\pm$2.8              \\ \hline
    $\downarrow$ one error      &04.6$\pm$1.8  &07.0$\pm$2.0        &{\bf 03.8$\pm$1.8}       &05.1$\pm$3.1      &03.7$\pm$2.4      &03.7$\pm$1.7  &04.6$\pm$2.6 &05.3$\pm$4.4        \\ \hline
    $\downarrow$ coverage      &16.2$\pm$0.9    &17.0$\pm$1.9     &13.9$\pm$1.6            &{\bf 12.4$\pm$1.6}      &13.4$\pm$1.6      &21.7$\pm$3.6 &13.2$\pm$1.6 &13.4$\pm$1.3 \\ \hline
   \end{tabular}
   \vspace{-0.1in}
\end{table*}
For our proposed approach, the optimal tuning parameters found are window size $T_w=10$, the regularization term $\lambda_r=10^{-4}$ and the cardinality parameter $\bar N_n=10$ \mycolor{as shown in Figure \ref{fig:realNn}, which presents the performance of our proposed WSCADL approach on varying cardinality parameter $\bar N_n$ for the AASP dataset. Setting the cardinality parameter less than or larger than the optimal value reduces the accuracy.} \mycolor{For all other discriminative dictionary learning algorithms, the parameters values are tuned with cross-validation. The SRC algorithm uses all training examples as dictionary. In LCKSVD1 and LCKSVD2, DLSI and FDDL, 10 dictionary words per each class are used so that the total number of dictionary atoms is 160. In COPAR and LRSDL algorithms, 10 dictionary words per class are used with 5 shared dictionary atoms. However, in the proposed WSCADL algorithm, we assign total of 16 dictionary words therefore only 1 dictionary word is learned to predict each class.} \anothercolor{The proposed model is limited to 16 words in total since the model uses a single word per class. Potential extensions to allow more words per class may be considered as future work.} 

\begin{figure}[ht]
\vspace{-1cm}
\captionsetup[subfigure]{aboveskip=-15pt,belowskip=-1pt}
  \centering
  \begin{subfigure}[b]{0.49\columnwidth}
  \centering
   \resizebox{1\textwidth}{!}{
  \includegraphics{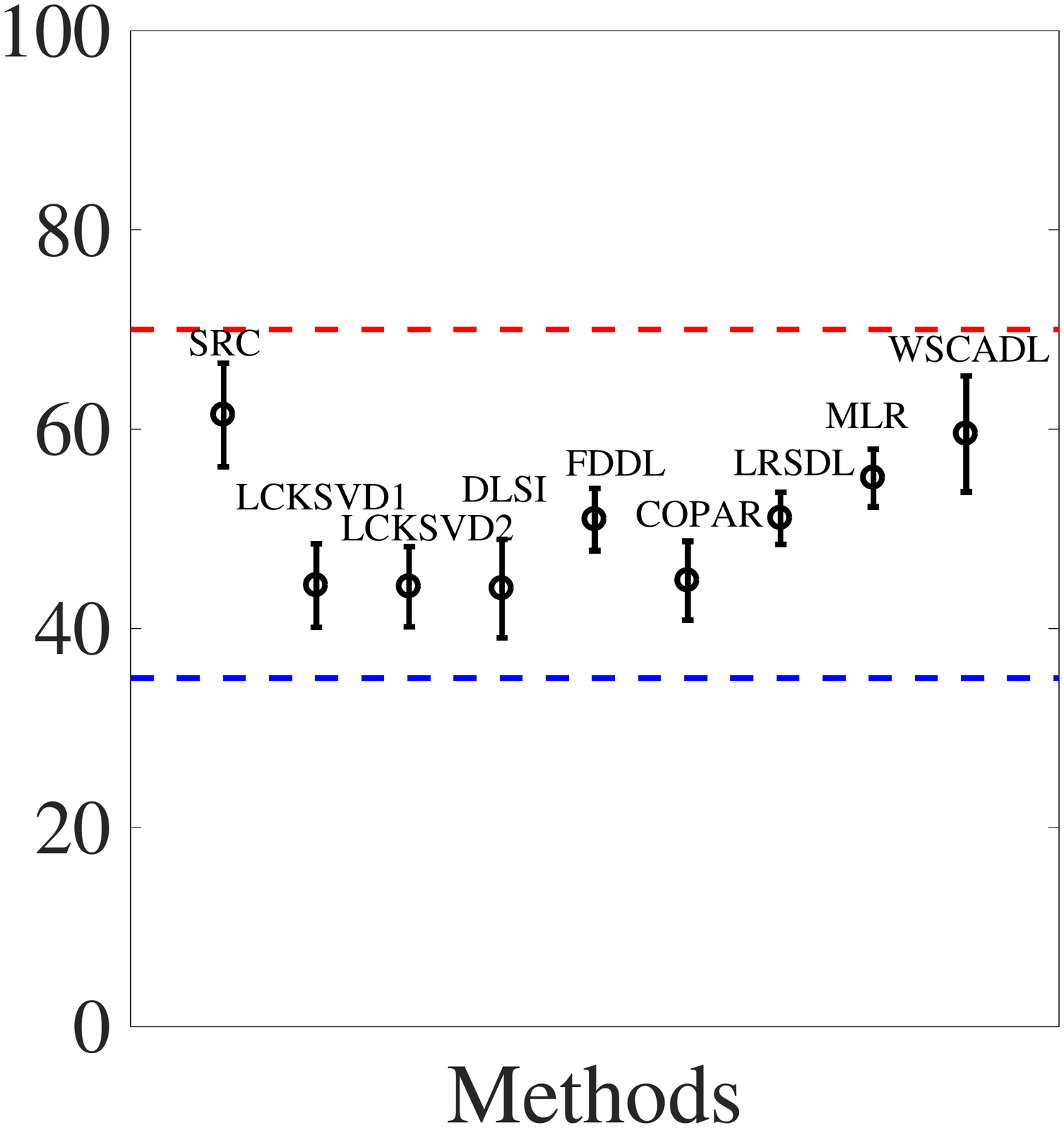}}
  \caption{Top 1}
    \end{subfigure}
    \vspace{-0.2in}
  \begin{subfigure}[b]{0.49\columnwidth}
  \centering
   \resizebox{1\textwidth}{!}{
  \includegraphics{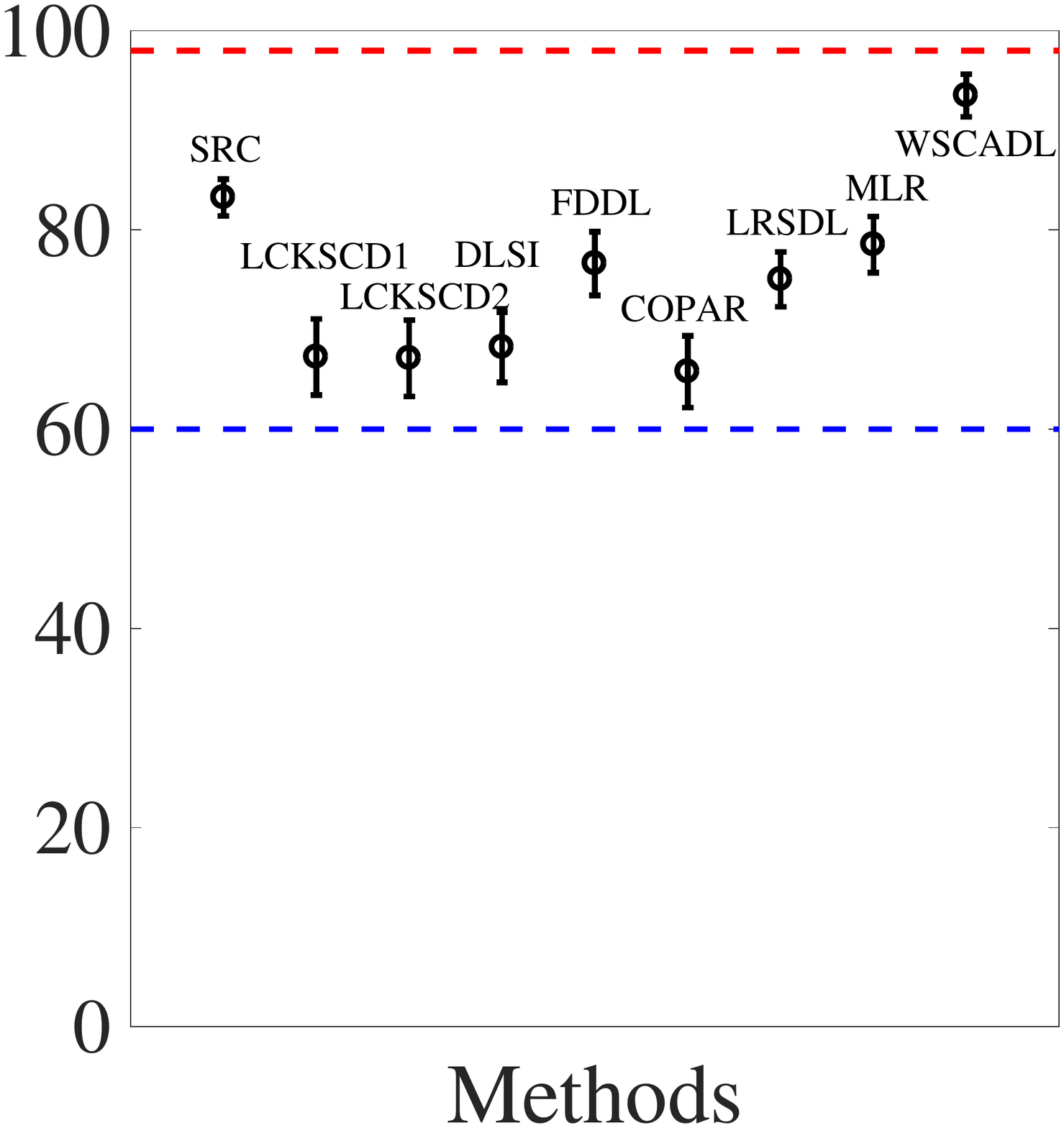}}
  \caption{Top 3}
    \end{subfigure}
     \begin{subfigure}[b]{0.49\columnwidth}
  \centering
   \resizebox{1\textwidth}{!}{
  \includegraphics{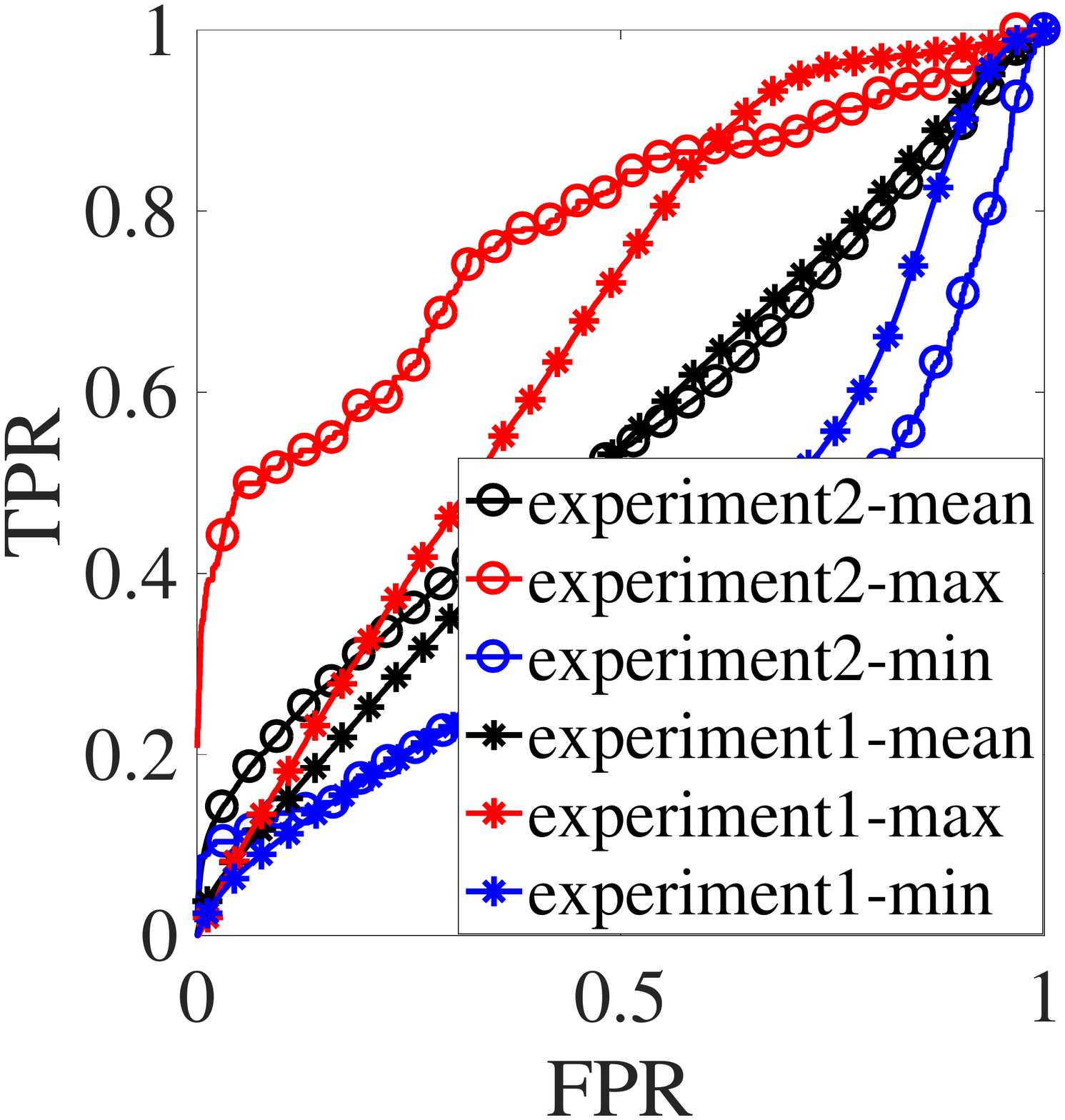}}
  \caption{Instance level ROC}
    \end{subfigure}
  \begin{subfigure}[b]{0.49\columnwidth}
  \centering
   \resizebox{1\textwidth}{!}{
  \includegraphics{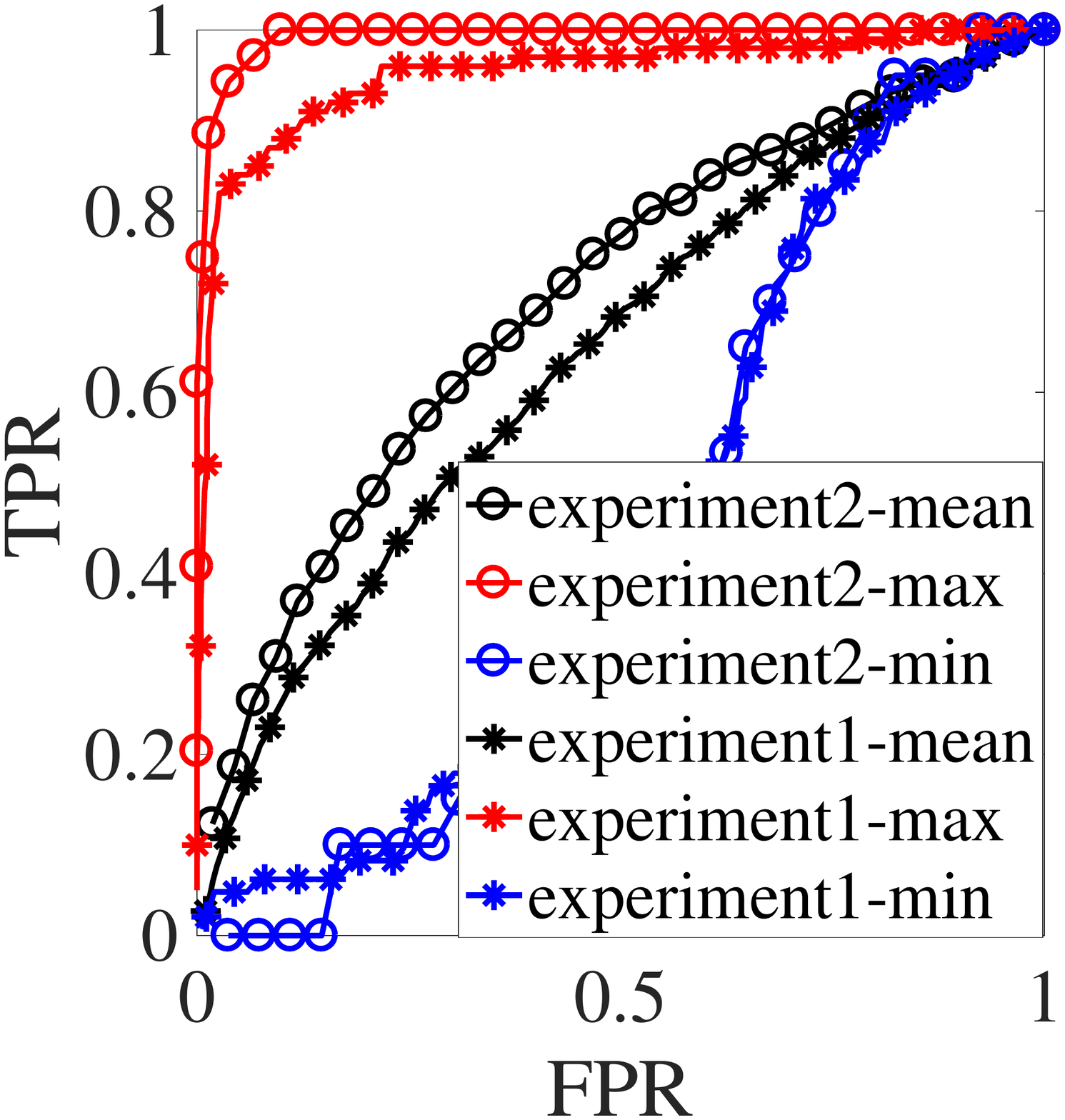}}
  \caption{signal ROC}
    \end{subfigure}
  \caption{Classification accuracy (\%) for the office live training data with mean and standard deviation over 5 MC runs with (a) selecting top 1 class and (b) selecting top 3 classes. Detection ROCs for (c) time instance level and (d) signal of both experiments.} \vspace{-0.1in}
  \label{fig:aaspresult}
 \end{figure}

Figure~\ref{fig:aaspresult}(a) shows that our proposed method outperforms other discriminative dictionary learning approaches except SRC. Additionally, our approach outperforms all others on predicting whether the true class is among the ranked three classes as shown in Figure~\ref{fig:aaspresult}(b). The instance and signal label detection receiver operating curves (ROCs) for the proposed method are shown in Figure~\ref{fig:aaspresult}(c) and (d), and the resulting AUCs are shown in Table~\ref{table:aasp_aucs}). 
The average and maximum detection AUCs across $16$ classes for instance and signal are slightly higher in experiment $2$ than experiment $1$, while the minimum AUCs are lower. The detection AUC for the best performing class in experiment $2$ is close to $100\%$, which indicates that WSCADL is able to discover that class perfectly for each test recording. Moreover, the potential of the proposed approach is demonstrated using experiment 2, in which only weak-supervision is provided. Despite this limiting setting, the average AUCs in experiment 2 are comparable or higher than the average AUCs reported in experiment 1 in which a single label per example is provided. This illustrates the potential in the label-economic weak-supervision setting and the potential of the proposed approach under this setting.

\noindent {\bf HJA bioacoustic classification:}
We compared our proposed WSCADL approach with the GDL-LR approach that both are dictionary learning based approaches, and with methods that perform segmentation and multi-instance multi-label (MIML) leaning approaches: MLR~\cite{pham2017dynamic}, SIM~\cite{briggs2012acoustic}, MIMLfast (short for Mfast)~\cite{huang2013fast}, and LSB-CMM (short for LSB)~\cite{liu2012conditional}. MIMLSVM (short for M-SVM)~\cite{zhou2006multi} and MIMLNN (short for M-NN)~\cite{zhou2012multi,zhang2007multi}.

We evaluated all of the approaches using multi-label evaluation metrics from~\cite{zhou2012multi}. 
\anothercolor{The results indicate that the proposed WSCADL approach outperforms GDL-LR for all metrics considered. Additionally, the proposed approach shows a slight advantage in terms of rank loss and average precision over the other MIML algorithms. For one error and Hamming loss the proposed approach is comparable in performance to the other MIML approaches. Our approach is outperformed by some of the alternative MIML approaches in terms of coverage.}
The results from column MLR to M-NN in Table~\ref{tab:hja_acc} are directly extracted from Table 4 in~\cite{pham2017dynamic}. 
\anothercolor{Note that the alternative MIML approaches from MLR to M-NN involve a process of converting spectrograms into a bag-of-words using segmentation and feature extraction for each segment while the proposed WSCADL approach and GDL-LR are directly applied to the raw spectrograms. We suspect that the disadvantage observed with the alternative MIML approaches (based on the bag-of-words representation) is due to error propagation from the segmentation and feature extraction steps, which are not jointly optimized for MIML classification.}

\section{conclusion}
\label{sec:conc}
In this work, we developed a novel probabilistic model that aims at learning a convolutive analysis dictionary under the weak-supervision setting. We incorporated cardinality constraints as observations to enforce sparsity of the signal label to determine the location of the patterns-of-interest from a given class. For the model parameter estimation, we developed the EM update rules and introduced novel chain and tree reformulations of the proposed graphical model to facilitate efficient probability calculations during the inference. In particular, under cardinality constraints that are expressed as a fraction of the signal length, we showed that the computational complexity for the chain reformulation is quadratic in the signal length and nearly-linear for the tree reformulation, which is verified in a numerical runtime comparison. 
As a sanity check, we demonstrated that the proposed discriminative approach performs comparably to a generative alternative on data that follows the generative paradigm. However, when the data follows a discriminative model, our approach outperforms the generative approach. Additionally, we showed that the proposed approach yields competitive and sometimes superior performance in terms of accuracy or AUC on real-world datasets when compared to either state-of-the-art approaches for dictionary learning or to alternative (i.e., non dictionary-based) solutions in the weak-supervision setting.

\appendices

\section{Derivation of auxiliary function}
 \label{append:aux}
 In the EM algorithm, the auxiliary function is given by the expectation of the complete data log-likelihood over the hidden variables and conditioned on the observed data. Therefore, the auxiliary function is formulated as:
 \[Q({\mathbold \theta}, {\mathbold \theta}^{i})=E_{{\bf H}|{\bf D};{\mathbold \theta}^{i},{\mathbold \phi}}[\log P({\mathcal D}, {\mathcal H} ; {\mathbold \theta}, {\mathbold \phi}))].\]
 Applying the natural logarithmic operation on the complete data likelihood, we have $\log P({\mathcal D}, {\mathcal H} ; {\mathbold \theta}, {\mathbold \phi})=\log P(\mathcal{X})+$
 \begin{eqnarray*}
   &&\hspace{-0.7cm} \sum_{n=1}^N  \log [\overbrace{\mathbb{I}_{(Y_{n}  = \cup_{t=-\Delta}^{T_n-1+\Delta} \{y_n(t)\})} +  \mathbb{I}_{(Y_{n} \cup \{0\} = \cup_{t=-\Delta}^{T_n-1+\Delta} \{y_n(t)\})}}^{P(Y_n|y_n)}]\\
   && \hspace{-0.7cm}+\log(\overbrace{\mathbb{I}_{(\sum_{t=-\Delta}^{T_n-1+\Delta} \mathbb{I}(y_{n}(t)\neq 0)\leq \bar N_n)})}^{P(I_n=1|y_n;\bar N_n)} +\hspace{-0.2cm}\sum_{t=-\Delta}^{T_n-1+\Delta} \hspace{-0.2cm}\log P (y_n(t) | x_{n},{\mathbold \theta}).
 \end{eqnarray*}
 Since the hidden data is only associated with each time instance label signal $y_1,\ldots,y_N$, the expectation of $P(I_n=1|y_n;\bar N_n)$ and $P(Y_n|y_n)$ are constant. Therefore the auxiliary function is computed as: 
 \[Q({\mathbold \theta}, {\mathbold \theta}^{i})\hspace{-0.1cm}=\hspace{-0.1cm}\sum_{n=1}^N\hspace{-0.1cm}\sum_{t=-\Delta}^{T_n-1+\Delta}\hspace{-0.2cm}E_{y_n(t)|{\bf D};{\mathbold \theta}^{i},{\mathbold \phi}}[\log P (y_n(t) | x_{n},{\mathbold \theta})]+ const.\]
 Since $\log P (y_n(t) | x_{n},{\bf w},{\bf b})={\mathbb I}_{(y_n(t)=c)}({\bf w}_{c}^T{\bf x}_{nt}+b_c)-\log (\sum_{u=0}^C e^{{\bf w}_{u}^T{\bf x}_{nt}+b_u})$,
 the final formulation of the auxiliary function is
 \begin{eqnarray*}
 Q({\mathbold \theta}, {\mathbold \theta}^{i})
 \hspace{-0.3cm}&=&\hspace{-0.3cm}\sum_{n=1}^N\sum_{t=-\Delta}^{T_n-1+\Delta} [\sum_{c=0}^C P(y_n(t)=c |{\bf D}; \bar N_n, {\mathbold \theta}^{i}) \cdot \\
 && {\bf w}_{c}^T{\bf x}_{nt}+b_c- \log (\sum_{u=0}^C e^{{\bf w}_{u}^T{\bf x}_{nt}+b_u})] + const.
 \end{eqnarray*}

 \section{Computational complexity analysis}
 \subsection{E-step chain inference}
%
%
 \vspace{0.1cm}
 \noindent {\bf Time complexity is (${\cal O}(\sum_{n=1}^N|Y_n|2^{|Y_n|}\bar N_nT_n)$):}
 In the chain inference with both forward and backward message passing, each update of forward and backward message requires running over all possible values of $y_n(t)$ and $({\mathbb L},l)$, therefore, the computational complexity is ${\cal O}((|Y_n|+1)2^{|Y_n|}\min(t, \bar N_n))$. Since each time step is only depend on the previous time step, the overall computational complexity is 
 \[{\cal T}^c_n(t) = {\cal T}^c_n(t-1) + {\cal O}((|Y_n|+1)2^{|Y_n|}\min(t, \bar N_n)).\]
 After solving this recursive formula, we have ${\cal T}^c_n = {\cal O}(|Y_n|2^{|Y_n|}\bar N_nT_n)$. Therefore, the overall chain inference needs a computational complexity of $\sum_{n=1}^N{\cal T}^c_n = {\cal O}(\sum_{n=1}^N |Y_n|2^{|Y_n|}\bar N_nT_n)$. {\bf The space complexity is (${\cal O}(2^{|Y_n|} T_n \bar N_n)$).}

 \subsection{E-step tree inference}
 \vspace{0.1cm}
 \noindent {\bf Time complexity is (${\cal O}(\sum_{n=1}^N 4^{|Y_n|}(\log_2 \bar N_n)^2T_n)$):}
 In the tree inference on both forward and backward message passing, each update of forward and backward message requires running over all possible values of $(Y^j_{n(2t-1)},N^j_{n(2t-1)})$ and $(Y^j_{n(2t)},N^j_{n(2t)})$, therefore, the computational complexity is ${\cal O}(4^{|Y_n|}(\min(\bar N_n, 2^{L-j})+1)^2)$.

However, the updates of the forward and backward messages on the tree for controlling cardinality parameter $\bar N_n$ are operating in a convolutive nature. When $\bar N_n$ is large, we rely on FFT and Inverse of FFT to speedup the convolution such that the convolution complexity will become ${\cal O}((\min(\bar N_n, 2^{L-j})+1)\log(\min(\bar N_n+1, 2^{L-j})+1))$. 
Since current instance on tree level $j$ only depend on previous two parents' at $j+1$, the recursive formula of the overall computational complexity is 
 \[{\cal T}^{\text{tr}(j)}_n(t) = {\cal T}^{\text{tr}(j+1)}_n(2t) +{\cal T}^{\text{tr}(j+1)}_n(2t-1)+{\cal O}(4^{|Y_n|}X\log X),\]
 where $X=\min(\bar N_n, 2^{L-j})+1$ and $1\leq t \leq 2^j, 1\leq j\leq L.$ 
 After solving this recursive formula, we have ${\cal T}^{\text{tr}}_n = {\cal O}(4^{|Y_n|}(\log_2 \bar N_n)^2T_n)$. The overall computational complexity of tree approach is $\sum_{n=1}^N {\cal T}^{\text{tr}}_n= \sum_{n=1}^N{\cal O}(4^{|Y_n|}(\log_2 \bar N_n)^2 T_n)$. 
{\bf The space complexity is (${\cal O}(2^{|Y_n|} T_n \log_2 \bar N_n)$).}
\ifCLASSOPTIONcaptionsoff
  \newpage
\fi

\bibliographystyle{IEEEtran}
\bibliography{WSCADL}
\vspace{-0.4in}
\begin{IEEEbiography}[{\includegraphics[width=1in,height=1.25in,clip,keepaspectratio]{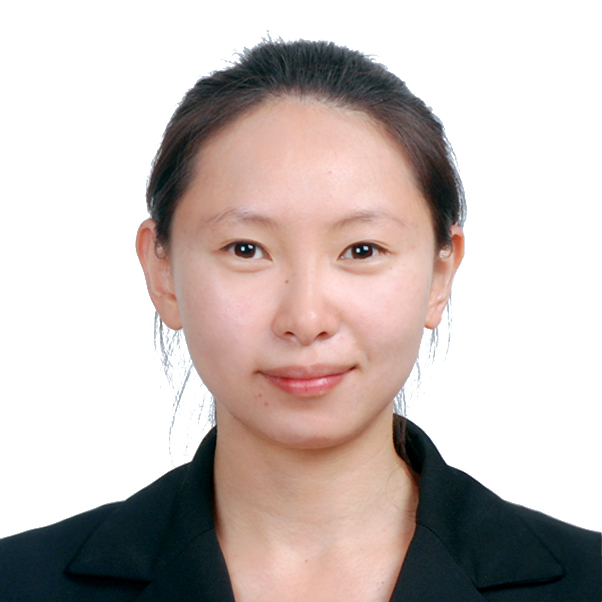}}]{Zeyu You}
Zeyu You received her B.S. in Electronics and Information Technology from Huazhong University of Science and Tech (2008) and her M.S. in Electrical and Computer Engineering from Oregon State University (2014). She is currently a Ph.D. Candidate in Electrical and Computer Engineering at Oregon State University, Corvallis, OR. Her current research interests include motif discovering, analysis dictionary learning and multi-instance multi-label learning, and applications to bioacoustics.
\end{IEEEbiography}
\vspace{-0.4in}
\begin{IEEEbiography}[{\includegraphics[width=1in,height=1.25in,clip,keepaspectratio]{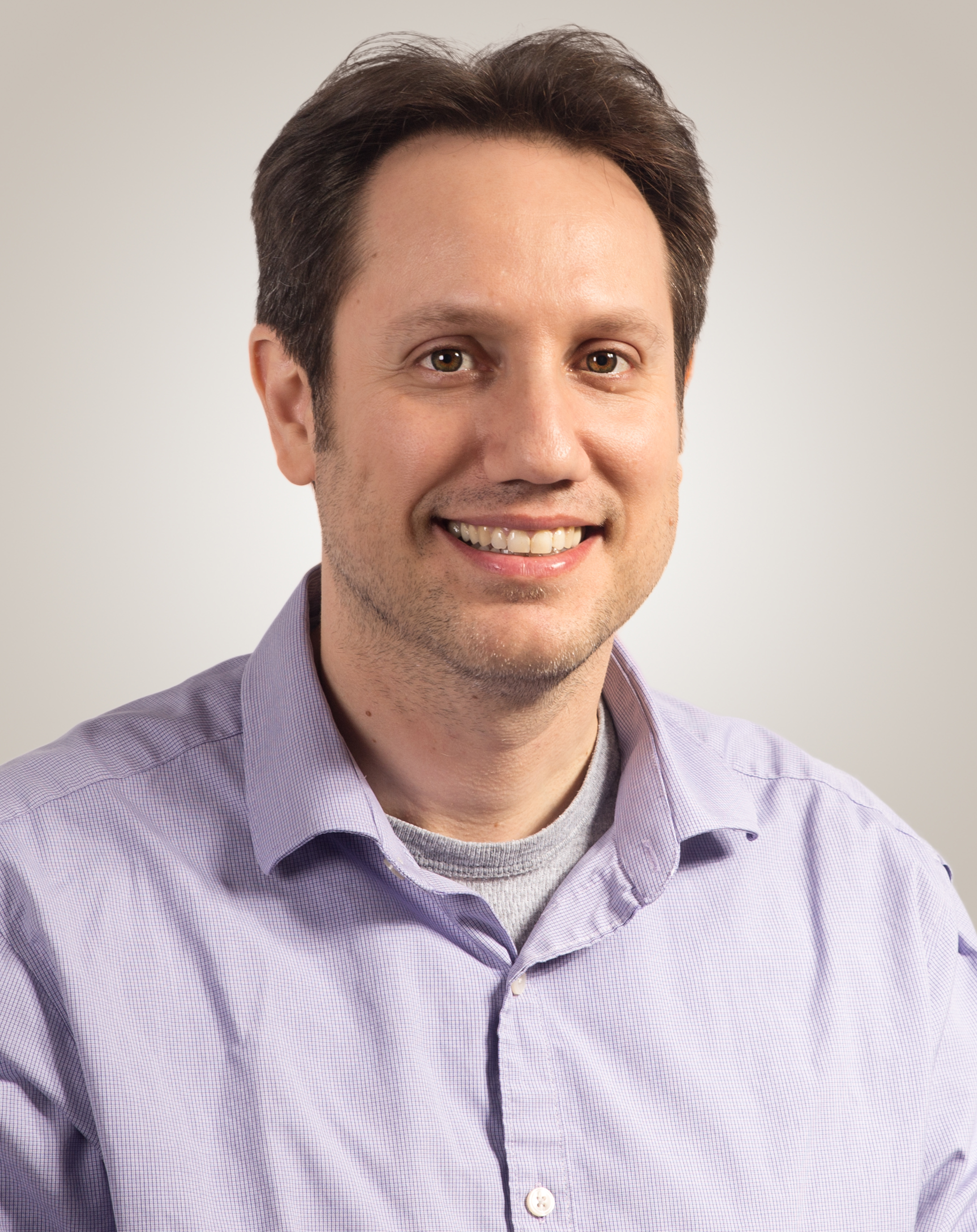}}]{Raviv Raich}
Raviv Raich (S’98–M’04-SM'17) received the B.Sc. and M.Sc. degrees from Tel Aviv University, Tel-Aviv, Israel, in 1994 and 1998, respectively, and the Ph.D. degree from the Georgia Institute of Technology, Atlanta, in 2004, all in electrical engineering. Between 1999 and 2000, he was a Researcher with the Communications Team, Industrial Research, Ltd., Wellington, New Zealand. From 2004 to 2007, he was a Postdoctoral Fellow with the University of Michigan, Ann Arbor. Dr. Raich has been with the School of Electrical Engineering and Computer Science, Oregon State University, Corvallis, as an Assistant Professor (2007-2013) and is currently an Associate Professor (2013-present). His research interests are in statistical signal processing and machine learning. Dr. Raich served as an Associate Editor for the IEEE Transactions On Signal Processing in 2011-2014. He currently serves as the chair of the Machine Learning for Signal Processing (MLSP) Technical Committee of the IEEE Signal Processing Society.
\end{IEEEbiography}
\vspace{-0.4in}
\begin{IEEEbiography}[{\includegraphics[width=1in,height=1.25in,clip,keepaspectratio]{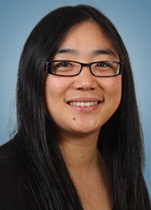}}]{Xiaoli Fern}
Xiaoli Fern received her Ph.D. degree in Computer Engineering from Purdue University, West Lafayette, IN, in 2005 and her M.S. degree from Shanghai Jiao Tong University (SJTU), Shanghai China in 2000. Since 2005, she has been with the School of Electrical Engineering and Computer Science, Oregon State University, Corvallis, OR where she is currently an associate professor. She received an NSF Career Award in 2011. Dr. Fern is currently serving as action editor for the Machine Learning Journal and regularly serves on the program committee for top-tier international conferences on machine learning and data mining including NIPS, ICML, ECML, AAAI, KDD, ICDM, SIAM SDM. Her general research interest is in the areas of machine learning and data mining.
\end{IEEEbiography}
\vspace{-0.4in}
\begin{IEEEbiography}[{\includegraphics[width=1in,height=1.25in,clip,keepaspectratio]{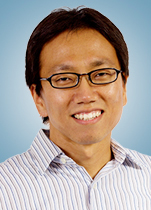}}]{Jinsub Kim}
Jinsub Kim received his Ph.D. in electrical and computer engineering from Cornell University. From September 2013 to June 2014, Jinsub Kim was a postdoctoral associate at the School of ECE at Cornell University. He joined the School of EECS in Oregon State University in August 2014.
His research interest spans statistical signal processing, statistical learning, optimization methods, and their applications to security and smart energy systems. 
\end{IEEEbiography}



\title{Supplementary Material of Weakly-supervised  Dictionary Learning }
\emptythanks
 \maketitle


 \section{Derivation of complete data likelihood}
 \label{append:llh}
 Given the observed data and the hidden data, we perform the complete data likelihood as:
 \begin{eqnarray*}
  && \hspace{-1cm} P({\mathcal D}, {\mathcal H} ; {\mathbold \theta}, {\mathbold \phi})=\\
  &&\hspace{-1cm}P(\mathcal{X},\mathcal{Y}, I_1=1,\ldots, I_N=1, y_1,\ldots, y_N; {\bf w},{\bf b},\bar N_1,\ldots, \bar N_N).
  \end{eqnarray*}
 Using the probability rule of $P(A,B)=P(A|B)P(B)$ and the independence assumption of each observed data point $(x_n,Y_n,I_n=1)$, the complete data likelihood can be further computed as:
  \[P({\mathcal D}, {\mathcal H} ; {\mathbold \theta}, {\mathbold \phi})=P(\mathcal{X}) \prod_{n=1}^N  P(Y_n,y_n, I_n =1| x_n; {\bf w},{\bf b}, \bar N_n).\]
  Apply the probabilistic graphic structure in Fig.~\ref{fig:gm}, we have
  \begin{eqnarray*}
     P({\mathcal D}, {\mathcal H} ; {\mathbold \theta}, {\mathbold \phi})=&\hspace{-0.3cm}P(\mathcal{X}) \prod_{n=1}^N P(I_n=1|y_n;\bar N_n) 
    P(Y_n|y_n)\\
    & ~~~~~~~~~~~~~~~~~~~~~~~\cdot P(y_n|x_n; {\bf w},{\bf b}).
  \end{eqnarray*}
  Plug in the model formulation in (2) 
  and (3) 
  and due to conditional independence assumption of the each time instance label, we arrive the final formulation of the complete data likelihood as:
 \begin{eqnarray*}
   &&\hspace{-0.7cm}P({\mathcal D}, {\mathcal H} ; {\mathbold \theta}, {\mathbold \phi})=\\
   &&\hspace{-0.3cm}P(\mathcal{X}) \prod_{n=1}^N  [\mathbb{I}_{(Y_{n}  = \cup_{t=-\Delta}^{T_n-1+\Delta} y_n(t))} +  \mathbb{I}_{(Y_{n} \cup \{0\} = \cup_{t=-\Delta}^{T_n-1+\Delta} y_n(t))}]\\
  && \hspace{-0.3cm}\mathbb{I}_{(\sum_{t=-\Delta}^{T_n-1+\Delta} \mathbb{I}(y_{n}(t)\neq 0)\leq \bar N_n)} \prod_{t=-\Delta}^{T_n-1+\Delta} P (y_n(t) | x_{n},{\bf w},{\bf b}).
 \end{eqnarray*}
 \section{Derivation of forward message passing on chain}
 \label{append:fmc}
 The derivation of the chain forward message passing is based on the definition of the forward message on the chain model $\alpha_t({\mathbb L},l)=P(Y_n^{t}={\mathbb L}, N_n^t=l|x_n; {\mathbold \theta}^i)$ and the marginal probability 
 \begin{eqnarray*}
 &P(Y_n^{t}={\mathbb L}, N_n^t=l|x_n; {\mathbold \theta}^i)=\sum_{Y_n^{t-1}}\sum_{N_n^{t-1}}\sum_{y_n(t)}\\
 &~~~~~~~P(Y_n^{t}={\mathbb L},N_n^t=l,Y_n^{t-1},N_n^{t-1},y_n(t)|x_n; {\mathbold \theta}^i).
 \end{eqnarray*}
 The forward message passing update rule can be formulated by marginalizing the previous state variables $(Y^{t-1}_n,N^{t-1}_n)$ and the current time instance $y_n(t)$. Rely on the v-structure on the update step of the chain structure in Fig. 4(a) 
 and the chain rule of the joint probability ($P(A,B,C)=P(A|B,C)P(B)P(C)$) such that 
 \begin{eqnarray*}
 &P(Y_n^{t}={\mathbb L},N_n^t=l,Y_n^{t-1},N_n^{t-1},y_n(t)|x_n; {\mathbold \theta}^i)\\
 &=P(Y_n^{t}={\mathbb L}, N_n^t=l|Y_n^{t-1},N_n^{t-1},y_n(t))\cdot\\
 &~~~P(y_n(t)|x_{nt};{\bf w}^i,{\bf b}^i)P(Y_n^{t-1},N_n^{t-1} | x_n; {\mathbold \theta}^i),
 \end{eqnarray*}
 we have
 \begin{eqnarray*}
 \alpha_t({\mathbb L},l)\hspace{-0.3cm}&=&\hspace{-0.3cm}\sum_{\mathbb A\in 2^{Y_n}}\sum_{a=0}^{t-1}\sum_{c=0}^C P(Y_n^{t}={\mathbb L}, N_n^t=l|Y_n^{t-1}={\mathbb A}\\
  && ,N_n^{t-1}=a,y_n(t)=c)P(y_n(t)=c|x_{nt};{\bf w}^i,{\bf b}^i)\\
  && ~~~P(Y_n^{t-1}={\mathbb A},N_n^{t-1}=a | x_n; {\mathbold \theta}^i).
  \end{eqnarray*}
 According to (9) 
 and (10) 
 , the conditional probability follows a deterministic rule such that $P(Y_n^{t}={\mathbb L}, N_n^t=l|Y_n^{t-1}={\mathbb A},N_n^{t-1}=a,y_n(t)=c)={\mathbb I}({\mathbb L}={\mathbb A}\cup \{c\}){\mathbb I}(l=a +  {\mathbb I}(c\neq 0))$. Therefore, the update rule of the forward message passing is:
  \begin{eqnarray*}
 \alpha_t({\mathbb L},l)\hspace{-0.3cm}&=&\hspace{-0.3cm}\sum_{\mathbb A\in 2^{Y_n}}\sum_{a=0}^{t-1}\sum_{c=0}^C {\mathbb I}({\mathbb L}={\mathbb A}\cup \{c\}){\mathbb I}(l=a +  {\mathbb I}(c\neq 0))\\
  && \cdot P(y_n(t)=c|x_{n};{\bf w}^i)\alpha_{t-1}({\mathbb A},a)\\
  \end{eqnarray*}
 Due to the constraints that ${\mathbb L}={\mathbb A}\cup \{c\}$ and $l=a +  {\mathbb I}(c\neq 0)$, for a fixed value of ${\mathbb L}$ and $l$, ${\mathbb A}$ and $a$ can only have one value for a particular class $c$. Thus the update rule of the forward message can be further simplified as:
  \begin{eqnarray*}
 \alpha_t({\mathbb L},l)\hspace{-0.3cm}&=&\hspace{-0.3cm}P(y_n(t)=0|x_{n};{\bf w}^i)\alpha_{t-1}({\mathbb L},l)\\
  && +\sum_{c=1}^C P(y_n(t)=c|x_{n};{\bf w}^i){\mathbb I}(l \neq 0)\\
  && [\alpha_{t-1}({\mathbb L},l-1) + {\mathbb I}(c\in {\mathbb L})\alpha_{t-1}({\mathbb L}_{\setminus c},l)].
 \end{eqnarray*}
 \section{Derivation of backward message passing on chain}
 \label{append:bmc}
 The derivation of the chain backward message passing is based on the definition of the backward message on the chain model $\beta_{t-1}({\mathbb L},l)=P(Y_n, I_n=1|Y_n^{t-1}={\mathbb L}, N_n^{t-1}=l, x_n; {\mathbold \theta}^i,\bar N_n)$ and the marginal probability
 \begin{eqnarray*}
 &P(Y_n, I_n=1| Y_n^{t-1}, N_n^{t-1}, x_n; {\mathbold \theta}^i,\bar N_n)=\sum_{Y_n^{t}}\sum_{N_n^{t}}\sum_{y_n(t)}\\
 &~~~~~~~P(Y_n, I_n=1,Y_n^{t},N_n^{t},y_n(t)| Y_n^{t-1},N_n^{t-1},x_n; {\mathbold \theta}^i,\bar N_n).
 \end{eqnarray*}
 Rely on the v-structure on the update step of the chain structure in Fig. 4(b) 
 and the chain rule of the conditional probability ($P(A,B|C)=P(A|B,C)P(B|C)$), we have 
 \begin{eqnarray*}
 &\hspace{-1cm}P(Y_n, I_n=1,Y_n^{t},N_n^{t},y_n(t)| Y_n^{t-1},N_n^{t-1},x_n; {\mathbold \theta}^i,\bar N_n)\\
 &=P(Y_n, I_n=1|Y_n^{t},N_n^{t},Y_n^{t-1},N_n^{t-1},y_n(t), x_n; {\mathbold \theta}^i,\bar N_n)\cdot\\
 &~~~P(Y_n^{t}, N_n^t|Y_n^{t-1},N_n^{t-1},y_n(t))P(y_n(t)|x_{nt};{\bf w}^i,{\bf b}^i).
 \end{eqnarray*}
 Given the current time joint state node $(Y_n^{t},N_n^{t})$, the observed node $(Y_n, I_n)$ is independent of the previous joint state node $(Y_n^{t-1},N_n^{t-1})$ and the current time instance $y_n(t)$, so $P(Y_n, I_n=1|Y_n^{t},N_n^{t},Y_n^{t-1},N_n^{t-1},y_n(t), x_n; {\mathbold \theta}^i,\bar N_n)=P(Y_n, I_n=1|Y_n^{t},N_n^{t}, x_n; {\mathbold \theta}^i,\bar N_n)$. Combining the above two equations, we obtain the update rule of the backward message passing as:
 \begin{eqnarray*}
  &&\hspace{-1cm}\beta_{t-1}({\mathbb L},l) \\
  &=&\hspace{-0.3cm}\sum_{\mathbb A\in 2^{Y_n}}\sum_{a=0}^t\sum_{c=0}^C P(Y_n, I_n=1 | Y_n^{t}={\mathbb A},N_n^{t}=a, {\bf X}_n; \bar N_n,{\bf w})\\
   && P(Y_n^{t}={\mathbb A}, N_n^{t}=a|Y_n^{(t-1)}={\mathbb L},N_n^{t-1}=l,y_n(t)=c)\\ 
   && P(y_n(t)=c|x_{n};{\bf w}^i,{\bf b}^i)
 \end{eqnarray*}
 Since $P(Y_n^{t}={\mathbb L}, N_n^t=l|Y_n^{t-1}={\mathbb A},N_n^{t-1}=a,y_n(t)=c)={\mathbb I}({\mathbb L}={\mathbb A}\cup \{c\}){\mathbb I}(l=a +  {\mathbb I}(c\neq 0))$ and each one of ${\mathbb A}, a$ is only limited to one value for a particular class $c$, therefore, the update rule of the forward message passing is:
   \begin{eqnarray*}
   &&\hspace{-1cm}\beta_{t-1}({\mathbb L},l) \\
 &=&\hspace{-0.3cm} \sum_{c=0}^C \beta_t({\mathbb L} \cup \{c\neq 0\}, l+{\mathbb I}_{(c\neq 0)})P(y_n(t)=c | x_{n},{\bf w}^i,{\bf b}^i).
 \end{eqnarray*}
 \section{Derivation of joint probability on chain}
 \label{append:jpc}
 To calculate the joint probability $P( {y}_{n}(t)=c, Y_n, I_n =1| x_n; {\mathbold \theta}^i, \bar N_n)$, we apply a conditional rule that 
 \begin{eqnarray*}
  && \hspace{-0.7cm}P( {y}_{n}(t)=c, Y_n, I_n =1| x_n; {\mathbold \theta}^i, \bar N_n)=\\
   &&\hspace{-0.6cm}P(Y_n, I_n=1 | y_n(t)=c,x_n; {\mathbold \theta}^i,\bar N_n)p(y_n(t)=c|x_{n};{\bf w}^i,{\bf b}^i).
   \end{eqnarray*}
 Once each time instance label $y_n(t)$ is known, the observed state node $(Y_n, I_n)$ is independent of the observed signal $x_n$ and parameter ${\mathbold \theta}$, so \[P( Y_n, I_n| {y}_{n}(t)=c, x_n; {\mathbold \theta}^i, \bar N_n)=P( Y_n, I_n| {y}_{n}(t)=c; \bar N_n).\]
 Since $P( Y_n, I_n| {y}_{n}(t)=c; \bar N_n)$ can be obtained by marginalizing out the joint state nodes of both $(Y_n^t, N_n^t)$ and $(Y_n^{t-1}, N_n^{t-1})$, 
 \begin{eqnarray*}
 &P( Y_n, I_n| {y}_{n}(t)=c; \bar N_n)=\sum_{Y_n^t}\sum_{N_n^t}\sum_{Y_n^{t-1}}\sum_{N_n^{t-1}}\\
 &~~~~~~P( Y_n, I_n,Y_n^t, N_n^t,Y_n^{t-1}, N_n^{t-1}| {y}_{n}(t)=c; \bar N_n) 
 \end{eqnarray*}
 Apply the chain rule of the conditional probability ($P(A,B|C)=P(A|B,C)P(B|C)$),
 \begin{eqnarray*}
 &P( Y_n, I_n,Y_n^t, N_n^t,Y_n^{t-1}, N_n^{t-1}| {y}_{n}(t)=c; \bar N_n)\\
 &~~~~~~=P(Y_n, I_n | Y_n^{t},N_n^{t}, Y_n^{t-1}, N_n^{t-1}, {y}_{n}(t), x_n; {\mathbold \theta}^i,\bar N_n)\\
 &~~~~~~P(Y_n^{t-1},N_n^{t-1} | x_n; {\mathbold \theta}^i)p(y_n(t)=c|x_{n};{\bf w}^i) 
 \end{eqnarray*}
 Given the current time joint state node $(Y_n^{t},N_n^{t})$, the observed node $(Y_n, I_n)$ is independent of the previous joint state node $(Y_n^{t-1},N_n^{t-1})$ and the current time instance $y_n(t)$, so $P(Y_n, I_n=1|Y_n^{t},N_n^{t},Y_n^{t-1},N_n^{t-1},y_n(t), x_n; {\mathbold \theta}^i,\bar N_n)=P(Y_n, I_n=1|Y_n^{t},N_n^{t}, x_n; {\mathbold \theta}^i,\bar N_n)=\beta_t(Y_n^{t},N_n^{t})$.
 Combining the above equations, applying the deterministic rule $P(Y_n^{t}={\mathbb L}, N_n^t=l|Y_n^{t-1}={\mathbb A},N_n^{t-1}=a,y_n(t)=c)={\mathbb I}({\mathbb L}={\mathbb A}\cup \{c\}){\mathbb I}(l=a +  {\mathbb I}(c\neq 0))$ and applying the definition of the forward message $P(Y_n^{t-1},N_n^{t-1} | x_n; {\mathbold \theta}^i)=\alpha_{t-1}(Y_n^{t-1},N_n^{t-1})$, the joint probability is performed as:
 \begin{eqnarray*}
  && \hspace{-0.8cm}P( {y}_{n}(t)=c, Y_n, I_n =1| x_n; {\mathbold \theta}^i, \bar N_n)\\
  &=&\hspace{-0.3cm}\sum_{\mathbb A\in 2^{Y_n}}\sum_{a=0}^{t-1}\sum_{\mathbb L\in 2^{Y_n}}\sum_{l=0}^t {\mathbb I}({\mathbb A}={\mathbb L}\cup \{c\}){\mathbb I}(a=l +  {\mathbb I}(c\neq 0))\\
  &&\hspace{-0.3cm} \alpha_{t-1}({\mathbb L},l)\beta_t({\mathbb A},a)\\
  &=&\hspace{-0.3cm}\sum_{\mathbb L\in 2^{Y_n}} \sum_{l=0}^{\bar N_n^*} \beta_t({\mathbb L} \cup \{c\neq 0\}, l+{\mathbb I}(c\neq 0)) \alpha_{t-1}({\mathbb L}, l))\\
 &&\hspace{-0.3cm}p(y_n(t)=c|x_{n};{\bf w}^i),
 \end{eqnarray*}
 where $\bar N_n^*=\min(\bar N_n-{\mathbb I}(c\neq 0), t)$.
 \section{Derivation of forward message passing on tree}
 \label{append:fmt}
 The forward message passing update on tree can be first applied with the definition of the forward message on tree $\alpha_t^{j-1}({\mathbb L},l)=P(Y_{nt}^{j-1}={\mathbb L}, N_{nt}^{j-1}=l| x_n; {\mathbold \theta}^i)$ and the marginal probability
 \begin{eqnarray*}
 &P(Y_{nt}^{j-1}, N_{nt}^{j-1}| x_n; {\mathbold \theta}^i)=\sum_{Y_{n(2t-1)}^{j}}\sum_{N_{n(2t-1)}^{j}}\sum_{Y_{n(2t)}^{j}}\sum_{N_{n(2t)}^{j}}\\
 &~~P(Y_{nt}^{j-1}, N_{nt}^{j-1},Y_{n(2t-1)}^{j},N_{n(2t-1)}^{j},Y_{n(2t)}^{j},N_{n(2t)}^{j}| x_n; {\mathbold \theta}^i)
 \end{eqnarray*}
 According to the v-structure of the update step in Fig. 6(a)
, the joint probability can be decomposed as:
 \begin{eqnarray*}
 &P(Y_{nt}^{j-1}, N_{nt}^{j-1},Y_{n(2t-1)}^{j},N_{n(2t-1)}^{j},Y_{n(2t)}^{j},N_{n(2t)}^{j}| x_n; {\mathbold \theta}^i)\\
 &= P(Y_{nt}^{j-1}, N_{nt}^{j-1}|Y_{n(2t-1)}^{j},N_{n(2t-1)}^{j},Y_{n(2t)}^{j},N_{n(2t)}^{j})\cdot\\
  &~~~~ P(Y_{n(2t-1)}^{j},N_{n(2t-1)}^{j} | x_n; {\mathbold \theta}^i)P(Y_{n(2t)}^{j},N_{n(2t)}^{j}| x_n; {\mathbold \theta}^i)
 \end{eqnarray*}
 Due to the deterministic rule between $(Y_{nt}^{j-1}, N_{nt}^{j-1})$ and $(Y_{n(2t-1)}^{j},N_{n(2t-1)}^{j}),(Y_{n(2t)}^{j},N_{n(2t)}^{j})$ as proposed in (15) 
and (16), 
$P(Y_{nt}^{j-1}, N_{nt}^{j-1}|Y_{n(2t-1)}^{j},N_{n(2t-1)}^{j},Y_{n(2t)}^{j},N_{n(2t)}^{j}) = {\mathbb I}(Y_{nt}^{j-1}=_{n(2t-1)}^{j}\cup Y_{n(2t)}^{j}){\mathbb I}(N_{nt}^{j-1}=N_{n(2t-1)}^{j} + N_{n(2t)}^{j})$. 
 Combining the above the equations, we obtain the update rule of the forward message passing on tree as:
 \begin{eqnarray*}
 \alpha^{j-1}_t({\mathbb L},l)\hspace{-0.3cm}&=&\hspace{-0.3cm}\sum_{\mathbb A\in 2^{Y_n}}\sum_{a=0}^{\bar N_n^{**}}\sum_{\mathbb E\in 2^{Y_n}}\sum_{e=0}^{\bar N_n^{**}}{\mathbb I}({\mathbb L}={\mathbb A}\cup {\mathbb E}){\mathbb I}(l=a + e)\\
  && \cdot \alpha^j_{2t-1}({\mathbb A},a)\alpha^j_{2t}({\mathbb E},e)\\
 \hspace{-0.3cm}&=&\hspace{-0.3cm} \sum_{\mathbb A\subseteq {\mathbb L}}\sum_{a=0}^l \alpha^j_{2t-1}({\mathbb A},a)\alpha^j_{2t}({\mathbb L}\setminus {\mathbb A},l-a),
 \end{eqnarray*}
 where $\bar N_n^{**}=\min(\bar N_b, 2^{L-j})+1$.
 \section{Derivation of backward message passing on tree}
 \label{append:bmt}
 Given the definition of the backward message on tree $\beta_{2t-1}^{j}({\mathbb A},a)=P(Y_n,I_n=1|Y_{n(2t-1)}^{j}={\mathbb A}, N_{n(2t-1)}^{j}=a,  x_n; {\mathbold \theta}^i,\bar N_n)$ and $\beta_{2t}^{j}({\mathbb E},e)=P(Y_n,I_n=1|Y_{n(2t)}^{j}={\mathbb E}, N_{n(2t)}^{j}=e,  x_n; {\mathbold \theta}^i,\bar N_n)$, the backward message passing update on tree can be derived based on marginal probabilities:
 \begin{eqnarray*}
 &\hspace{-2cm}P(Y_n,I_n=1|Y_{n(2t-1)}^{j}, N_{n(2t-1)}^{j},x_n; {\mathbold \theta}^i,\bar N_n)\\
 &~~=\hspace{-0.1cm}\sum_{Y_{n(2t)}^{j}}\sum_{N_{n(2t)}^{j}}\sum_{Y_{nt}^{j-1}}\sum_{N_{nt}^{j-1}}P(Y_n,I_n=1,Y_{n(2t)}^{j},\\
 &~~N_{n(2t)}^{j},Y_{nt}^{j-1},N_{nt}^{j-1}|Y_{n(2t-1)}^{j}, N_{n(2t-1)}^{j},  x_n; {\mathbold \theta}^i,\bar N_n)
 \end{eqnarray*}
 and 
 \begin{eqnarray*}
 &\hspace{-3.5cm}P(Y_n,I_n=1|Y_{n(2t)}^{j}, N_{n(2t)}^{j},  x_n; {\mathbold \theta}^i,\bar N_n)\\
 &~~=\hspace{-0.1cm}\sum_{Y_{n(2t-1)}^{j}}\hspace{-0.3cm}\sum_{N_{n(2t-1)}^{j}}\hspace{-0.3cm}\sum_{Y_{nt}^{j-1}}\hspace{-0.1cm}\sum_{N_{nt}^{j-1}}P(Y_n,I_n=1,Y_{n(2t-1)}^{j},\\
 &~~N_{n(2t-1)}^{j},Y_{nt}^{j-1},N_{nt}^{j-1}|Y_{n(2t)}^{j}, N_{n(2t)}^{j},  x_n; {\mathbold \theta}^i,\bar N_n).
 \end{eqnarray*}
 According to the v-structure of the update step in Fig. 6(b), 
 the joint probabilities can be decomposed as:
 \begin{eqnarray*}
 &\hspace{-2.8cm}P(Y_n,I_n=1,Y_{n(2t)}^{j},N_{n(2t)}^{j},Y_{nt}^{j-1},N_{nt}^{j-1}\\
 &|Y_{n(2t-1)}^{j}, N_{n(2t-1)}^{j},  x_n; {\mathbold \theta}^i,\bar N_n)\\
 &=P(Y_{nt}^{j-1},N_{nt}^{j-1}|Y_{n(2t)}^{j},N_{n(2t)}^{j},Y_{n(2t-1)}^{j}, N_{n(2t-1)}^{j})\\
 &P(Y_n,I_n=1|Y_{nt}^{j-1},N_{nt}^{j-1},  x_n; {\mathbold \theta}^i,\bar N_n)\\
 &P(Y_{n(2t)}^{j},N_{n(2t)}^{j}|x_n; {\mathbold \theta}^i)
 \end{eqnarray*}
 and
 \begin{eqnarray*}
 &\hspace{-2cm}P(Y_n,I_n=1,Y_{n(2t-1)}^{j},N_{n(2t-1)}^{j},Y_{nt}^{j-1},N_{nt}^{j-1}\\
 &|Y_{n(2t)}^{j}, N_{n(2t)}^{j},  x_n; {\mathbold \theta}^i,\bar N_n)\\
 &=P(Y_{nt}^{j-1},N_{nt}^{j-1}|Y_{n(2t)}^{j},N_{n(2t)}^{j},Y_{n(2t-1)}^{j}, N_{n(2t-1)}^{j})\\
 &P(Y_n,I_n=1|Y_{nt}^{j-1},N_{nt}^{j-1},  x_n; {\mathbold \theta}^i,\bar N_n)\\
 &P(Y_{n(2t-1)}^{j},N_{n(2t-1)}^{j}|x_n; {\mathbold \theta}^i).
 \end{eqnarray*}
 Due to the deterministic rule that
 \begin{eqnarray*}
 &\hspace{-2cm}P(Y_{nt}^{j-1}, N_{nt}^{j-1}|Y_{n(2t-1)}^{j},N_{n(2t-1)}^{j},Y_{n(2t)}^{j},N_{n(2t)}^{j}) \\
 &= {\mathbb I}(Y_{nt}^{j-1}=_{n(2t-1)}^{j}\cup Y_{n(2t)}^{j}){\mathbb I}(N_{nt}^{j-1}=N_{n(2t-1)}^{j} + N_{n(2t)}^{j}),
 \end{eqnarray*}
 we derive the update of the backward message passing update rule by combining the above equations as:
 \begin{eqnarray*}
 \beta^j_{2t-1}({\mathbb A},a) \hspace{-0.3cm}&=&\hspace{-0.3cm}\sum_{\mathbb L\in 2^{Y_n}}\sum_{l=0}^{\bar N_n^{**}} \sum_{\mathbb E\in 2^{Y_n}}\sum_{e=0}^{\bar N_n^{**}} {\mathbb I}({\mathbb L}={\mathbb A}\cup {\mathbb E}){\mathbb I}(l=a + e) \\
  && \beta^{j-1}_t({\mathbb L},l)\alpha^j_{2t}({\mathbb E},e)\\
 &=&\hspace{-0.3cm} \sum_{\mathbb E\in 2^{Y_n}}\sum_{e=0}^{\bar N_n^{**}} \beta^{j-1}_t({\mathbb A}\cup {\mathbb E},a+e)\alpha^j_{2t}({\mathbb E},e).
 \end{eqnarray*}
 and
 \begin{eqnarray*}
 \beta^j_{2t}({\mathbb E},e) \hspace{-0.3cm}&=&\hspace{-0.3cm}\sum_{\mathbb L\in 2^{Y_n}}\sum_{l=0}^{\bar N_n^{**}} \sum_{\mathbb E\in 2^{Y_n}}\sum_{e=0}^{\bar N_n^{**}} {\mathbb I}({\mathbb L}={\mathbb A}\cup {\mathbb E}){\mathbb I}(l=a + e)\\
  && \beta^{j-1}_t({\mathbb L},l)\alpha^j_{2t-1}({\mathbb A},a)\\
 &=&\hspace{-0.3cm} \sum_{\mathbb A\in 2^{Y_n}}\sum_{a=0}^{\bar N_n^{**}} \beta^{j-1}_t({\mathbb A}\cup {\mathbb E},a+e)\alpha^j_{2t-1}({\mathbb A},a).
 \end{eqnarray*}
 \section{AASP spectrograms in the training data}
%
%
 
 
 
%
 \begin{figure}[h]
   \centering
   \begin{subfigure}[b]{1\columnwidth}
   \centering
    \resizebox{0.9\textwidth}{!}{
   \includegraphics{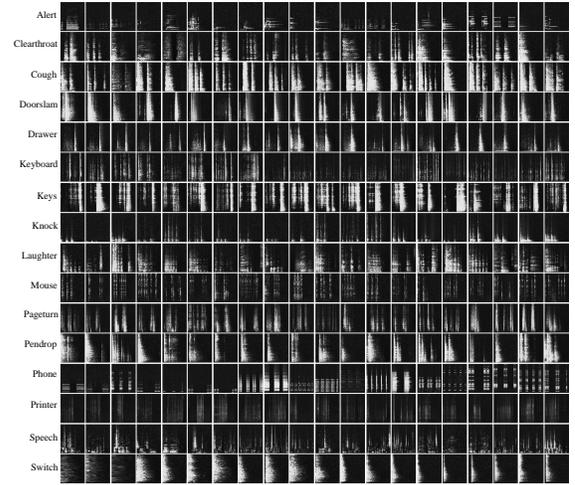}}
   \label{fig:aaspdata}
   \end{subfigure}
   \caption{Overview of Office live scene sound transformed spectrograms in AASP challenge}
  \end{figure}
%
 
 
 

\end{document}